\documentclass[a4paper,12pt]{article}
\pdfoutput=1 

\usepackage{jheppub}
\usepackage{graphicx}
\usepackage{axodraw}
\usepackage{epsfig}
\usepackage{amsmath, amsthm, amssymb}
\usepackage{amsfonts}
\usepackage{subfig}
\usepackage{color}
\usepackage{xspace}
\usepackage[dvipsnames]{xcolor}
\usepackage{paralist}
\usepackage{multirow}
\usepackage{paralist}

\newcommand{\beq}{\begin{equation}}
\newcommand{\eeq}{\end{equation}}
\newcommand{\bea}{\begin{eqnarray}}
\newcommand{\eea}{\end{eqnarray}}
\newcommand{\bfig}{\begin{figure}}
\newcommand{\efig}{\end{figure}}
\newcommand{\bc}{\begin{center}}
\newcommand{\ec}{\end{center}}

\newcommand{\eps}{\varepsilon}

\newcommand{\br}{{\rm BR}}

\newcommand{\lsim}{\lesssim}

\def\nuebar{{\rm \bar{\nu}_e}}
\def\nue{{\rm \nu_e}}

\def\s2tw{{\rm sin ^2 \theta_{W}}}

\def\sq2{\sqrt{2}}
\newcommand{\be}{\begin{equation}}
\newcommand{\ee}{\end{equation}}

\preprint{SMU-HEP-16-09}

\title{Neutrino Masses and Absence of Flavor Changing Interactions in the 2HDM from Gauge Principles}

\author[a]{Miguel D. Campos,}
\author[b,c]{D.\ Cogollo,}
\author[a]{Manfred Lindner,}
\author[d]{T. Melo,}
\author[a]{Farinaldo S.\ Queiroz,}
\author[a]{Werner Rodejohann}

\affiliation[a]{Max Planck Institut f\"ur Kernphysik, Saupfercheckweg 1, 69117 Heidelberg, Germany}

\affiliation[b]{Departamento de F\'isica, Universidade Federal de Campina Grande,
Caixa Postal 10071, 58109-970, Campina Grande, PB, Brazil}

\affiliation[c]{CFTP, Departamento de F\'{\i}sica, Instituto Superior T\'{e}cnico, Universidade de Lisboa}

\affiliation[d]{Departamento de F\'{\i}sica, Universidade Federal da Para\'{\i}ba, Jo\~ao Pessoa, Brazil}

\emailAdd{farinaldo.queiroz@mpi-hd.mpg.de}

\abstract{
	We propose several Two Higgs Doublet Models with the addition of an Abelian gauge group which  free the usual framework from flavor changing neutral interactions and explain neutrino masses through the seesaw mechanism. We discuss the kinetic and mass-mixing gripping phenomenology which encompass several constraints coming from atomic parity violation, the muon anomalous magnetic moment, rare meson decays, Higgs physics, LEP precision data, neutrino-electron scattering, low energy accelerators and LHC probes. }
	

\begin{document} 
\maketitle
\flushbottom

\clearpage
\section{Introduction}
\label{sec:1}

The discovery of a $125$~GeV spin-0 scalar announced by the ATLAS \cite{Aad:2012tfa} and CMS \cite{Chatrchyan:2012xdj} collaborations is a major triumph for the Standard Model (SM). The determination of the 
scalar sector of particle physics may however not be completed, as there are many extensions of the SM that 
require additional scalar particles, such as Higgs triplets, singlets or doublets. The $\rho$ parameter provides 
here a direct constraint on such models, and the value obtained from electroweak precision data of $\rho=1 \pm 0.0082$ \cite{Wang:2014wua} favors for instance small additional vacuum expectation values or scalar doublet 
representations with hypercharge $0,\pm 1$. In this paper we will 
study models with an additional Higgs doublet that has identical SM quantum numbers as the usual one. Such Two Higgs 
Doublet Models (2HDM) are in fact typical in a variety of SM extensions \cite{Branco:2011iw}. \\

The 2HDM framework has been proved to be a hospitable environment for axion models  \cite{Dasgupta:2013cwa,Mambrini:2015sia,Alves:2016bib}, baryogenesis \cite{Funakubo:1993jg,Davies:1994id,Cline:1995dg}, collider physics \cite{Aoki:2009ha,Bai:2012ex,Barger:2013ofa,Dumont:2014wha}, supersymmetry \cite{Haber:1984rc}, lepton flavor violation \cite{Lindner:2016bgg,Davidson:2016utf}, and flavor anomalies \cite{Misiak:2017bgg}, and a natural environment for new Abelian gauge groups \cite{Ko:2013zsa,Ko:2014uka,Crivellin:2015mga,Huang:2015wts,Wang:2016vfj}. Albeit, the 2HDM framework in its general form is plagued with Flavor Changing Neutral Interactions (FCNI). To cure this FCNI problem, 
an ad-hoc discrete symmetry is usually evoked. Furthermore, neutrino masses, one of the major observational evidences for new physics, are typically not addressed in 2HDM. \\

In this work we discuss a gauge solution to the FCNI problem which in addition naturally can incorporate Majorana 
neutrino masses. The idea is to add a gauged Abelian $U(1)_X$ symmetry to the 2HDM and find anomaly-free models that 
effectively lead to the usual 2HDM classes that have no FCNI. 
Anomaly-free models are also possible when right-handed neutrinos are added to the particle content. Their mass terms generate Majorana masses for the light neutrinos.  
Tracing the absence of dangerous flavor physics and the presence of neutrino masses to the same anomaly-free 
gauge origin is an attractive approach within 2HDM that deserves careful study. A whole class of models is generated by the 
idea. A new vector gauge boson that has mass and kinetic-mixing with the SM $Z$ boson is present, and 
we investigate its phenomenology in a limit which resembles often studied dark photon models. 
In particular, we address several constraints coming from low energy as well as high energy probes, including atomic parity violation, the muon anomalous magnetic moment, electron-neutrino scattering, and new physics searches at the LHC and several other MeV-GeV colliders such as BaBar.\\ 

Our work is structured as follows: In Section \ref{sec:2HDM}, we shortly review the 2HDM framework before we augment it in 
Section \ref{sec:U(1)X} with gauged Abelian symmetries and study the constraints on the models from anomaly 
cancellation, including right-handed neutrinos. In Section \ref{sec:pheno} the models are confronted with 
various phenomenological constraints before we conclude in Section \ref{sec:concl}. Some details are delegated to appendices. \\


\section{The 2HDM Framework\label{sec:2HDM}}

In the Standard Model, one scalar doublet accounts for the masses of all charged fermions and gauge bosons. However, extended scalar sectors are also possible. Among the various constraints on such cases, the $\rho$ parameter is particularly 
well constrained by electroweak precision data \cite{Olive:2016xmw}; it is defined as,
\begin{equation}
\rho = \frac{{\displaystyle \sum_{i=1}^n} \left[
I_i \left( I_i+1 \right) - \frac{1}{4}\, Y_i^2 \right] v_i}
{{\displaystyle \sum_{i=1}^n}\, \frac{1}{2}\, Y_i^2 v_i}. 
\label{jduei}
\end{equation} 
Here $I_i$ and $Y_i$ are the isospin and hypercharge of a scalar representation with vev $v_i$. The value 
$\rho=1$ is not altered by the addition of scalar doublets under $SU(2)$ with hypercharge $Y=\pm 1$, or scalar singlets with $Y=0$.  
Therefore, enlarging the Standard Model with a scalar doublet under $SU(2)$ is a natural and popular framework, the 
so-called the Two Higgs Doublet Model (2HDM) \cite{Lee:1973iz}. \\

In the 2HDM the most general potential for two doublets with hypercharge $Y = 1$, gauge invariant and renormalizable, is given by, 
\begin{equation}
\begin{split}
V \left( \Phi _1 , \Phi _2 \right)  = &  m_{11} ^2 \Phi _1 ^\dagger \Phi _1 + m_{22} ^2 \Phi _2 ^\dagger \Phi _2 - \left( m_{12} ^2 \Phi _1 ^\dagger \Phi _2 + h.c. \right) + \frac{\lambda _1}{2} \left( \Phi _1 ^\dagger \Phi _1 \right) ^2\\
& +  \frac{\lambda _2}{2} \left( \Phi _2 ^\dagger \Phi _2 \right) ^2 + \lambda _3 \left( \Phi _1 ^\dagger \Phi _1 \right) \left( \Phi _2 ^\dagger \Phi _2 \right) + \lambda _4 \left( \Phi _1 ^\dagger \Phi _2 \right) \left( \Phi _2 ^\dagger \Phi _1 \right)\\
& + \left[ \frac{\lambda _5}{2} \left( \Phi _1 ^\dagger \Phi _2 \right) ^2 + \lambda _6 \left( \Phi _1 ^\dagger \Phi _1 \right) \left( \Phi _1 ^\dagger \Phi _2 \right) + \lambda _7 \left( \Phi _2 ^\dagger \Phi _2 \right) \left( \Phi _1 ^\dagger \Phi _2 \right) + h.c. \right].
\end{split}
\label{generalpotential}
\end{equation}
The Yukawa Lagrangian reads 
\begin{equation}
\begin{split}
- \mathcal{L} _{Y _{\text{2HDM}}} & = y ^{1d} \bar{Q} _L \Phi _1 d_R + y ^{1u} \bar{Q} _L \widetilde \Phi _1 u_R + y ^{1e} \bar{L} _L \Phi _1 e_R  \\
& + y ^{2d} \bar{Q} _L \Phi _2 d_R + y ^{2u} \bar{Q} _L \widetilde \Phi _2 u_R + y ^{2e} \bar{L} _L \Phi _2 e_R + h.c., 
\end{split}
\label{generalyukawa}
\end{equation}
where
\begin{equation}
\Phi _i = \begin{pmatrix} \phi ^+ _i \\ \left( v_i + \rho _i + i\eta _i \right)/ \sqrt{2}\end{pmatrix}.
\end{equation}

Having two Higgs doublets generating masses for all fermions leads in general to the presence of FCNI at tree level, subjecting the model to tight bounds from flavor probes \cite{Atwood:1996vj}. The easy solution \cite{Paschos:1976ay,Glashow:1976nt}  
to this issue is the evocation of an ad-hoc $Z_2$ symmetry, where in particular,
\begin{eqnarray}
\Phi _1 & \rightarrow - \Phi _1,~
\Phi _2 & \rightarrow + \Phi _2,
\label{z2parity}
\end{eqnarray}
also known as the Natural Flavor Conservation (NFC) criterion. \\

Assuming CP conservation, the transformations in Eq.\ \eqref{z2parity} yield a new scalar potential,
\begin{equation}
\begin{split}
\label{pot_2hdm_z2}
V \left( \Phi _1 , \Phi _2 \right) &= m_{11} ^2 \Phi _1 ^\dagger \Phi _1 + m_{22} ^2 \Phi _2 ^\dagger \Phi _2 - m_{12} ^2 \left( \Phi _1 ^\dagger \Phi _2 + \Phi _2 ^\dagger \Phi _1 \right) + \frac{\lambda _1}{2} \left( \Phi _1 ^\dagger \Phi _1 \right) ^2  \\
&+ \frac{\lambda _2}{2} \left( \Phi _2 ^\dagger \Phi _2 \right) ^2 + \lambda _3 \left( \Phi _1 ^\dagger \Phi _1 \right) \left( \Phi _2 ^\dagger \Phi _2 \right) + \lambda _4 \left( \Phi _1 ^\dagger \Phi _2 \right) \left( \Phi _2 ^\dagger \Phi _1 \right)  \\
&+ \frac{\lambda _5}{2} \left[ \left( \Phi _1 ^\dagger \Phi _2 \right) ^2 + \left( \Phi _2 ^\dagger \Phi _1 \right) ^2 \right] .
\end{split}
\end{equation} 
Here the $m_{12}$ term softly violates the condition in Eq.\ \eqref{z2parity}, in order to avoid domain walls. The discrete symmetry Eq.\ \eqref{z2parity} will eliminate some of the terms in the general Yukawa Lagrangian Eq.\ \eqref{generalyukawa} avoiding also the FCNI. Which terms will be eliminated depends upon the parity assignment of the fermions under $Z_{2}$. We can, for example, consider that all fermions are even under $Z_{2}$ transformation. In this case $\mathcal{L} _{Y _{\text{2HDM}}}$ becomes,
\begin{equation}
\label{2hdm_tipoI}
- \mathcal{L} _{Y _{\text{2HDM}}} = y_2 ^d \bar{Q} _L \Phi _2 d_R + y_2 ^u \bar{Q} _L \widetilde \Phi _2 u_R + y_2 ^e \bar{L} _L \Phi _2 e_R + h.c., 
\end{equation}
with only $\Phi _2$ coupling to fermions. This is the Type I 2HDM. 
Other choices of fermion parities are presented in Table \ref{tipos_2hdm_paridades};  the four 2HDM shown in the table are subject to different phenomenologies and constraints (see \cite{Branco:2011iw} for a review). \\

After this short summary of the general 2HDM framework, we will discuss how to base those 
flavor-safe models on a gauged $U(1)_X$.  

\begin{table}[t]
\centering
\begin{tabular*}{\columnwidth}{@{\extracolsep{\fill}}lllllllll@{}}
\hline 
\multicolumn{1}{@{}l}
\ \ \ \ \ \    Model        & $\Phi_1$ & $\Phi_2$ &$u_R$ & $d_R$ & $e_R$ & $Q_L$ & $L_{L}$ \\ \hline 
\ \ \ \ \ \    Type I        &    $-$   &    $+$   &  $+$ &  $+$  &  $+$  &  $+$  &   $+$   \\
\ \ \ \ \ \    Type II       &    $-$   &    $+$   &  $+$ &  $-$  &  $-$  &  $+$  &   $+$   \\
\ \ \ \ \ \    Lepton-specific        &    $-$   &    $+$   &  $+$ &  $+$  &  $-$  &  $+$  &   $+$   \\
\ \ \ \ \ \    Flipped      &    $-$   &    $+$   &  $+$ &  $-$  &  $+$  &  $+$  &   $+$   \\ \hline
\end{tabular*}
\caption{{\footnotesize  Different types of 2HDM according to the $Z_2$ parities of the SM fermions. In the type I only $\Phi_2$ couples to all SM fermions; In the type II $\Phi_2$ couples to up quarks and $\Phi_1$ couples to leptons and down quarks. In the third type of 2HDM, also known as lepton-specific, $\Phi_1$ couples to leptons while $\Phi_2$ couples to quarks. Lastly, in the fourth type, called flipped 2HDM, the scalar doublet $\Phi_1$ couples to down quarks while $\Phi_2$ couples to leptons and up quarks.}}
\label{tipos_2hdm_paridades}
\end{table}

\section{\label{sec:U(1)X}2HDM with $U(1)_X$ Symmetries}

A fundamental solution to the flavor problem in the 2HDM could come from well-established gauge principles. 
It is known that an Abelian gauge symmetry when spontaneously broken gives rise to a discrete symmetry, simply because the latter  is a subgroup of the former. The quantum numbers of the particles charged under the new $U(1)_X$ symmetry will dictate what is the remnant symmetry. It has been shown that the necessary $Z_2$ symmetry to cure 2HDM from FCNI can be generated from gauge principles \cite{Ko:2012hd} under certain conditions. In what follows we will review these conditions using general arguments and address the implications.

\subsection{Anomaly Cancellation}

In order to truly prevent FCNI in the 2HDM, and mimic the effect of the $Z_2$ symmetry at lower energies, the scalar doublets $\Phi_1$ and $\Phi_2$ have to transform differently under $U(1)_X$, reducing the scalar potential to 
\begin{equation}
\begin{split}
V \left( \Phi _1 , \Phi _2 \right) &= m_{11} ^2 \Phi _1 ^\dagger \Phi _1 + m_{22} ^2 \Phi _2 ^\dagger \Phi _2 + \frac{\lambda _1}{2} \left( \Phi _1 ^\dagger \Phi _1 \right) ^2 + \frac{\lambda _2}{2} \left( \Phi _2 ^\dagger \Phi _2 \right) ^2  \\
&+ \lambda _3 \left( \Phi _1 ^\dagger \Phi _1 \right) \left( \Phi _2 ^\dagger \Phi _2 \right) + \lambda _4 \left( \Phi _1 ^\dagger \Phi _2 \right) \left( \Phi _2 ^\dagger \Phi _1 \right) .
\end{split}
\label{pot_2hdm_U1}
\end{equation}

In addition, one needs to successfully generate fermion masses by properly choosing the transformations of the 
fermions under the $U(1)_X$ symmetry. The requirement that the scalar doublets transform differently still leaves enough freedom to construct several models, based on the specific charge assignments for the Standard Model particles. We shall see what kind models one can build using simply gauge invariance and anomaly cancellation.\\

Generally speaking, a local transformation shifts the fields as follows,
\begin{equation}
\begin{split}
\label{transf_u1}
L _L & \rightarrow L ' _L = e ^{i l \alpha (x)} L _L \\
Q _L & \rightarrow Q ' _L = e ^{i q \alpha (x)} Q _L \\
e _R & \rightarrow e ' _R = e ^{i e \alpha (x)} e _R \\
u _R & \rightarrow u ' _R = e ^{i u \alpha (x)} u _R \\
d _R & \rightarrow d ' _R = e ^{i d \alpha (x)} d _R \\
\Phi _1 & \rightarrow \Phi _1 ' = e ^{i h_1 \alpha (x)} \Phi _1 \\
\Phi _2 & \rightarrow \Phi _2 ' = e ^{i h_2 \alpha (x)} \Phi _2 ,
\end{split}
\end{equation}
where $l, q, e, u, d, h_1, h_2$ are the charges of the fields under $U(1)_{X}$. Once we write down a Yukawa Lagrangian and demand gauge invariance, the transformations in Eq.\ \eqref{transf_u1} are no longer arbitrary, and the charges under $U(1)_X$ will be interconnected. In the Type I 2HDM, on which we focus in this paper, where fermions couple only with $\Phi_2$, see 
Eq.\ (\ref{2hdm_tipoI}), the following $U(1)_{X}$ transformations apply:
\begin{equation}
\begin{split}
\label{2hdm_tipoI_u1}
-\mathcal{L} _{Y _{\text{2HDM}}} \rightarrow -\mathcal{L} _{Y _{\text{2HDM}}} ' &= e^{(- q + h_2 + d)i\alpha} y_2 ^d \bar{Q} _L \Phi _2 d_R + e^{(- q - h_2 + u)i\alpha} y_2 ^u \bar{Q} _L \widetilde \Phi _2 u_R \\
&+ e^{(- l + h_2 + e)i\alpha} y_2 ^e \bar{L} _L \Phi _2 e_R + h.c.
\end{split}
\end{equation}
The $U(1)_{X}$ invariance imposes the following conditions on the charges of the fields:
\begin{equation}
\begin{split}
\label{cargas_yukawa}
d - q + h_2 &= 0 \\
u - q - h_2 &= 0 \\
e - l + h_2 &= 0 .
\end{split}
\end{equation}
Notice that in this case couplings of fermions with $\Phi _1$ are forbidden by the $U(1)_{X}$ symmetry. These couplings would be allowed only if $h_1$ satisfies the same equations (\ref{cargas_yukawa}) as $h_2$, implying that  $h_1 = h_2$. However, since we require that $h_1 \neq h_2$, there is no value of $h_1$ satisfying these equations. Besides the constraints in Eqs.\ \eqref{cargas_yukawa}, anomaly freedom must also be respected.  The general constraints for an anomaly free $U(1)_{X}$ gauge symmetry are discussed in Appendix \ref{sec:app1}. It turns out that for the Type I 2HDM, the anomaly  cancellation can be achieved without addition of new fermions whenever the condition $u=-2d$ is respected.  To see this, we combine Equation \eqref{anomalycond2} ($l = -3q$) with the constraints from \eqref{cargas_yukawa} and write the charges of the fields as function of $u$ and $d$, 
\begin{equation}
\begin{split}
\label{expres_cargas_u_d}
&q = \frac{\left( u + d \right)}{2}, \\
l &= \frac{ - 3 \left( u + d \right)}{2}, \\
e &= - \left( 2 u + d \right), \\
&h_2 = \frac{\left( u - d \right)}{2} .
\end{split}
\end{equation}
It is then straightforward to prove that these charge assignments in Eq.\ \eqref{expres_cargas_u_d} satisfy the anomaly  conditions Eqs.\ \eqref{anomalycond1}-\eqref{anomalycond4}. However, for the cancellation of the $ \left[ U(1)_X \right] ^3 $ term, Eq.\ \eqref{anomalycond5}, we find,
\begin{equation}
\begin{split}
e ^3 + 3 u ^3 + 3 d ^3 - 2 l ^3 - 6 q ^3 &= \left[ - \left( 2 u + d \right) \right] ^3 + 3 u ^3 + 3 d ^3 - 2 \left[ \frac{ - 3 \left( u + d \right)}{2} \right] ^3  \\
& -6 \left[ \frac{\left( u + d \right)}{2} \right] ^3 \\
&= - \left( 2 u + d \right) ^3 + 3 u ^3 + 3 d ^3 + 6 \left( u + d \right) ^3 \\
&= u ^3 + 8 d ^3 + 6 u ^2 d + 12 u d ^2 \\
&= \left( u + 2d \right) ^3 .
\end{split}
\end{equation}
This anomaly is not canceled unless $u = -2d$, i.e.\ if the up and down quark charges under $U(1)_X$ are proportional  to their  electric ones.\\ 

Here is the point at which neutrino physics can enter: if we decide to keep $u$ and $d$ arbitrary, the most straightforward 
possibility is to add right-handed neutrinos (one per generation). If their charge $n$ is given by 
\begin{equation}
\label{carg_neutrino_right}
n = - \left( u + 2d \right) ,
\end{equation}
the $ \left[ U(1)_X \right] ^3 $ anomaly term is canceled because Eq.\ (\ref{anomalycond5}) becomes
\begin{equation}
n ^3 + e ^3 + 3 u ^3 + 3 d ^3 - 2 l ^3 - 6 q ^3 = - \left( u + 2d \right) ^3 + \left( u + 2d \right) ^3 = 0.
\end{equation}
Concerning the  $\Phi_1$ charge under $U(1)_X$, we have only demanded so far that $h_1 \neq h_2$ to respect the NFC criterion, and no relation between $h_1$ and $h_2$ exist. By adding a singlet scalar to generate a Majorana mass term for the neutrinos, necessary for the implementation of the seesaw mechanism, the values of $h_1$ and $h_2$ are no longer independent, as we will see next. \\

\begin{table}[!t]
\centering
{\color{blue} \bf Two Higgs Doublet Models free from FCNI}
\begin{tabular}{ccccccccc}
\hline 
Fields & $u_R$ & $d_R$ & $Q_L$ & $L_L$ & $e_R$ & $N_R$ & $\Phi _2$  & $\Phi_1$ \\ \hline 
Charges & $u$ & $d$ & $\frac{(u+d)}{2}$ & $\frac{-3(u+d)}{2}$ & $-(2u+d)$ & $-(u+2d)$ & $\frac{(u-d)}{2}$ & $\frac{5u}{2} +\frac{7d}{2}$  \\
$U(1)_{A}$ & $1$ & $-1$ & $0$ & $0$ & $-1$ & $1$ & $1$ & $-1$\\
$U(1)_{B}$ & $-1$ & $1$ & $0$ & $0$ & $1$ & $-1$ & $-1$ & $1$\\
$U(1)_{C}$ & $1/2$ & $-1$ & $-1/4$ & $3/4$ & $0$ & $3/2$ & $3/4$ & $9/4$\\
$U(1)_{D}$ & $1$ & $0$ & $1/2$ & $-3/2$ & $-2$ & $-1$ & $1/2$ & $5/2$\\ 
$U(1)_{E}$ & $0$ & $1$ & $1/2$ & $-3/2$ & $-1$ & $-2$ & $7/2$ & $-1/2$\\
$U(1)_{F}$ & $4/3$ & $2/3$ & $1$ & $-3$ & $-4$ & $-8/3$ & $1/3$ & $17/3$\\
$U(1)_{G}$ & $-1/3$ & $2/3$ & $1/6$ & $-1/2$ & $0$ & $-1$ & $-1/2$ & $-3/2$\\
$U(1)_{B-L}$ & $1/3$ & $1/3$ & $1/3$ & $-1$ & $-1$ & $-1$ & $0$ & $2$\\
\hline \hline
$U(1)_{Y}$ & $2/3$ & $-1/3$ & $1/6$ & $-1/2$ & $-1$ & $$ & $1/2$ & $\neq h_2$ \\
$U(1)_{N}$ & $0$ & $0$ & $0$ & $0$ & $0$ & $$ & $0$  &  $\neq h_2$\\
 \hline
\end{tabular}
\caption{The first block of models are capable of explaining neutrino masses and the absence of flavor changing interactions in the 2HDM type I, whereas the second block refer to models where only the flavor problem is addressed. The first block accounts for type I 2HDM in which right-handed neutrinos are introduced without spoiling the NFC criterion ($h_1 \neq h_2$). This is possible when $u \neq -2d$ (see Eq.\ \eqref{expres_cargas_u_d} and Eq.\ \eqref{h1charge}). Conversely, the second block shows Type I 2HDM with $u=-2d$. To preserve the NFC criterion, right-handed neutrinos can not be introduced while at the same time $h_1$ is kept as a free parameter. The $U(1)_N$ models leads to a fermiophobic $Z^{\prime}$ setup \cite{Alves:2013tqa}. The $U(1)_Y$ yields a "right-handed-neutrino-phobic" $Z^{\prime}$ boson. The $U(1)_{B-L}$ is the well-known model in which  the accidental baryon and lepton global symmetries are gauged.  The $U(1)_{C,G}$ models feature null couplings to right-handed charged leptons, whereas the $U(1)_{A,B}$ models have vanishing couplings to left-handed leptons. The $U(1)_{D}$ has null couplings to right-handed down-quarks. The $U(1)_{E,F}$  models induce $Z^\prime$ interactions to all fermions, but have rather exotic $U(1)_X$ charges. }
\label{cargas_u1_2hdm_tipoI}
\end{table}

\subsection{Neutrino Masses}

As aforementioned, in the conventional 2HDM neutrinos are massless. Similarly to the Standard Model one can simply add right-handed neutrinos and generate Dirac masses to the neutrinos. However, a compelling explanation for tiny neutrino masses arises via the seesaw mechanism \cite{Minkowski:1977sc,Mohapatra:1979ia,Lazarides:1980nt,Mohapatra:1980yp,Schechter:1980gr}. In order to realize the type I seesaw mechanism  one needs Dirac and Majorana mass terms for the neutrinos. This can be realized in our 2HDM framework by proper assignments of the quantum numbers, as we will demonstrate in what follows.  \\

Typically, a bare mass term is introduced for the right-handed neutrinos in the realization of the seesaw mechanism without explaining its origin. Here, we explain the neutrino masses by adding a scalar singlet $\Phi_s$, with charge $h_s$ under $U(1)_X$. The first consequence of introducing a new singlet scalar is the extension of the scalar potential which adds to Eq.\ \eqref{pot_2hdm_U1} the potential 
\begin{equation}
V _s = m_s ^2 \Phi _s ^\dagger \Phi _s + \frac{\lambda _s}{2} \left( \Phi _s ^\dagger \Phi _s \right) ^2 + \mu _1 \Phi _1 ^\dagger \Phi _1 \Phi _s ^\dagger \Phi _s + \mu _2 \Phi _2 ^\dagger \Phi _2 \Phi _s ^\dagger \Phi _s + \left( \mu \Phi _1 ^\dagger \Phi _2 \Phi _s + h.c. \right) ,
\label{potscalarsinglet}
\end{equation}
where 
\begin{equation*}
\Phi _s = \frac{1}{\sqrt{2}} \left( v_s + \rho _s + i \eta _s \right).
\end{equation*}
All these terms are straightforwardly invariant under $U(1)_X$ except for the last term which requires $h_s= h_1-h_2$. That said, the Yukawa Lagrangian involving the neutrinos reads 
\begin{equation}
\mathcal{-L} \supset  y^{D}_{ij} \bar{L} _{iL}\widetilde \Phi _2 N_{jR} + Y^{M}_{ij}\overline{(N_{iR})^{c}}\Phi_{s}N_{Rj}\,,
\label{yukawaneutrinosphis}
\end{equation}
which leads to the usual type I seesaw mechanism equation
\begin{equation}
\left(\nu \, N\right)
\left(\begin{array}{cc}
0 & m_D\\
m_D^T & M_R\\
\end{array}\right)\left(\begin{array}{c}
\nu \\
N \\
\end{array}\right)
\end{equation}
with $m_\nu = -m_D^T \frac{1}{M_R}m_D$ and $m_N = M_R$, as long as $M_R \gg m_D$, where $m_D= \frac{y^D v_2}{2\sqrt{2}}$ and $M_R= \frac{y^M v_s}{2\sqrt{2}}$. We take $v_s$ to be at the TeV scale, and in this case 
$y^D \sim 10^{-4}$ and $y^M \sim 1$ lead to  $m_\nu \sim 0.1$~eV  in agreement with current data \cite{Ade:2015xua}. In this scenario right-handed neutrinos have masses at around $300-400$~GeV, although smaller right-handed neutrino masses are also possible. 

Let us now take a closer look at Eq.\ \eqref{yukawaneutrinosphis}. Gauge invariance of the first term requires 
\begin{equation}
- l - h_2 + n = 0.
\label{eq:cond}
\end{equation}
Using Eq.\ \eqref{expres_cargas_u_d} and Eq.\ \eqref{carg_neutrino_right} we get 
\begin{equation}
\begin{split}
- l - h_2 + n &= - \left[ \frac{ - 3 \left( u + d \right)}{2} \right] - \left[ \frac{\left( u - d \right)}{2} \right] - \left( u + 2d \right) = 0.
\end{split}
\end{equation}
Therefore, the condition in Eq.\ \eqref{eq:cond} is automatically fulfilled. However, the Majorana mass term in Eq.\ \eqref{yukawaneutrinosphis} is gauge invariant if $2n+h_s=0$, which implies from Eq.\ \eqref{carg_neutrino_right} that $h_s= 2u+4d$. Using that $h_s=h_1 -h_2$ from the term $\mu \Phi^{\dagger}_{1}\Phi_{2}\Phi_{s}$ in the scalar potential Eq.\  \eqref{potscalarsinglet}, we get 
\begin{equation}
h_{1}=\dfrac{5u}{2}+\dfrac{7d}{2}.
\label{h1charge}
\end{equation}
Now the $\Phi_1$ charge under $U(1)_X$ is generally determined so that neutrino masses are generated. If we happened to choose $u=d=1/3$, then $h_s = h_{1} =2$, $h_{2}=0$, and the $U(1)_{X}$ symmetry is identified to be $U(1)_{B-L}$ symmetry, which is spontaneously broken when $\Phi_{s}$ gets a vacuum expectation value. 
Various other choices of the charges are possible, see Table \ref{cargas_u1_2hdm_tipoI} for a list. From the list, the $U(1)_{B-L}$, $U(1)_N$ have been previously investigated in the literature in different contexts \cite{Davoudiasl:2012qa,Davoudiasl:2012ag,Davoudiasl:2013aya,Lee:2010hf,Lee:2013fda,Lee:2014tba,Davoudiasl:2014kua,Batell:2014mga,Rodejohann:2015lca,Kaneta:2016vkq,Batell:2016zod,Alves:2016cqf}.\\

The spontaneous symmetry breaking pattern from high to low energy goes as follows: (i) the vev $v_s$ sets the scale which the $U(1)_X$ symmetry is broken, say TeV; (ii) then $v_2$  breaks the $SU(2)_L \otimes U(1)_Y$ group to Quantum Electrodynamics. As for the $v_1$ scale, there is some freedom, but it should be either comparable to $v_2$ or smaller, as long as  $v^2= v_2^2+v_1^2$, where $v=246$~GeV, since $M_W^2=g^2 v^2/4$ (see Appendix \ref{sec:app2}). In the regime in which $v_s > v_2 > v_1$ one needs to tune down the $g_X$ coupling in order to have a $Z'$ boson that is lighter than 
the SM $Z$, which is the regime we will focus in here.

In summary, the introduction of a new gauge symmetry with the charge assignments as exhibited in Table \ref{cargas_u1_2hdm_tipoI} leads to a compelling solution to the flavor problem in the Type I 2HDM, while successfully generating fermion masses. In particular, neutrino masses are explained via the seesaw mechanism. A similar reasoning, respecting the NFC criterion ($h_1 \neq h_2$), can be applied to other types of 2HDM preventing them of FCNI. Nevertheless, the addition of extra chiral fermions is required to preserve them free of anomalies. Therefore, we focus in this paper on 
2HDM of Type I, see Table \ref{cargas_u1_2hdm_tipoI}.\\


Now that we have reviewed the theoretical motivations for introducing an Abelian symmetry to the framework of the 2HDM we discuss in model detail the spectrum of the gauge bosons and neutral currents. \\


\subsection{Physical Gauge Bosons and Neutral Currents}

We emphasize that we are including all renormalizable terms allowed guided by gauge invariance. Therefore,  kinetic mixing between the two Abelian groups is present.  To understand the impact of kinetic mixing in the determination of the physical gauge boson we should start off writing down the kinetic terms of the gauge bosons. Note that throughout, the kinetic mixing parameter should fulfill $\epsilon \ll 1$ to be consistent with precision electroweak constraints. That said, the most general gauge Lagrangian associated to these groups is \cite{Babu:1997st,Langacker:2008yv,Gopalakrishna:2008dv}:
\begin{equation}
\mathcal{L} _{\rm gauge} =  - \frac{1}{4} \hat{B} _{\mu \nu} \hat{B} ^{\mu \nu} + \frac{\epsilon}{2\, cos \theta_W} \hat{X} _{\mu \nu} \hat{B} ^{\mu \nu} - \frac{1}{4} \hat{X} _{\mu \nu} \hat{X} ^{\mu \nu},
\label{Lgaugemix1}
\end{equation}
with the following covariant derivative
\begin{equation}
D_\mu = \partial _\mu + ig T^a W_\mu ^a + ig ' \frac{Q_{Y}}{2} \hat{B}_\mu + ig_X \frac{Q_X}{2} \hat{X}_\mu .
\label{Dcovgeral}
\end{equation}

Here $T^{a}$, $W_{\mu}^{a}$ and $g$ are the generators, gauge bosons and gauge coupling constant of $SU(2)_{L}$ respectively; $\hat{X}_\mu$ and $\hat{B}_\mu$ the $U(1)_{X}$ and $U(1)_{Y}$ gauge bosons, $g_X \ (Q_{X})$ is the $U(1)_{X}$ coupling constant (charge) and $g^{\prime} \ (Q_{Y})$
is $U(1)_{Y}$ coupling constant (charge). The hats means that they are non-physical, i.e.\ yet to be diagonalized, fields. As usual $\hat{B}_{\mu\nu}= \partial_\mu \hat{B}_\nu-\partial_\nu \hat{B}_\mu$ and $\hat{X}_{\mu\nu}= \partial_\mu \hat{X}_\nu-\partial_\nu \hat{X}_\mu$. \\

One first performs a so-called GL(2,R) rotation in order to make the kinetic terms canonical, 
\begin{equation}
\begin{pmatrix}
X_{\mu} \\
B_{\mu}
\end{pmatrix}
=
\begin{pmatrix}
\sqrt{1-(\epsilon/\cos\theta_W)^2} & 0 \\
-\epsilon/\cos\theta_W & 1
\end{pmatrix}
\begin{pmatrix}
\hat{X}_{\mu} \\
\hat{B}_{\mu}
\end{pmatrix}.
\label{kinectmixing}
\end{equation}
Therefore $\hat{B}_\mu= \eta_X X_\mu + B_\mu$, and $\hat{X}_\mu= X_\mu$, where
\begin{equation}
\eta_X = \frac{\epsilon/\cos\theta_W}{\sqrt{1 - (\epsilon/\cos\theta_W)^2}} \simeq \epsilon/\cos\theta_W,
\end{equation}
since we are taking $\epsilon/\cos\theta_W \ll 1$ throughout. Thus, the covariant derivative now reads,
\begin{equation}
D_\mu = \partial _\mu + ig T^a W_\mu ^a + ig ' \frac{Q_Y}{2} B_\mu + \frac{i}{2} \left( g_X Q_X + g^{\prime} \frac{\epsilon}{\cos\theta_W} Q_Y\right) X_\mu .
\label{Dcovdiagonal}
\end{equation}which is from where we derive the gauge boson masses. \\

The general formalism of diagonalizing the neutral gauge boson mass matrix is 
delegated to Appendix \ref{sec:app2}. The gauge boson mixing is parametrized in terms of $\epsilon_Z$ and $\epsilon$, coming from the contributions of the second Higgs doublet and the kinetic mixing between the $U(1)$ groups respectively (see below and  \eqref{angulo_xi1}). In the regime in which the new vector boson is much lighter than the SM $Z$ boson, we get two 
mass eigenstates; one identified as the SM $Z$ boson, labeled $Z^0$ with, $m_{Z^0}^2= \frac{g^2 v^2}{4\cos_W^2}$ and the  $Z^{\prime}$ boson with, 
\begin{eqnarray}
m_{Z^\prime}^2=  \frac{v_s^2}{4} g_X^2 q_X^2 + \frac{g_X^2 v^2 \cos^2\beta \sin^2\beta}{4}(Q_{X1} - Q_{X2})^2, 
\label{Eq:MZprimegeneral1}
\end{eqnarray}where $q_X$, $Q_{X1}$, $Q_{X2}$ are the charges under $U(1)_X$ of the singlet scalar, Higgs doublets $\Phi_1$ and $\Phi_2$ respectively, $\tan\beta=v_2/v_1$, $v=\sqrt{v_1^2+v_2^2}=246$~GeV, $v_s$ sets the $U(1)_X$ scale of spontaneous symmetry breaking, and $g_X$ is the coupling constant of the $U(1)_X$ symmetry. \\

It will be useful to write the $Z^\prime$ mass in a compact form as (see Appendix \ref{sec:app3}) by defining $\tan \beta_{d} = \dfrac{v_s}{v_1}$ as follows, 
\begin{equation}
m_{Z^\prime}=  \dfrac{g_X v \cos^2\beta}{\delta},
\label{Zprimedelta}
\end{equation}
where 
\begin{equation}
\delta = \dfrac{2\cos \beta \cos \beta_d}{\sqrt{q_X^2 +\cos^2\beta_d\left(\sin^2\beta(Q_{X1} - Q_{X2})^2-q_X^2\right)}}.
\label{deltinhageral1}
\end{equation}

The mixing angle for the diagonalization of the gauge bosons, $\xi$, in this general setup can be parametrized as follows (see \eqref{xisimplifiedEq}),
\begin{equation}
\xi \equiv \epsilon_Z + \epsilon \tan\theta_W,
\end{equation}where, 
\begin{equation}
 \epsilon_Z \equiv \dfrac{g_{X}}{g_{Z}}(Q_{X1}\cos^{2}\beta+Q_{X2}\sin^{2}\beta).
\end{equation}
For instance, in the $B-L$ model one has, 
\begin{equation}
\epsilon_Z=2 \frac{g_X}{g} \cos^2 \beta,
\label{Eq.epsZ}
\end{equation}and with the use of Eq.\ \eqref{Zprimedelta} we get, 
\begin{equation}
\delta=\dfrac{m_Z}{m_{Z^{\prime}}}\epsilon_{z},
\label{deltaepsilonz1}
\end{equation}which agrees with \cite{Lee:2013fda}, validating our findings.\\

Having obtained the physical fields we can rewrite the neutral current Lagrangian 
(see Appendices \ref{sec:app2} and \ref{sec:app4}): 
\begin{equation}
\begin{split}
\mathcal{L_{\rm NC}} = &- e J ^\mu _{\rm em} A_\mu - \frac{g}{2\cos\theta_W}  J ^\mu _{NC} Z_\mu - \left( \epsilon e J^\mu _{em} + \epsilon _Z \frac{g}{2\cos\theta_W}  J^\mu _{NC} \right) Z' _\mu \\
&+ \frac{1}{4} g_X \sin \xi \left[ \left( Q_{Xf} ^R + Q_{Xf} ^L \right) \bar{\psi} _f \gamma ^\mu \psi _f + \left( Q_{Xf} ^R - Q_{Xf} ^L \right) \bar{\psi} _f \gamma ^\mu \gamma _5 \psi _f \right] Z_\mu \\
&- \frac{1}{4} g_X \cos \xi \left[ \left( Q_{Xf} ^R + Q_{Xf} ^L \right) \bar{\psi} _f \gamma ^\mu \psi _f - \left( Q_{Xf} ^L - Q_{Xf} ^R \right) \bar{\psi} _f \gamma ^\mu \gamma _5 \psi _f \right] Z' _\mu,
\end{split}
\label{www1}
\end{equation}
where $Q^R_X$ ($Q^L_X$) are the left-handed (right-handed) fermion charges under $U(1)_X$. We emphasize that Eq.\ \eqref{www1} is the general neutral current for 2HDM augmented by a $U(1)_X$ gauge symmetry. \\

Again, it is important to validate our results with the existing literature. For instance, in the $U(1)_{B-L}$ model we get 
\begin{equation}
\begin{split}
\mathcal{L_{\rm NC}} = &- e J ^\mu _{\rm em} A_\mu - \frac{g}{2\cos\theta_W} J ^\mu _{NC} Z_\mu - \left( \epsilon e J^\mu _{em} + \epsilon _Z \frac{g}{2\cos\theta_W}  J^\mu _{NC} \right) Z' _\mu \\
&- \frac{g_X}{2} Q_{Xf} \left[ \bar{\psi} _f \gamma ^\mu \psi _f \right] Z' _\mu,
\end{split}
\label{zzgeralcoma1}
\end{equation}
where $Q_{Xf}=-1$ for charged leptons and $Q_{Xf}=1/3$ for quarks, with $g_X$ and $\epsilon_Z$ related by Eq.\ \eqref{Eq.epsZ}, in agreement with \cite{Klasen:2016qux}.\\

Now that we have obtained the neutral current for a generic $U(1)_X$ model in the context of the 2HDM we will address the relevant constraints these $U(1)_X$ models are subject to.


\subsection{$Z^{\prime}$ Decays}

We have introduced a multitude of Abelian gauge groups in the context of the 2HDM that address two major issues in the original 2HDM framework, namely the absence of neutrino masses and the presence of flavor changing interactions. Abelian groups generally give rise to neutral gauge bosons which are subject to a rich phenomenology that we plan to explore in what follows.  Before doing so, some general remarks are in order: 
\begin{itemize}

\item[{\bf (i)}] The kinetic mixing ($\epsilon$) as well as the mass mixing ($\epsilon_Z$) parameters are required to be smaller than $10^{-3}$ to be consistent with a variety of constraints that we will discuss. 
 
\item[{\bf (ii)}] We will focus on the regime $m_{Z^{\prime}} \ll m_Z$, say $m_{Z^{\prime}}=1\,{\rm MeV}-10\,{\rm GeV}$. Some comments on different regimes 
will nevertheless be made whenever relevant.

\item[{\bf (iii)}] A light $Z^{\prime}$ can be achieved at the expense of tuning the gauge coupling $g_X$. 

\item[{\bf (iv)}] The phenomenology of our models will be dictated by either the kinetic mixing or the mass-mixing terms.

\end{itemize}
That said, some of the constrains we will investigate are based on dark photon searches. Notice that our models are a bit different than the dark photon model that has only the kinetic mixing term, due to the presence of mass-mixing and the non-vanishing $U(1)_X$ charges of the SM fermions. We remind the reader that only the models that simultaneously explain neutrino masses and free the 2HDM from flavor changing interactions are of interest throughout this work, as displayed in the first block of Table \ref{cargas_u1_2hdm_tipoI}. With this in mind we discuss the $Z^\prime$ decays in each one of the models.



\begin{figure}[h]
\begin{center}
    \includegraphics[width=0.45\columnwidth,height=5.252cm]{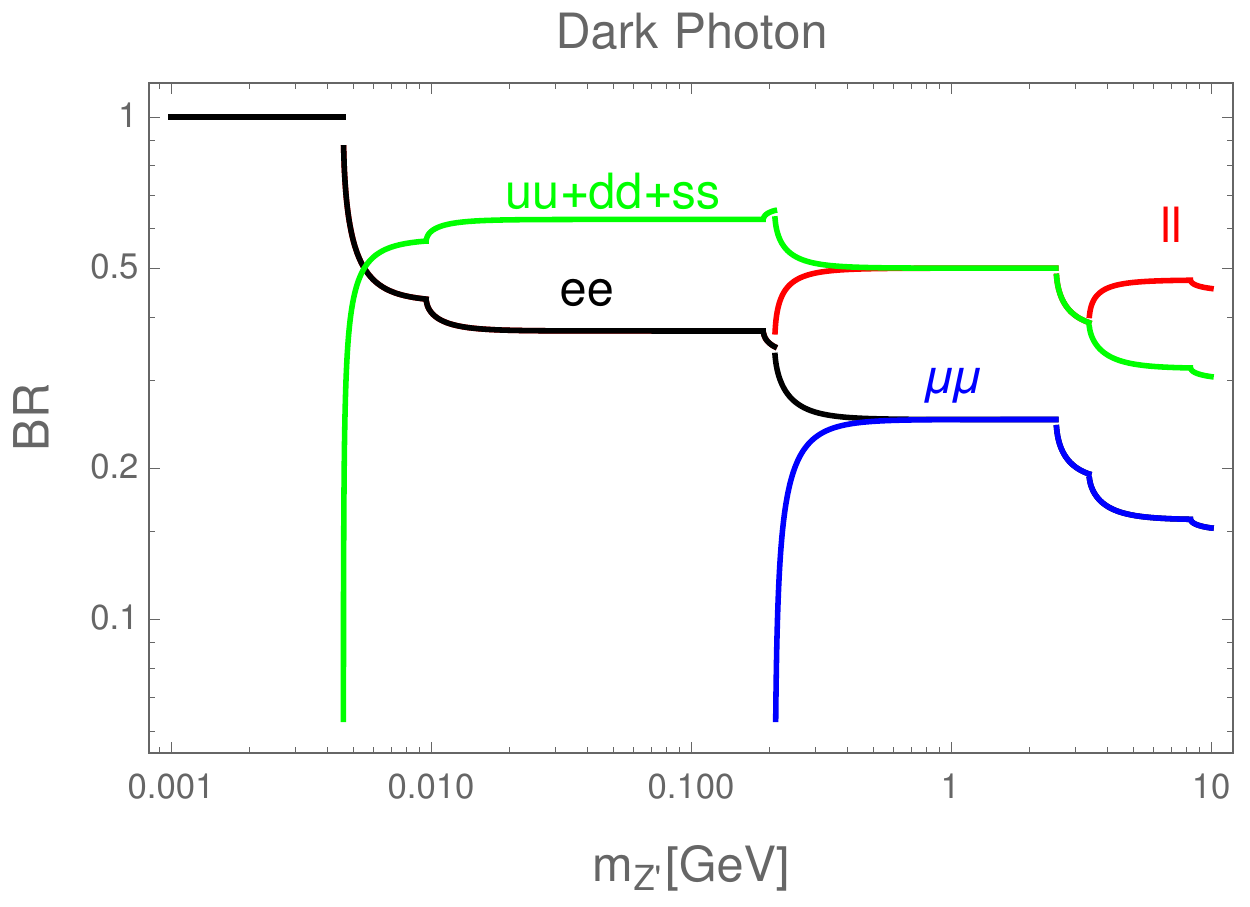}
	\includegraphics[width=0.45\columnwidth,height=5.252cm]{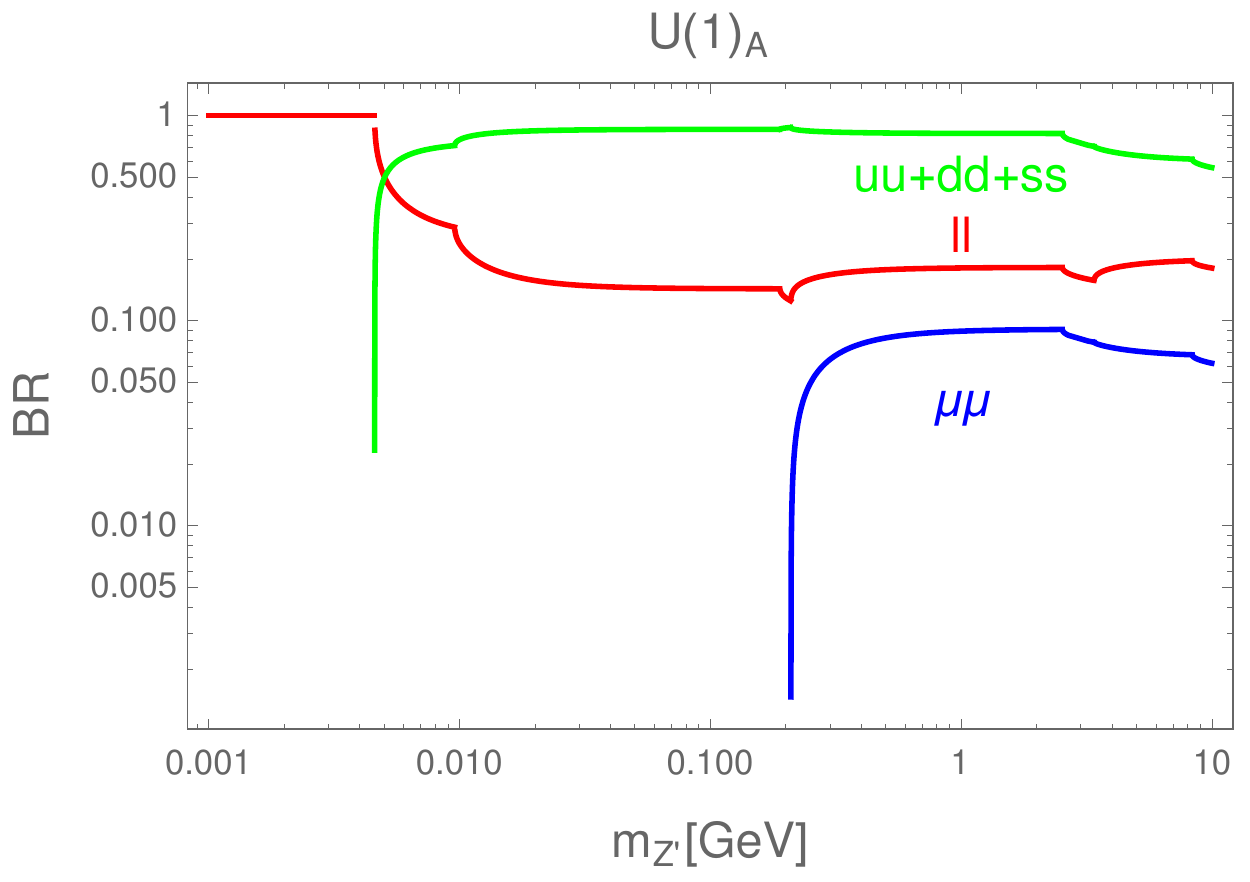}
	\includegraphics[width=0.45\columnwidth,height=5.252cm]{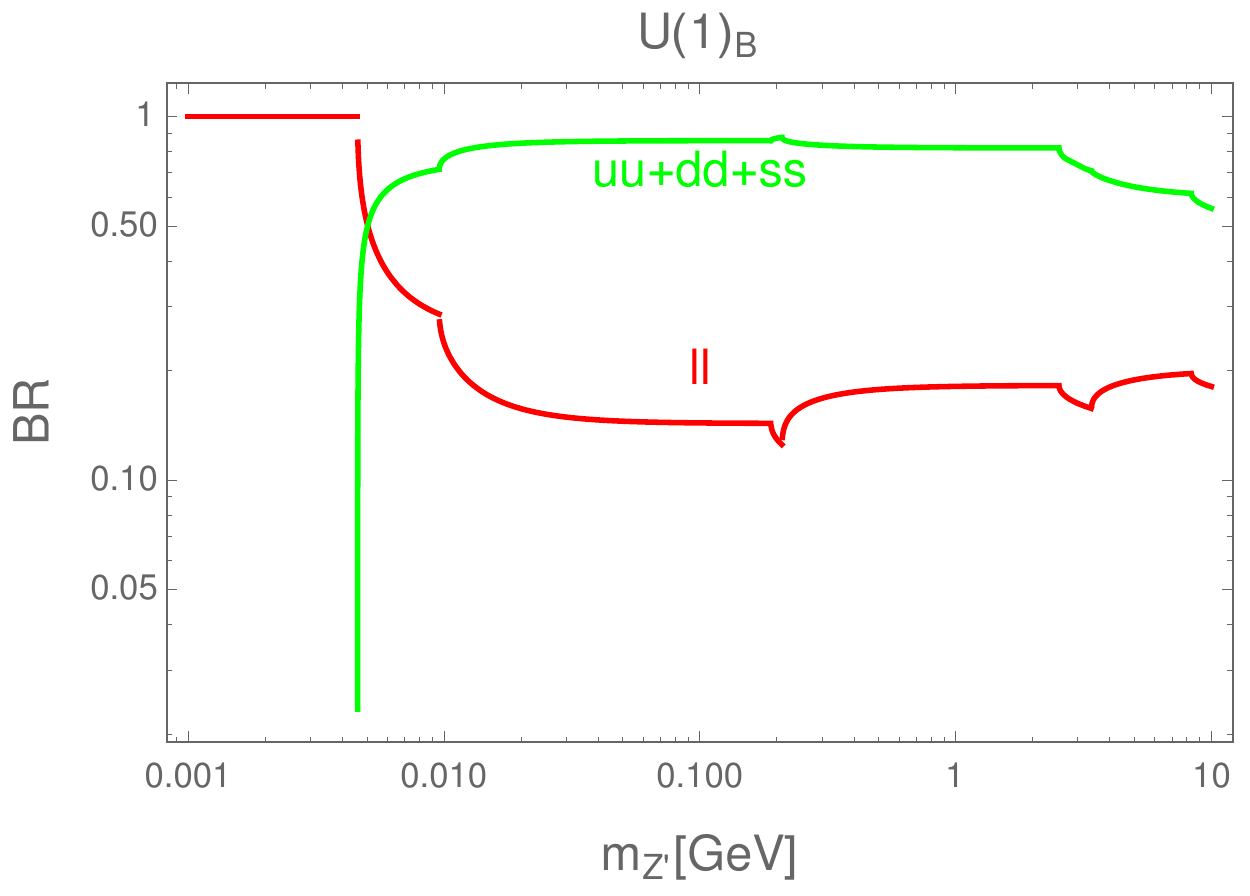}
    \includegraphics[width=0.45\columnwidth,height=5.252cm]{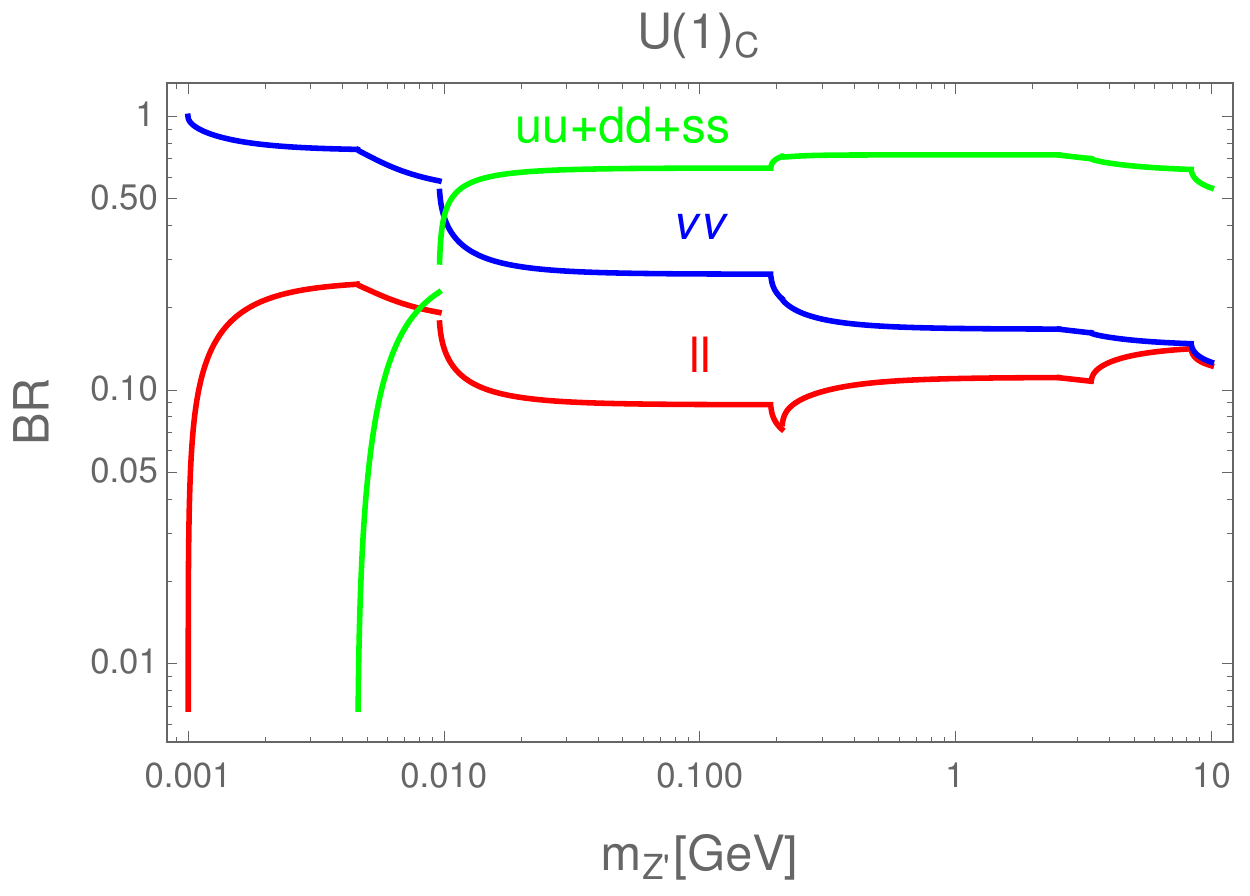}
	\includegraphics[width=0.45\columnwidth,height=5.252cm]{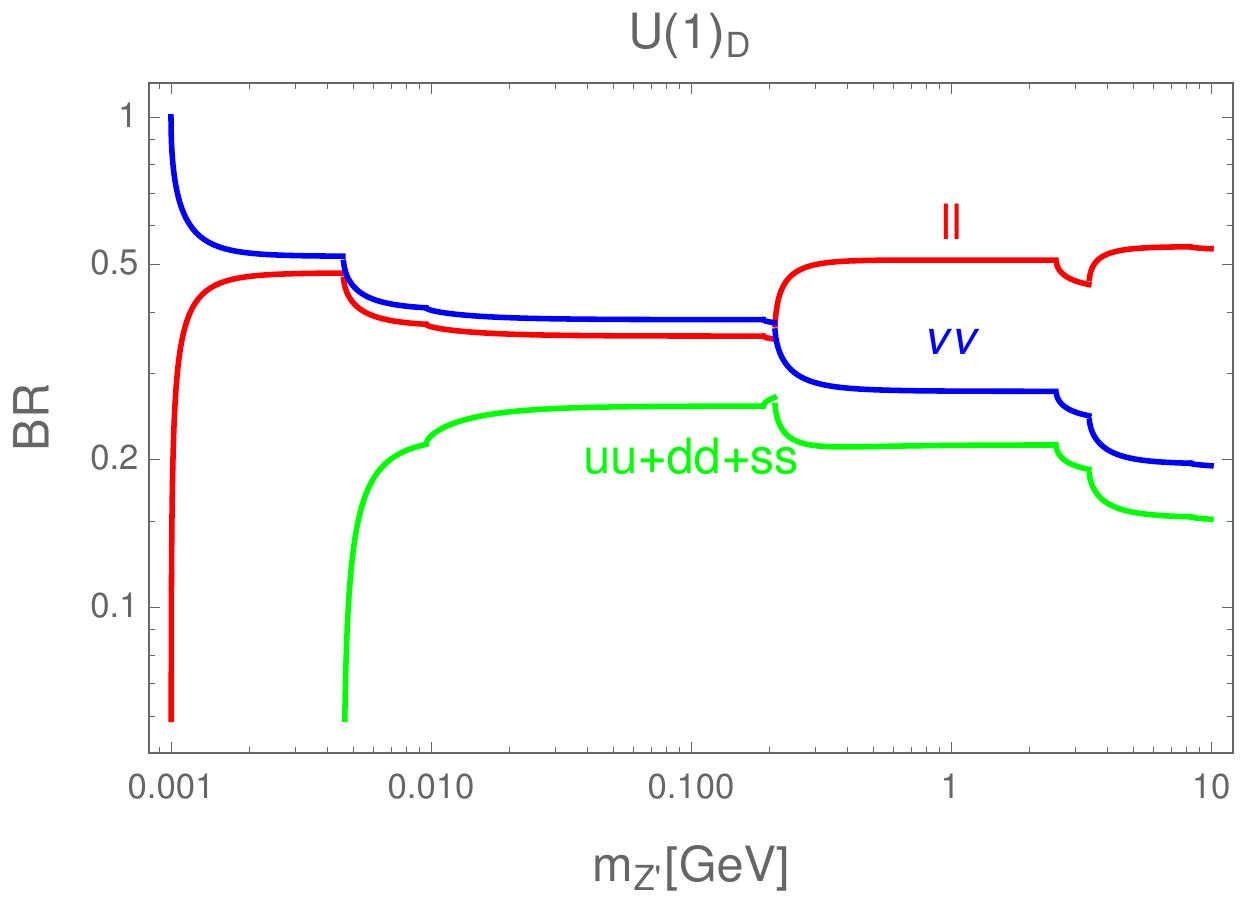}
    \includegraphics[width=0.45\columnwidth,height=5.252cm]{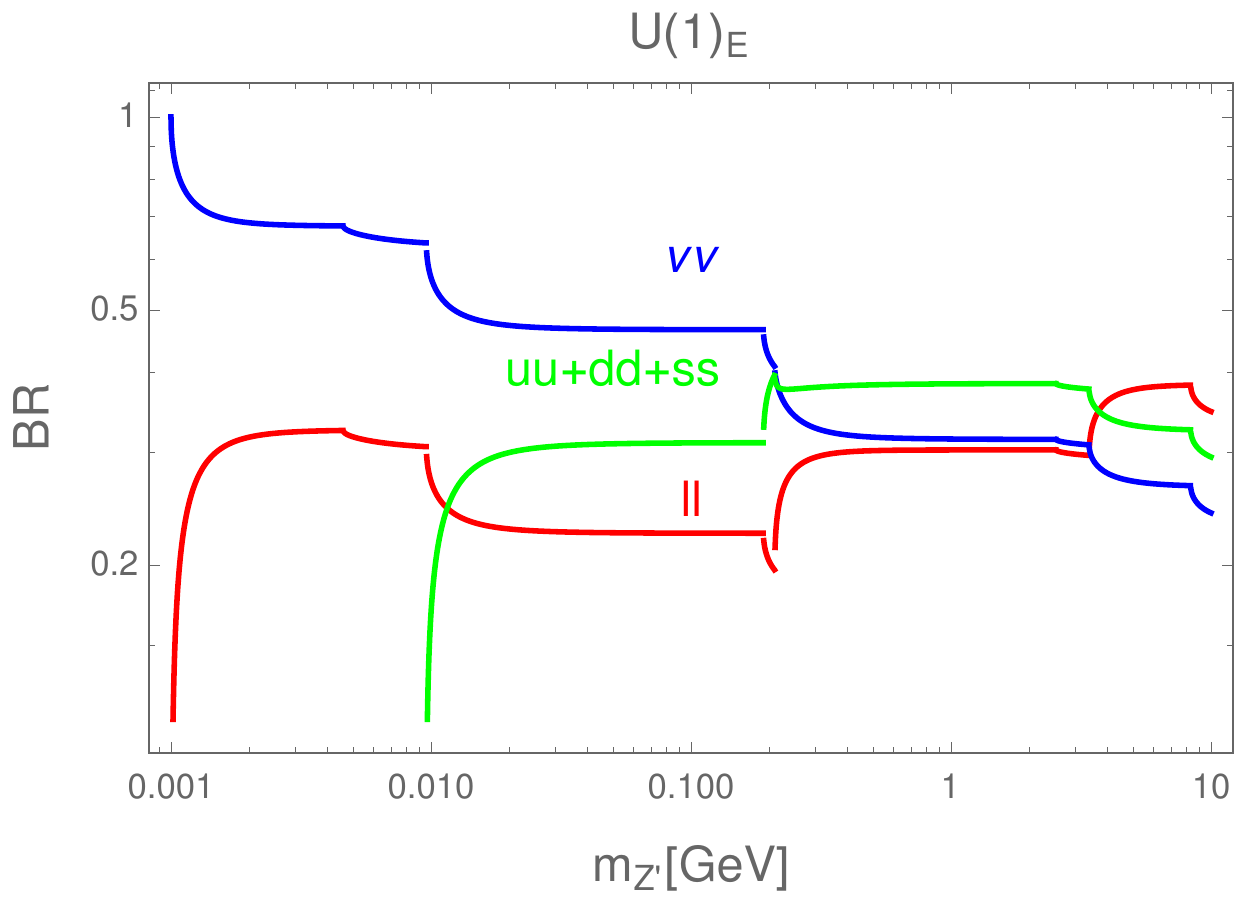}	
    \includegraphics[width=0.45\columnwidth,height=5.252cm]{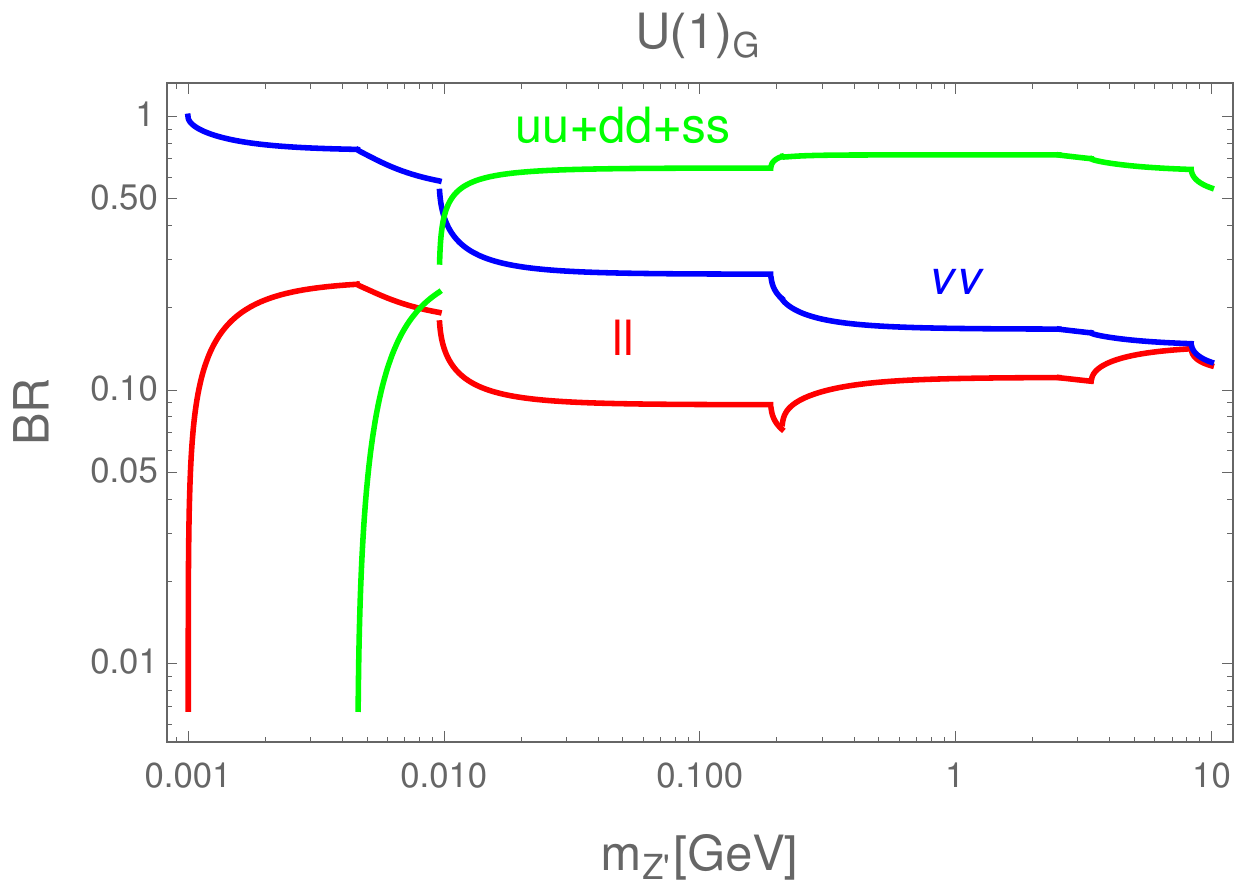}
    \includegraphics[width=0.45\columnwidth,height=5.252cm]{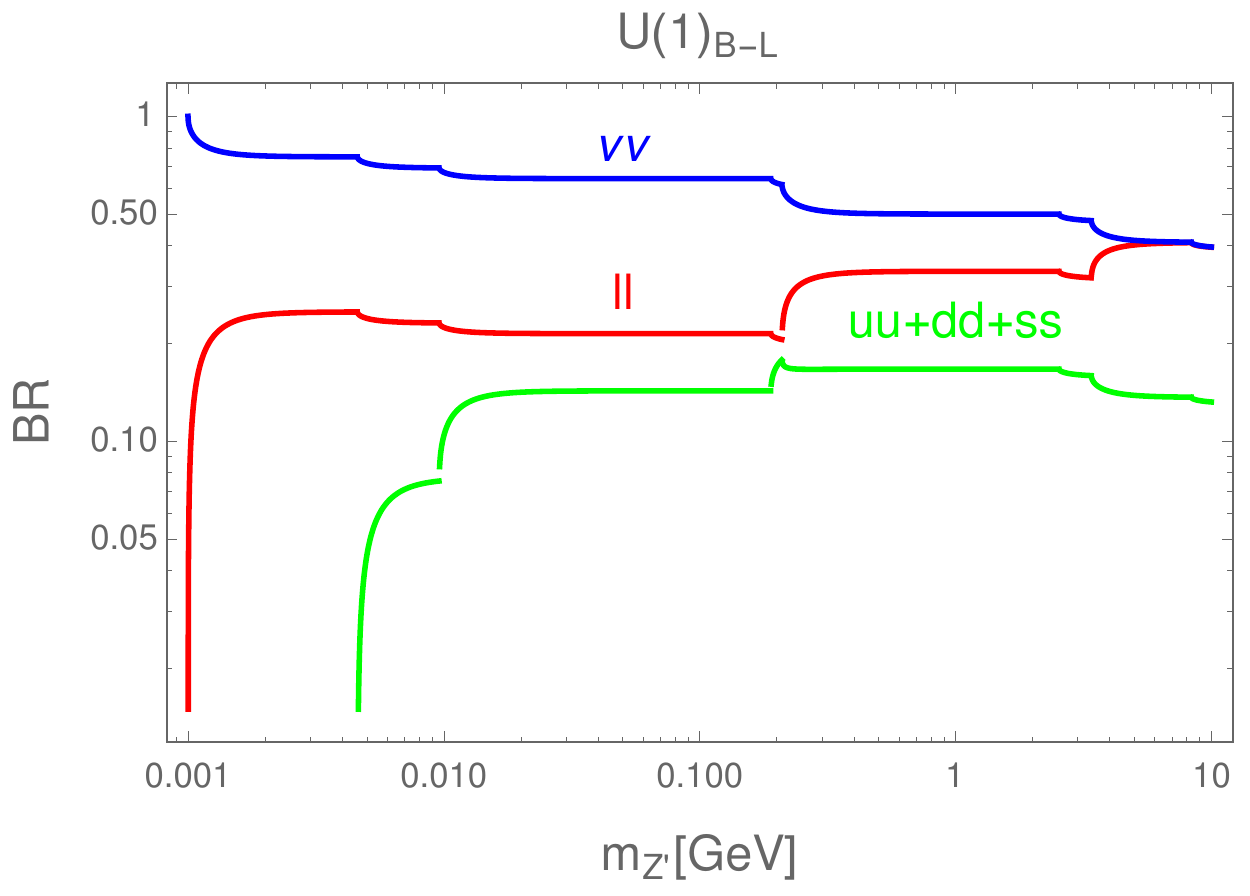}
\caption{Branching ratios as a function of the $Z^{\prime}$ mass for several $U(1)_X$ models under study.}
\label{Zpdecays2}
\end{center}
\end{figure}

\begin{itemize}


\item It is important to first mention the dark photon model. In such models the 
coupling of the dark photon $A^{\prime}$ with SM fermions $f$ goes as $\bar f \gamma^\mu f A^{\prime}_\mu$. 
The corresponding branching ratios are shown in Fig.\ \ref{Zpdecays2}. It is important to have a clear picture of the dark photon model because some of the bounds discussed in this work have the dark photon model as benchmark as we shall see when we address neutrino-electron scattering and low energy accelerator constraints. 

\item In the $U(1)_A$ model, the charged leptons and light quarks charges under $U(1)_A$ are the same but due to color multiplicity the $Z^{\prime}$ decays mostly into light quarks as shown in Fig.\ \ref{Zpdecays2}. As for the $U(1)_B$ model, the results are same. Notice that the label $B$ has nothing to do with baryon number. No decays into active neutrinos exist since the lepton doublet is uncharged under the new gauge group.

\item In the $U(1)_C$ model, the branching ratio into neutrinos is more relevant in comparison with previous models since now the lepton doublet has charge $3/4$ under 
the new gauge group. However, decays into light quarks are still the most relevant. The $U(1)_G$ model has a similar behavior.

\item In the $U(1)_D$ model, the branching ratio into leptons prevails. A similar feature happens in the $U(1)_{B-L}$ model, where $B$ and $L$ account for the baryon and lepton numbers. In the former, the branching ratios into charged fermions and neutrinos are very similar, but as soon the decay into muons becomes kinematically accessible the branching ratio into charged leptons increases. In the latter,  decays into neutrinos are always dominant in the mass region of interest, as a straightforward consequence of the baryon and lepton quantum numbers of the fermions.

\item In the $U(1)_E$ model, decays into neutrinos are dominant until the $Z^{\prime}$ mass approximates the strange quark and muon kinematic thresholds. 

\end{itemize}

Now that we have highlighted the properties of the $Z^{\prime}$ gauge boson for each $U(1)_X$ model we will discuss a variety of constraints going from mesons decays to low energy accelerators. 

\clearpage




\section{\label{sec:pheno}Phenomenological Constraints}

In this section we will span over the existing limits on the $U(1)_X$ models proposed previously. Our main goal is to estimate limits on the parameter space of these models and assess how relevant they are. A more dedicated study will be conducted elsewhere. We start with meson decays.

\subsection{Meson Decays}

\subsubsection{Rare $K$ Decays}

The main decays modes of the charged Kaon are $\mu \nu_\mu,\, \pi^+ \pi^0$ and   $\pi^+\pi^+\pi^0$ with branching ratios of $64\%,\, 21\%$ and $6\%$ respectively. Searches for rare meson decays such as $K^+ \to \pi^+ l^+ l^-$ have also been performed \cite{Appel:1999yq,Batley:2009aa}, which led to the experimental constraints \cite{Olive:2016xmw},
\bea
&&\br (K^+ \to \pi^+ e^+ e^-)_\text{exp}     = (3.00 \pm 0.09) \times 10^{-7} \label{eq:n2a},\\
&&\br (K^+ \to \pi^+ \mu^+ \mu^-)_\text{exp} = (9.4  \pm 0.6)  \times 10^{-8} \label{eq:n2b},\\
&&\br (K^+ \to \pi^+ \nu \bar\nu)_\text{exp} = (1.7  \pm 1.1)  \times 10^{-10}\label{eq:n2c}.
\eea

In a Two Higgs Doublet Model with $Z-Z^{\prime}$ mass mixing the branching ratio of $K^+ \rightarrow \pi^+ Z^{\prime}$ is estimated to be \cite{Hall:1981bc}, 
\beq
\br (K^+ \to \pi^+ Z^\prime) \simeq 4 \times 10^{-4} \, \delta^2,
\label{eq:n1}
\eeq  
where $\delta=\epsilon_{Z} m_Z/m_{Z^{\prime}}$ (see Appendix \ref{sec:app3}). Comparing Eq.\ \eqref{eq:n1} with Eqs.\ \eqref{eq:n2a}-\eqref{eq:n2c} we conservatively find that, 
\bea
&&\delta \lsim \frac{2\times 10^{-2}}{\sqrt{BR(Z^{\prime} \to l^+ l^-)}} ,\label{eq:n3a}\\
&&\delta \lsim \frac{7 \times 10^{-4}}{\sqrt{ BR(Z^{\prime} \to {\rm missing\; energy})}}\label{eq:n3b}.
\eea

These bounds should be used with care since they are not applicable to any $Z^{\prime}$ mass. For instance, the bound obtained in Eq.\ \eqref{eq:n2a} was obtained with a hard cut in the dilepton invariant mass, namely $m_{ee} > 140$~MeV \cite{Batley:2009aa}. Thus this limit is valid for $m_{Z^{\prime}} > 140$~ MeV.\\

In the $U(1)_{B-L}$ model, for instance, for $m_Z^{\prime} < 2 m_\mu$, the $Z^{\prime}$ decays with $\sim 75\%$ braching ratio into neutrinos and therefore Eq.\ \eqref{eq:n3b} should be used, giving stronger constraints. In the $U(1)_N$ model, on other hand, the situation strongly depends on the ratio $\epsilon/\epsilon_Z$. In particular, for $\epsilon/\epsilon_Z \gg 1$, the $Z^{\prime}$ decays mostly into charged leptons with Eq.\ \eqref{eq:n3a} yielding stronger limits, conversely for $\epsilon/\epsilon_Z < 1$, Eq.\ \eqref{eq:n3b} is more restrictive in agreement with \cite{Davoudiasl:2012ag}. Either way it is clear that rare kaon decays introduce an interesting pathway to probe new physics, specially low mass $Z^{\prime}$ gauge bosons  \cite{Buras:2015jaq,Crivellin:2016vjc,Christ:2016mmq,Ibe:2016dir,Chiang:2016cyf}.

\subsubsection{Rare $B$ Decays}

Similar to the $K$ mesons discussed previously rare $B$ decays offer a promising environment to probe new physics. In particular, the charged $B$ meson with mass of $5.3$~GeV, comprised of $u\bar{b}$, may possibly decay into $K^+ l^+ l^-$ \cite{Batell:2009jf,Freytsis:2009ct,Eigen:2015zva} or $K^+\nu\bar{\nu}$ \cite{Aubert:2008ps,Wei:2009zv}. Such decays have been measured to be \cite{Olive:2016xmw},
\bea
&& BR(B^+ \to K^+ \bar l^+ l^-)_\text{exp} < 4.5\times 10^{-7} ,\\
&& BR(B^+ \to K^+ \bar\nu \nu)_\text{exp} < 1.6 \times 10^{-5} .
\label{eq:BtoKnunu}
\eea
Having in mind that the mass mixing in the 2HDM induces \cite{Hall:1981bc,Freytsis:2009ct,Davoudiasl:2012ag}, 
\beq
\br (B \to K Z_d)\simeq 0.1 \delta^2 , 
\label{eq:BtoKZd}
\eeq implying that,
\bea
&&\delta \lsim \frac{2\times 10^{-3}}{\sqrt{BR(Z^{\prime} \to l^+ l^-)}} ,\label{eqB:n3a}\\
&&\delta \lsim \frac{1.2 \times 10^{-2}}{\sqrt{ BR(Z^{\prime} \to {\rm missing\; energy})}}\label{eqB:n3b}.
\eea

Comparing  Eqs.\ \eqref{eqB:n3a}--\eqref{eqB:n3b} with Eqs.\ \eqref{eq:n3a}--\eqref{eq:n3b} we can see the rare $B$ decays give rise to more stringent limits on the parameter $\delta$ when the $Z^{\prime}$ decays mostly into charged leptons. We highlight that the large factor in  Eq.\ \eqref{eq:BtoKZd} is result of the presence of the top quark in the Feynman diagram responsible for the $b \rightarrow s$ conversion, and consequently the $B \rightarrow K Z^\prime$ decay. \\

As for $Z^{\prime}$ decays into neutrino pairs, then precise measurements on Kaon decays offer the leading constraints. The constraints from meson decays are summarized in Table \ref{tab:meson}. 
We will now move to Higgs physics.

\begin{table}[t]
\centering
\begin{tabular}{|c|}
\hline
{\color{blue} $K$ Decays}\\
\hline
$\delta \lesssim 2 \times 10^{-2}/\sqrt{BR(Z^{\prime} \rightarrow l^+l^-)}$ \\
$\delta \lesssim 7 \times 10^{-4}/\sqrt{BR(Z^{\prime} \rightarrow {\rm missing\; energy})}$ \\
\hline
{\color{blue} $B$ Decays}\\
\hline
$\delta \lesssim 2 \times 10^{-3}/\sqrt{BR(Z^{\prime} \rightarrow l^+l^-)}$ \\
$\delta \lesssim 1.2 \times 10^{-2}/\sqrt{BR(Z^{\prime} \rightarrow {\rm missing\; energy})}$ \\
\hline
\end{tabular}
\caption{Summary of constraints on the model from meson decays.}
\label{tab:meson}
\end{table}

\begin{table}
\centering
\begin{tabular}{|c|c|}
\hline
vertex & coupling constant\\
\hline
$H\, t\bar{t},H\, b \bar{b} , H\, \tau \bar{\tau}$ & $\frac{\sin\alpha}{\sin\beta}$ \\
\hline
$H\, W W, H\, Z Z$ & $\cos(\beta-\alpha)$\\
\hline
$h\, t\bar{t},h\, b \bar{b} , h\, \tau \bar{\tau}$ & $\frac{\cos\alpha}{\sin\beta}$ \\
\hline
$h\, W W, h\, Z Z$ & $\sin(\beta-\alpha)$\\
\hline
\end{tabular}
\caption{Higgs and light scalar interactions in the 2HDM type I. The coupling  constants in the second column are the overall multiplicative factor in front of the SM couplings. In other words, when $\alpha=\beta$ the Higgs in the 2HDM type I interacts with fermions and gauge bosons identically to the SM Higgs.}
\label{tableHiggs}
\end{table}

\subsection{Higgs Physics}

\subsubsection{Higgs Properties}
\label{sec:Hdecay}

Our models are comprised of two Higgs doublets and a singlet scalar. In the limit in which the scalar doublets do not mix with the singlet, i.e.\ the regime in which the parameters $\mu_1$, $\mu_2$, $\mu$ in the potential 
(\ref{potscalarsinglet}) are suppressed, one finds \cite{Lee:2013fda} 
\begin{eqnarray}
m_s^2 & = & \lambda_s v_s^2,\nonumber \\
m_h^2  &= & \frac{1}{2}\left( \lambda_1 v_1^2 + \lambda_2 v_2^2 - \sqrt{(\lambda_1 v_1^2-\lambda_2 v_2^2)^2 +4(\lambda_3+\lambda_4)^2 v_1^2 v_2^2} \right), \label{massescalars}\\ 
m_H^2  &= & \frac{1}{2}\left( \lambda_1 v_1^2 + \lambda_2 v_2^2 + \sqrt{(\lambda_1 v_1^2-\lambda_2 v_2^2)^2 +4(\lambda_3+\lambda_4)^2 v_1^2 v_2^2} \right),\nonumber
\end{eqnarray}
where the $H$-$h$ mixing is given by 
\begin{eqnarray}
\left(\begin{array}{c}
H\\
h\\
\end{array}\right)=\left(\begin{array}{cc}
\cos\alpha & \sin\alpha\\
-\sin\alpha & \cos\alpha \\
\end{array}\right)\left(\begin{array}{c}
\phi_1\\
\phi_2\\
\end{array}\right)
\end{eqnarray}
with
\begin{equation}
\tan 2\alpha = \frac{2(\lambda_3+\lambda_4)v_1v_2}{\lambda_1 v_1^2 -\lambda_2 v_2^2}. 
\end{equation}

Notice that in the limit $\sin\alpha \sim 1$, $H \sim \phi_2$, i.e.\ the SM Higgs, and $h \sim \phi_1$. Moreover, it is clear from Eq.\ \eqref{massescalars} that in the 2HDM we are considering the SM-like Higgs is heavier than the light scalar $h$. Their interactions strength with SM particles is summarized in Table \ref{tableHiggs}. The couplings constants in the second column of the table are multiplicative factors appearing in front of the SM couplings. In other words, when $\alpha=\beta$ the Higgs in the 2HDM type I interacts with fermions and gauge bosons identically to the SM Higgs. Furthermore, the regime $\beta \sim \alpha$ renders the $h\, t \bar{t}$, $h \, b \bar{b}$ and $h\, \tau \bar{\tau}$ couplings governed by $\cot\beta$, whereas the $h\,WW$, $h\,ZZ$ interactions are  dwindled.

\subsubsection{Higgs Associated Production}
\begin{figure}[!t]
\begin{center}
    \includegraphics[width=0.4\columnwidth]{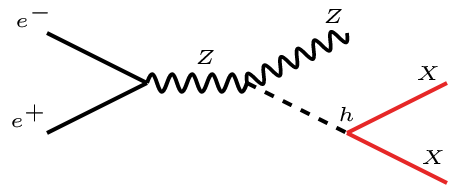}
\caption{Higgs associated production at LEP followed by its invisible decay, illustrated by $h \rightarrow XX$. }
\label{Higgsassociated}
\end{center}
\end{figure}

Several experiments have searched for scalars with similar properties to the SM Higgs at LEP. They were particularly focused on the associated production with the $Z$ boson, with the scalar decaying either into fermions or invisibly as displayed in Fig.\ \ref{Higgsassociated}. The light Higgs in the models under study, $h$, decays at tree-level into $Z^{\prime} Z^{\prime}$. Since the LEP searches did not cover fermions with very small invariant mass, i.e.\ stemming from a light $Z^{\prime}$, one should use the results from the invisible decay search. That said, the $Zh$ associated production search resulted into limits on the product of the production cross section strength and branching ratio, i.e.\ $\sigma (Zh)/\sigma(Z H_{SM}) BR(h\rightarrow \rm inv)$. \\

\begin{figure}[!t]
\begin{center}
    \includegraphics[width=0.49\columnwidth]{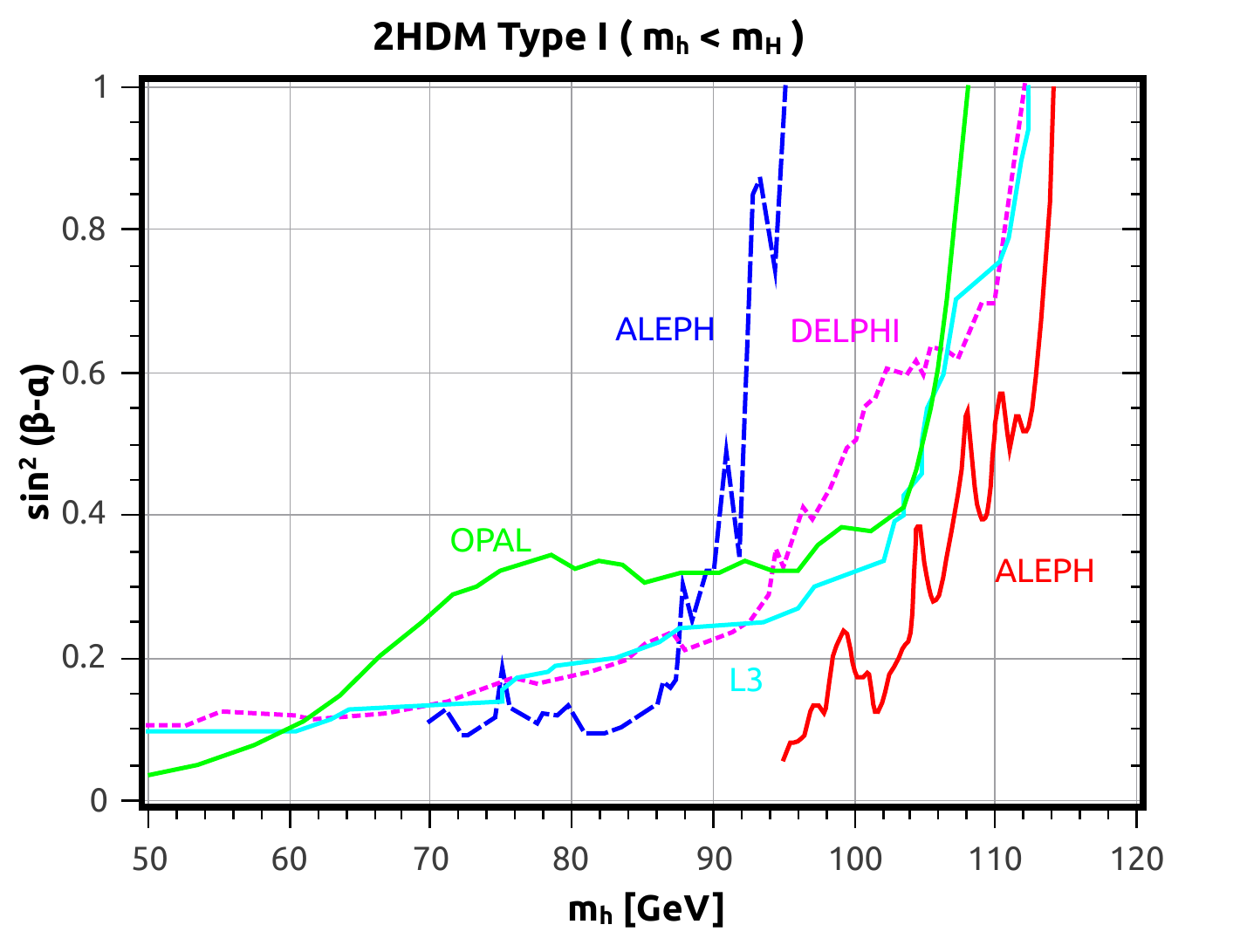}
	\includegraphics[width=0.49\columnwidth]{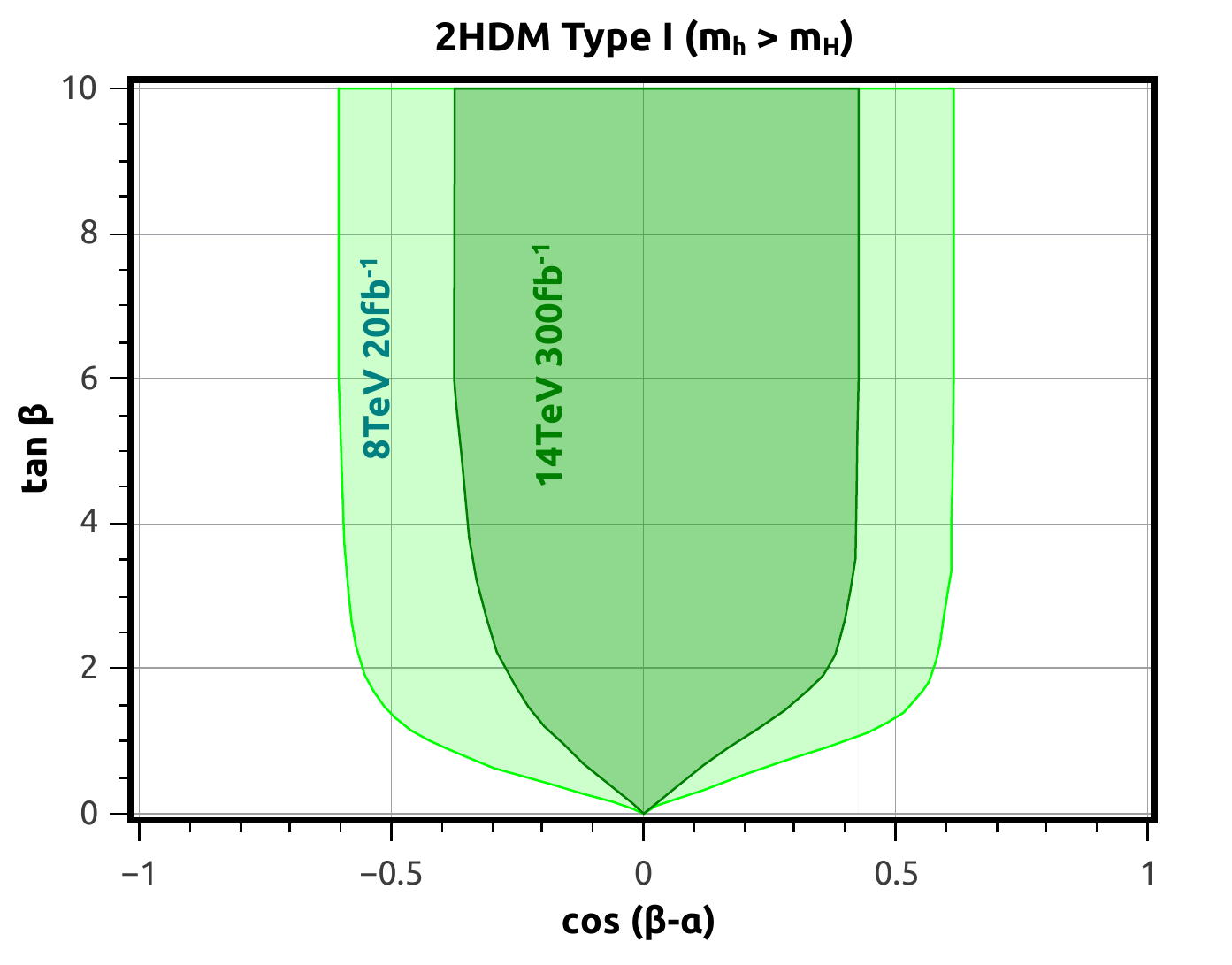}
\caption{Left-panel: Upper limits from invisible Higgs decay searches translated to the light Higgs mass $m_h$. Right-panel: Constraints from the LHC on the Higgs properties in the context of 2HDM type  I with $m_h  > m_H$, where $H$ is the SM Higgs, a  scenario which is opposite to what it is being considered here. }
\label{Hcouplings}
\end{center}
\end{figure}

Assuming $BR(h\rightarrow \rm inv)\simeq 1$ throughout, one can reinterpret the results from \cite{Barate:1999uc,Achard:2004cf,Abbiendi:2007ac}  for the light Higgs $h$, having in mind that the $h\, Z Z$ coupling goes with $\sin (\beta-\alpha)$, to place a bound on $\sin^2 (\beta-\alpha)$ as a function of the scalar mass as shown in the left panel of Fig.\ \ref{Hcouplings} \cite{Lee:2013fda}. From  Fig.\ \ref{Hcouplings}, one can conservatively conclude that $\sin^2(\beta-\alpha) \lesssim 0.1$, $\cos^2(\beta-\alpha) > 0.9$, independent of $\tan\beta$. Weaker limits are applicable depending on the light Higgs mass.\\

\begin{figure}[!t]
\begin{center}
    \includegraphics[width=0.49\columnwidth]{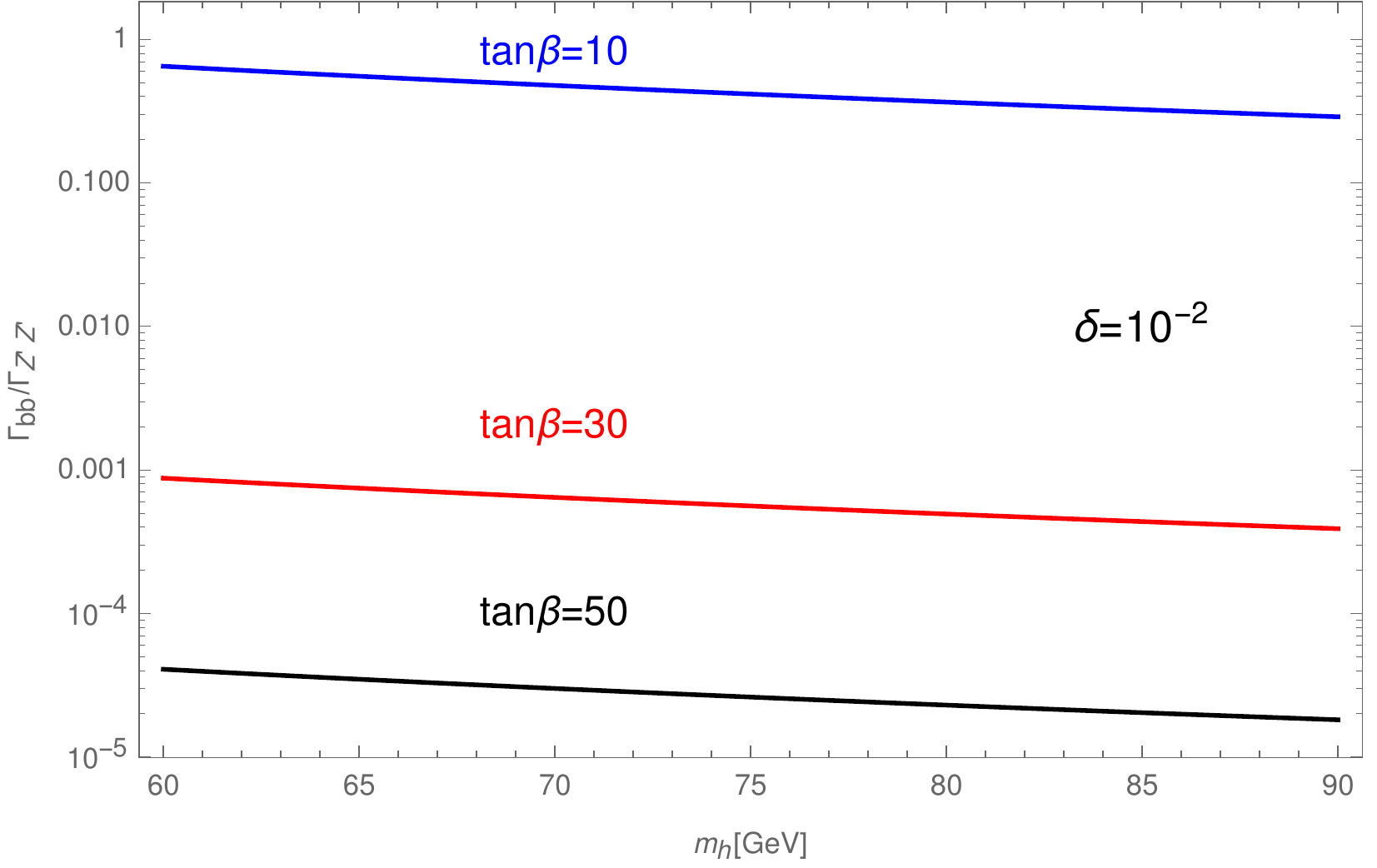}
        \includegraphics[width=0.49\columnwidth]{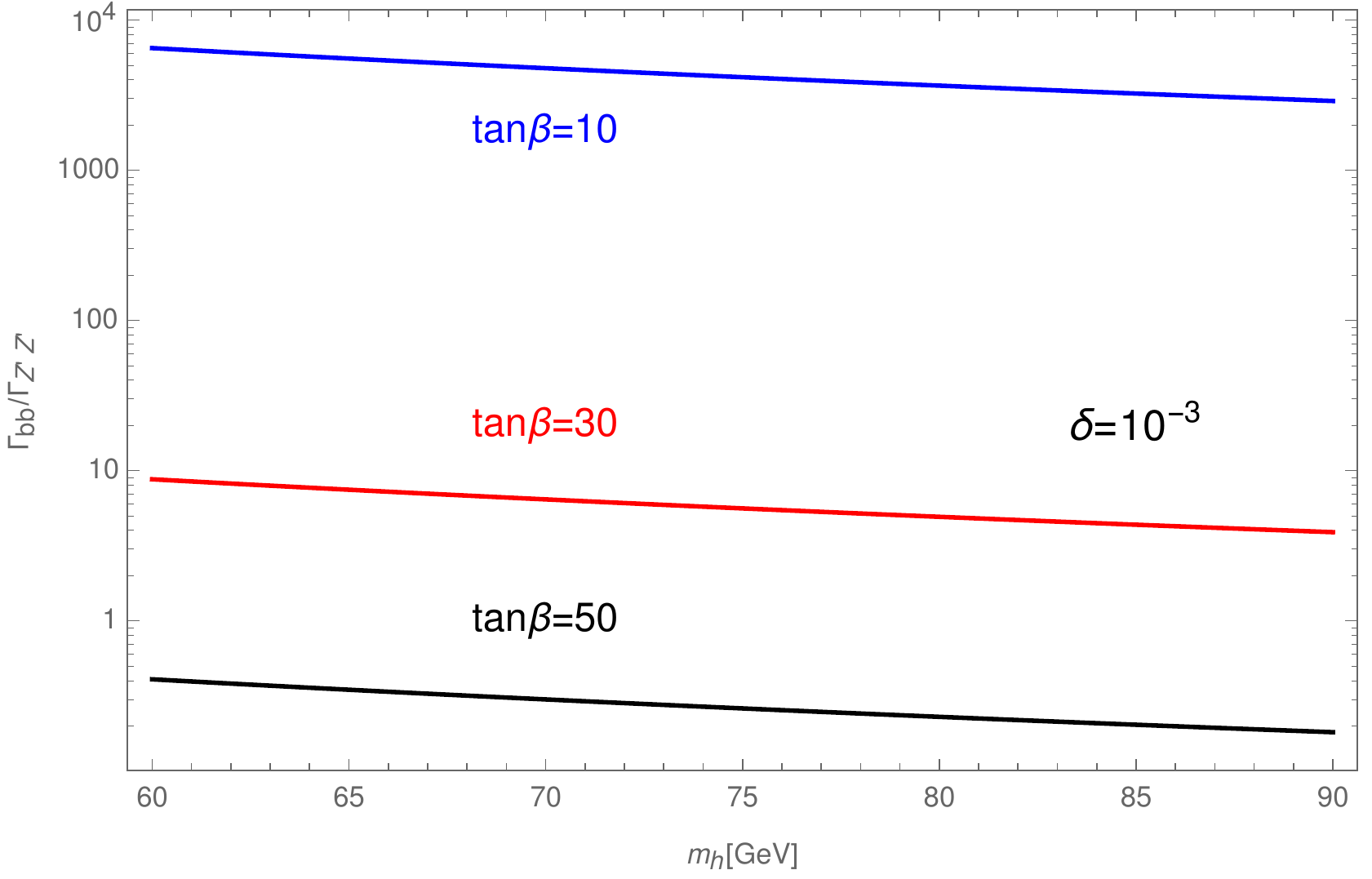}
\caption{Ratio of the light Higgs decay width for different values of $\tan\beta$. In the left panel we set $\delta=10^{-2}$, while in the right one $\delta=10^{-3}$. One can see that if the product $\delta\tan\beta$ is sufficiently small, the light Higgs decays dominantly into $Z^\prime Z^\prime$. In this regime, the limits presented in the left panel of Fig.\ \ref{Hcouplings} can be directly applied.}
\label{Higgsratio}
\end{center}
\end{figure}

However, the limit presented in Fig.\ \ref{Hcouplings} may be not robust because it relies on the assumption that $BR(h\rightarrow \rm inv)\simeq 1$. 
A simple check can be done by comparing the decay into $Z^{\prime}Z^{\prime}$ with the usually dominant $b\bar{b}$ mode that lead to the following decay rates,
\begin{equation}
\Gamma_{h\rightarrow Z^\prime Z^\prime}= \frac{g_Z^2}{128\pi}\frac{m_h^3}{m_Z^2}(\delta \tan\beta)^4 \left(\frac{\cos^3\beta \cos\alpha -\sin^3\beta\sin\alpha}{\cos\beta\sin\beta}\right)^2,
\end{equation}
\begin{equation}
\Gamma_{h\rightarrow b\bar{b}} = \frac{3 m_b^2 m_h}{8\pi v^2}\left(\frac{\cos\alpha}{\sin\beta}\right)^2.
\label{Eqhiggsratio}
\end{equation}
We thus conclude that the ratio reads
\begin{equation}
\frac{\Gamma_{h\rightarrow b\bar{b}}}{\Gamma_{h\rightarrow Z^\prime Z^\prime}}=\frac{12 m_b^2}{m_h^2}\frac{1}{(\delta \tan\beta)^4}\left( \frac{\cos\beta \sin\beta}{\cos^3\beta \cos\alpha - \sin^3\beta \sin\alpha } \right)^2 \left(\frac{\cos\alpha}{\sin\beta} \right)^2,
\end{equation}
which is displayed in Fig.\ \ref{Higgsratio}, where we plot this ratio for different values of $\tan\beta$ as a function of the light Higgs mass. In the left panel we fix $\delta=10^{-2}$, whereas in the right one $\delta=10^{-3}$. One can see that if the product $\delta\tan\beta$ is sufficiently small, the light Higgs decays dominantly into $Z^\prime Z^\prime$, as predicted by Eq.\ \eqref{Eqhiggsratio}, justifying our procedure in the derivation of Fig.\ \ref{Hcouplings}. A more detailed study regarding the light Higgs properties has been conducted elsewhere \cite{Lee:2013fda}. In this work, we are limited to discuss all relevant limits to the $U(1)_X$ models introduced above. \\

For completeness, in the right panel of Fig.\ \ref{Hcouplings}, we exhibit the limits from LHC data for the 2HDM type I, assuming that the light Higgs is instead the SM Higgs, i.e.\ $m_h > m_H$, whose interactions strength with gauge bosons is proportional $\cos(\beta-\alpha)$. The SM limit in this case is found when $\beta \sim \alpha +\pi/2$, where $H \sim \phi_1$ and $h \sim \phi_2$.  Fig.\ \ref{Hcouplings} is still useful to us though, because it shows that for $\tan \beta > 1$ larger deviations from the SM Higgs are allowed. Moreover, we indeed assume $v_2 > v_1$, thus $\tan\beta >1$, throughout.

\subsubsection{Higgs Decays}

\begin{table}
\centering
\begin{tabular}{|c|c|c|}
\hline
Higgs decay channel & branching ratio & error\\
\hline
$b\bar{b}$ & $5.84 \times 10^{-1}$ & 1.5\% \\
\hline
$c\bar{c}$ & $2.89 \times 10^{-2}$ & 6.5\% \\
\hline
g g &  $8.18 \times 10^{-2}$ & 4.5\% \\
\hline
$Z Z^{\ast}$ & $2.62 \times 10^{-1}$ & 2\% \\
\hline
$W W^{\ast}$ & $2.14 \times 10^{-1}$ & 2\%  \\
\hline
$\tau^+\tau^-$ & $6.27 \times 10^{-2}$ & 2\% \\
\hline
$\mu^+\mu^-$ & $2.18 \times 10^{-4}$ & 2\% \\
\hline
$\gamma\gamma$ & $2.27 \times 10^{-3}$& 2.6\% \\
\hline
$Z\gamma$ & $1.5 \times 10^{-3}$ & 6.7\% \\
\hline
$Z Z^{\ast}\rightarrow 4 \ell$ & $2.745 \times 10^{-4}$ & 2\% \\
\hline
 $Z Z^{\ast}\rightarrow 2 \ell 2\nu$ & $1.05 \times 10^{-4}$ & 2\% \\
\hline 
\end{tabular}
\caption{List of experimental limits on the branching ratio of the SM Higgs. The channel $ ZZ^{\ast}\rightarrow 2\ell 2\nu$ was obtained using the relation $BR(H\rightarrow Z Z^{\ast} \rightarrow 2\ell 2\nu)= BR(H \rightarrow ZZ^{\ast}) BR(Z \rightarrow 2\ell)BR(Z \rightarrow 2\nu)^2 $.}
\label{tableLHChiggs}
\end{table}

After the Higgs discovery the LHC has turned into a Higgs factory and today we have at our disposal much better measurements of the Higgs branching ratio (see Table \ref{tableLHChiggs}). Since we are mostly interested in the regime in which the $Z^{\prime}$ is light enough for the Higgs to decay into, some channels are of great interest for our purposes, namely 
$H \to Z Z^* \to 4\ell$ and $H \to Z Z^* \to 2\ell 2\nu$. In the context of 2HDM it has been shown that in the limit in which the $Z^{\prime}$ gauge boson is much lighter than the $Z$ boson we get \cite{Lee:2013fda},
\beq
\Gamma (H \to Z Z^{\prime})\\
= \frac{g_Z^2}{64\pi}\frac{(M_H^2-M_Z^2)^3}{M_H^3 M_Z^2} \delta^2 \tan\beta^2 \sin^2(\beta-\alpha),
\label{eq:20}
\eeq 
and 
\beq
\Gamma (H \to Z^\prime Z^{\prime})
= \frac{g_Z^2}{128\pi}\frac{M_H^3}{M_Z^2} \delta^4 \tan\beta^4 \left( \frac{\cos^3\beta \sin\alpha +\sin^3\beta \cos\alpha}{\cos\beta \sin\beta} \right)^2.
\label{eq:21}
\eeq 

One can now use precision measurements on Higgs properties summarized in Table \ref{tableLHChiggs} to constrain the model. We will focus on the decay into $ZZ^{\prime}$ since $\delta$ is supposed to be small to obey meson decay constraints\footnote{In some regions of the parameter space with sufficiently large $\tan\beta$ the decay $Z^{\prime}Z^{\prime}$ might become relevant as discussed in \cite{Lee:2013fda}.}.  Enforcing the branching ratio $\Gamma (H \rightarrow Z Z^{\prime} \rightarrow 4 \ell)/\Gamma_{\rm total}$ with $\Gamma_{\rm total}=4.1\, {\rm MeV}$, to match the measured value within the error bars as indicated in the Table \ref{tableLHChiggs} we obtain,
\begin{equation}
\delta^2 \leq \frac{4.6\times10^{-6}}{{ BR(Z^{\prime} \rightarrow l^+l^-)\sin^2(\beta-\alpha)\tan\beta^2}}.
\label{Higgsdelta}
\end{equation}

To have an idea on how competitive this constraint is compared to previous discussions we shall plug in some numbers. Taking $\sin^2(\beta-\alpha)=0.01$ and $\tan\beta=10$, we get
\begin{equation}
\delta \leq \frac{0.002}{\sqrt{  BR(Z^{\prime} \rightarrow l^+l^-)}}, 
\end{equation}which is comparable to the bound stemming from Kaon decays. We emphasize that this bound is applicable to all $U(1)_X$ models under study here.  One need now to simply choose a model and substitute the respective branching ratio into charged leptons as  provided by Fig.\ \ref{Zpdecays2}.

\subsection{$Z$ Decays}
\label{sec:3B}
In the models we are investigating both the light Higgs $h$ and the $Z^{\prime}$ can be much lighter than the $Z$, kinematically allowing the decay $Z \rightarrow h Z^{\prime}$. In the limit that the $Z^{\prime}$ mass is very small compared to the $Z$ mass we find,
\bea
\Gamma (Z \to h Z') & = & ({\cal C}_{h-Z-Z'})^2 \frac{m_Z}{64\pi m_{Z'}^2} \left( 1 - \frac{m_h^2}{m_Z^2} \right)^3.
\eea 
where (see Appendix \ref{sec:app6})
\begin{equation}
{\cal C}_{h-Z-Z^\prime}= g_Z g_X v \cos\beta \sin\beta \cos(\beta-\alpha).
\end{equation}

Knowing that we can write down the $Z^{\prime}$ mass as a function of $\delta$, as derived in Appendix \ref{sec:app3}, we get
\bea
\Gamma (Z \to h Z') 
&= & \frac{g_Z^2 m_Z}{64 \pi} \left( \delta \tan\beta \right)^2 \cos^2 (\beta-\alpha) \left( 1 - \frac{m_h^2}{m_Z^2} \right)^3 .
\label{EqZhZp}
\eea

\begin{figure}
\centering
\includegraphics[scale=0.4]{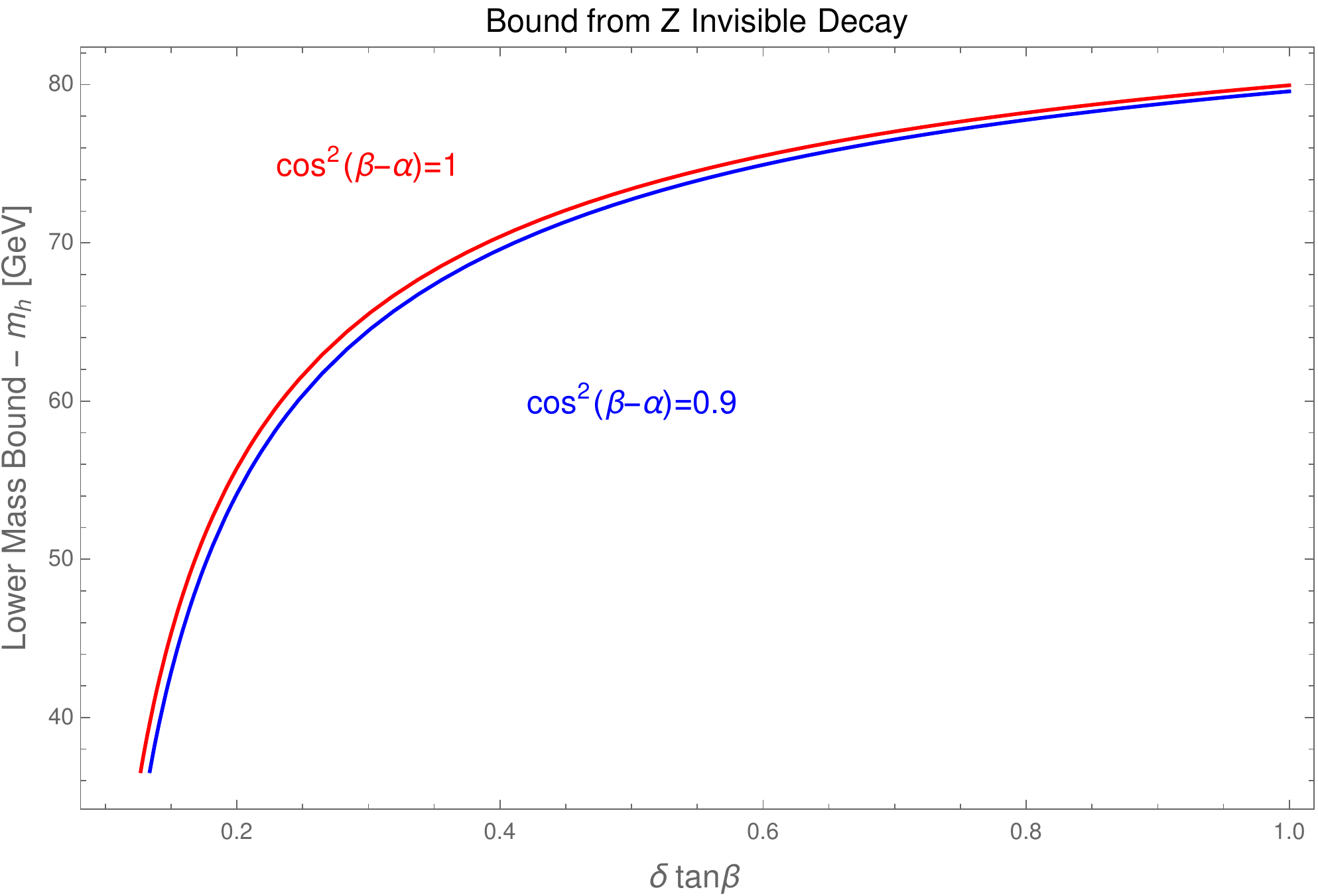}
\caption{Lower mass bound on the light Higgs stemming from the LEP precision measurement on the $Z$ width.
}\label{figmhbound}
\end{figure}

We highlight that the exact expression for this decay depends on the $\Phi_1$ charge under $U(1)_X$. Eq.\ \eqref{EqZhZp} is valid for the $B-L$ model for instance, and it agrees with \cite{Lee:2013fda}. Anyways, knowing that the total decay width of the $Z$ is $\Gamma_Z= 2.4952\pm 0.0023$~GeV \cite{Carena:2003aj}, one can conservatively enforce the new physics decay to be within the error bars of the measured value. One can use this to place a lower mass limit on $m_h$ as a function of $ \delta \tan\beta$ taking $\cos^2(\beta-\alpha) \sim 0.9-1$ as shown in Fig.\ \ref{figmhbound}.\\

One can conclude that for sufficiently small $\delta \tan\beta$ the bounds from LEP substantially weaken. We have seen in the previous sections that $\delta < 10^{-2} -10^{-3}$, and since we are interested in the limit of large $\tan\beta$, say $\tan\beta \sim 10$, then the light Higgs in the $U(1)_X$ models under study can be arbitrarily light as long as a fine-tuning in Eq.\ (\ref{massescalars}) is invoked. It has been noted that if $\sin\alpha$ is different from unity, $m_h$ cannot be lighter than $m_H/2$, otherwise the heavy Higgs, i.e.\ the SM Higgs, would decay dominantly into $hh$ in strong disagreement with data \cite{Ferreira:2012my}. Thus this very light Higgs scenario is only possible in the limit  $\sin\alpha= 1$. 

\subsection{Charged Higgs Searches}

A pleasant feature of our framework is that in contrast to canonical 2HDM 
there is no pseudoscalar Higgs particle\footnote{There is a pseudoscalar associated with a Higgs singlet, which remains decoupled as we assume no mixing between the doublets and singlet. When sizable mixing is introduced, the remaining pseudoscalar would have small couplings to the SM particles and it could be in principle detectable.\\
}, as this degree of freedom becomes the longitudinal polarization of the $Z'$.
However, it certainly does have a charged scalar, $H^\pm$, which is orthogonal to the longitudinal component of the $W^\pm$.  The phenomenology of the charged Higgs in the ordinary Type I model was recently discussed in Ref.~\cite{Branco:2011iw}.\\

In 2HDM type I, the coupling of the charged Higgs to fermions is suppressed by a factor of $\tan\beta$. In the models under study, the charged Higgs mass is found to be $m_{H^+}^2 = \frac{\lambda_4}{2} v^2$. This mass determines which final state is  dominant in its decays \cite{Searches:2001ac,Jung:2010ik,Branco:2011iw,Aoki:2011wd,Aad:2012tj}. In this work we will adopt $\lambda_4 \sim 1$, and this case the  $hW$, $HW$ and $t\bar{b}$ decays are the dominant ones and are found to be described by  \cite{Djouadi:1995gv,Akeroyd:1998dt} 
 \beq
\Gamma (H^\pm \to h W^\pm) = \frac{\cos^2(\beta - \alpha)}{16 \pi v^2} \frac{1}{m_{H^\pm}^3} \lambda^{3/2}(m_{H^\pm}^2,m_h^2,m_W^2) \label{eq:HhW}
\eeq
with $\lambda(x,y,z) \equiv x^2 + y^2 + z^2 - 2xy - 2yz - 2zx$, 
\beq
\Gamma (H^\pm \to H W^\pm) = \frac{\sin^2(\beta - \alpha)}{16 \pi v^2} \frac{1}{m_{H^\pm}^3} \lambda^{3/2}(m_{H^\pm}^2,m_H^2,m_W^2) ,
\eeq
and the decay width into $t \bar b$ is given by
\beq
\Gamma (H^\pm \to t \bar b) \simeq \frac{3 m_{H^\pm}}{8 \pi v^2} \frac{m_t^2}{\tan^2\beta} \left(1 - \frac{m_t^2}{m_{H^\pm}^2}\right)^2
\eeq where we have taken $V_{tb} = 1$.

The constraints coming from charged Higgs bosons searches are not very restrictive and in the limit of large $\tan\beta$ as  assumed in this work, charged Higgs searches do not yield competitive limits and thus ignored henceforth. For a detailed discussion see \cite{Ramos:2013wea}.

\subsection{Atomic Parity Violation}
\label{sec3}

The search  for Atomic Parity Violation (APV) provides a promising pathway to probe new physics, especially the existence of neutral light bosons. It is known that for $m_{Z^\prime} \sim 0.1-1$~GeV, existing limits exclude $\epsilon^2 > 10^{-6}$ \cite{Batell:2014mga}. As we shall see in what follows, APV 
offers an orthogonal and complementary probe for new physics depending on the parameter $\delta$.\\

Anyways, this parity violation is two fold: (i)  it can be induced via the non-zero SM fermion charges under the $U(1)_X$ symmetry; (ii) it can arise via the $Z^\prime-Z$ mass mixing. That said, let us first review how one can constrain $U(1)_X$ models via atomic parity violation.  Using effective field theory  APV is parametrized as \cite{Bouchiat:2004sp}
\begin{equation}
\label{leff}
\begin{array}{ccl}
{-\ \mathcal L}_{\rm eff}&=&
\frac{g^2+g'^2}{m_Z^{\,2}}  
 \frac{1}{4}\bar e\gamma_\mu\gamma_5e 
\left[ \,(\,\frac{1}{4}-\frac{2}{3}\,s_{W}^2\,)\ \,\bar u\gamma^\mu u\, +\,
(\,-\,\frac{1}{4}+\frac{1}{3}\,s^2\,)\ \,\bar d\,\gamma^\mu d\,
\right]\\
&&-\frac{f_{Ae}}{m_{Z'}^{\,2}}\bar e\,\gamma_\mu\gamma_5\,e
\left[f_{Vu}\ \bar u\gamma^\mu u\,+\,f_{Vd}\ \bar d\,\gamma^\mu d\,\right].
\end{array}
\end{equation}

\begin{table*}[tb]
\begin{tabular}{|c|r|l|l|}
\hline
~~Experiment~~   & ~$\left<Q\right>$~~~~~ & ~$\sin^2\theta_W$($m_Z$)~ & ~~~~Bound on dark $Z$ ~($90\%$ CL)~~ \\
\hline
~Cesium APV~     & ~$2.4 ~\rm MeV$~ & ~~$0.2313(16)$~  & ~$\eps^2 < \frac{39 \times 10^{-6}}{\delta^2} \left( \frac{m_{Z_d}}{m_Z} \right)^2 \frac{1}{K(m_{Z_d})^2}$~ \\
\hline
~E158 (SLAC)~    & ~$160 ~\rm MeV$~ & ~~$0.2329(13)$~  & ~$\eps^2 < \frac{62 \times 10^{-6}}{\delta^2} \left( \frac{(160 ~\rm MeV)^2 + m_{Z_d}^2}{m_Z \, m_{Z_d}} \right)^2$~ \\
\hline
~Qweak (JLAB)~   & ~$170 ~\rm MeV$~ & ~~$\pm 0.0007$~  & ~$\eps^2 < \frac{7.4 \times 10^{-6}}{\delta^2} \left( \frac{(170 ~\rm MeV)^2 + m_{Z_d}^2}{m_Z \, m_{Z_d}} \right)^2$~ \\
\hline
~Moller (JLAB)~  & ~$75 ~\rm MeV$~  & ~~$\pm 0.00029$~ & ~$\eps^2 < \frac{1.3 \times 10^{-6}}{\delta^2} \left( \frac{(75 ~\rm MeV)^2 + m_{Z_d}^2}{m_Z \, m_{Z_d}} \right)^2$~ \\
\hline
~MESA (Mainz)~   & ~$50 ~\rm MeV$~  & ~~$\pm 0.00037$~ & ~$\eps^2 < \frac{2.1 \times 10^{-6}}{\delta^2} \left( \frac{(50 ~\rm MeV)^2 + m_{Z_d}^2}{m_Z \, m_{Z_d}} \right)^2$~ \\
\hline
\end{tabular}
\caption{Existing (Cesium, E158) and projected constraints on the kinetic mixing parameter as a function of the mass mixing parameter $\delta$ and the $Z^{\prime}$ mass. All masses are in MeV, hence $m_Z =91 000$ MeV.
}
\label{tab:1}
\end{table*}
\begin{figure}
\centering
\includegraphics[scale=0.5]{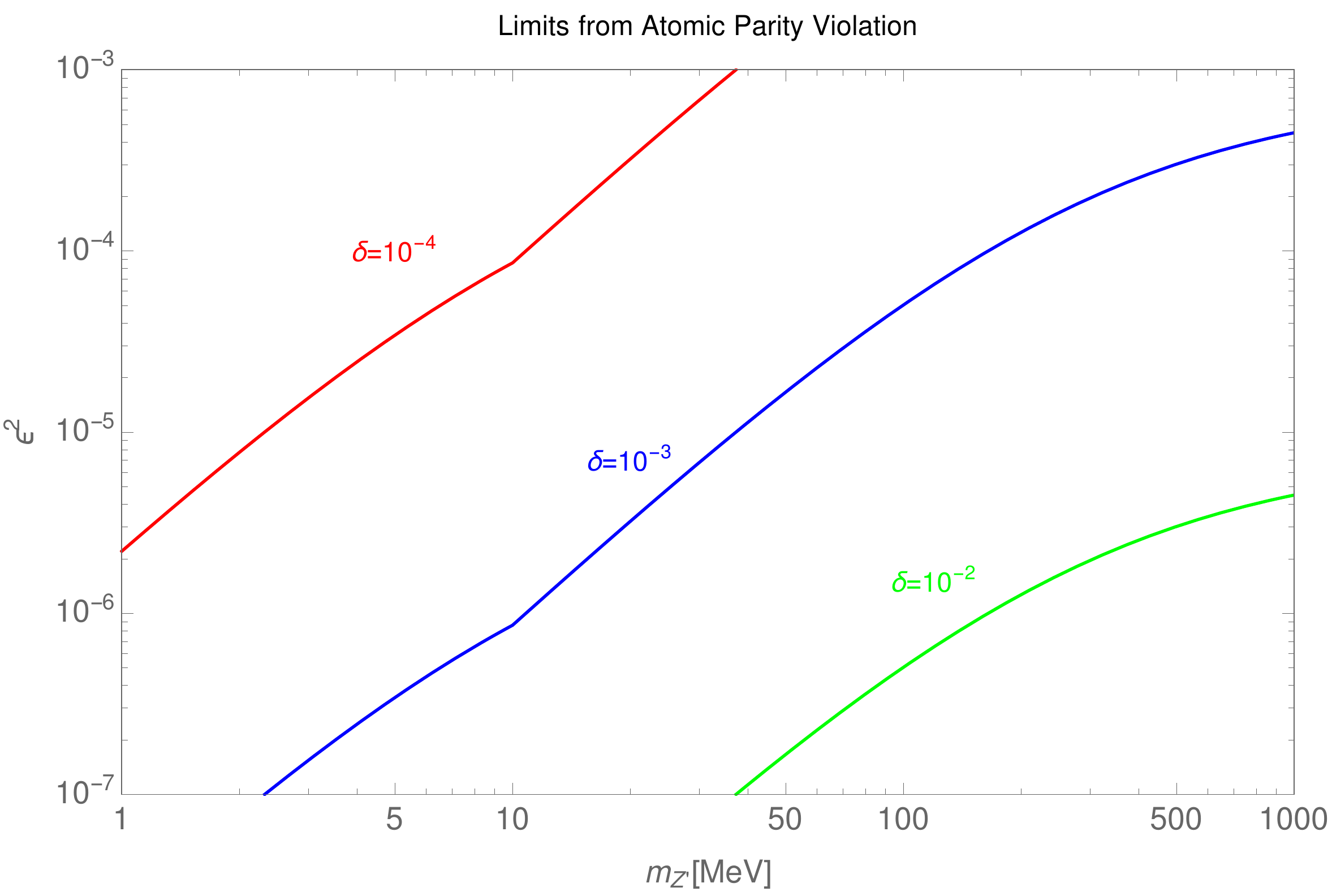}
\caption{Upper limits on the kinetic mixing as a function of the $Z^\prime$ mass for different values of the mass mixing parameter $\delta$ according to the first line of Table \ref{tab:1}. }
\label{Fig:APV}
\end{figure}
Here $f_{xy}$ are effective couplings to be derived below for the different models. \\

The Lagrangian involves the product of the $Z$ and $Z^{\prime}$ axial vector currents of the electron with the vector neutral currents of the quarks. Remembering that the vector part of the $Z$ weak neutral current is associated with the $Z$ weak charge, we get from Eq.\ \eqref{leff}
\begin{eqnarray}
Q_{Z} & = & (2Z+N)\left(\dfrac{1}{4}-\dfrac{2}{3}\sin^{2} \theta_{W}\right)+(Z+2N)\left(-\dfrac{1}{4}+\dfrac{1}{3}\sin^{2}\theta_{W}\right),\\\nonumber
	& = & \dfrac{1}{4}\left[Z(1-4\sin^{2}\theta_{W})-N\right]=\dfrac{1}{4}Q_{W}^{\rm SM}(Z,N).
\end{eqnarray}
The quantity $Q_{W}^{\rm SM}$ is usually referred to as weak charge of a nucleus of $Z$ protons and $N$ neutrons. Similarly, the quark contribution to the charge $Q_{Z^{\prime}}$ associated with the vector part of the $Z^{\prime}$ current is found to be
\begin{eqnarray}
\label{qzcharge}
Q_{Z^\prime} & = & (2Z+N)f_{Vu}+(Z+2N)f_{Vd}\\
	& = & (2f_{Vu}+f_{Vd})z + (f_{Vu}+2f_{Vd})N.
\end{eqnarray}
The effective Lagrangian Eq.\ \eqref{leff} implies the following parity violation Hamiltonian density for the electron field in the vicinity of the nucleus\footnote{$\delta(\vec{r})$ can be replaced by the nuclear density $\rho(\vec{r})$ to take into account finite size effects of the nucleus. For a more detailed discussion about APV see reference \cite{Bouchiat:2004sp}.}
\begin{equation}
\label{heff}
\begin{aligned}
{\cal H}_{\rm eff}=&e^\dagger(\vec r\,)\,\gamma_5\,e(\vec r\,)
\left[\frac{g^2+g'^2}{4\,m_Z^{\,2}}\frac{1}{4}Q_{W}^{\rm SM}
-\frac{g^2+g'^2}{4\,m_{Z'}^{2}}\epsilon_{Z}^{2}\left(1-\dfrac{l-e}{Q_{x1}\cos^2_\beta+Q_{x2}\sin^2_\beta}\right)Q_{Z'}\right]\delta(\vec r)\\
=&e^\dagger(\vec r)\gamma_5e(\vec r)\frac{G_F}{2\sqrt 2}Q_W^{\rm \,eff}(Z,\,N)\delta(\vec r),
\end{aligned}
\end{equation}
where
\begin{equation}
Q_{W}^{\rm eff}=Q_{W}^{\rm SM}-4\delta^2Q_{Z'}{\left(1-\dfrac{l-e}{Q_{x1}\cos^2_\beta+Q_{x2}\sin^2_\beta}\right)},
\label{qeffective}
\end{equation} 
using that $\epsilon_{Z}=\frac{M_{Z'}}{M_Z}\delta$. We remind the reader that $l$ and $e$ are the charges of the left-handed and right-handed electron under $U(1)_X$.\\

Notice that the effective weak charge of the nucleus $Q_{W}^{\rm eff}$ includes in addition to the standard contribution 
$Q_{W}^{ \rm SM}$ an additional $Z^{\prime}$ contribution. In order to know $Q_{W}^{\rm eff}$, it is necessary to calculate $Q_{Z^{\prime}}$. To do so, we need to specify from Eq.\ \eqref{www1} $f_{Vu}$ and $f_{Vd}$ associated to the $Z^{\prime}$ boson,
\begin{equation}
\begin{split}
f_{Vu} &= \left[ \frac{1}{4} - \frac{2}{3}\sin ^2 \theta _W(1-\frac{\epsilon \cos\theta_W}{\epsilon_{Z}\sin\theta_W})+\dfrac{1}{4}\dfrac{q+u}{Q_{x1}\cos^2_\beta+Q_{x2}\sin^2_\beta} \right] ,
\end{split}
\label{zprimeudifferent2}
\end{equation}
\begin{equation}
\begin{split}
f_{Vd} &= \left[ -\frac{1}{4} + \frac{1}{3}\sin ^2 \theta _W(1-\frac{\epsilon \cos\theta_W}{\epsilon_{Z}\sin\theta_W})+\dfrac{1}{4}\dfrac{q+d}{Q_{x1}\cos^2_\beta+Q_{x2}\sin^2_\beta} \right],\end{split}
\label{zprimeddifferent3}
\end{equation}where $q(u)$ is the charge of the left-handed (right-handed) quark field under $U(1)_X$.\\

Substituting \eqref{zprimeudifferent2} and \eqref{zprimeddifferent3} into \eqref{qeffective} we obtain the following general expression for $\Delta Q_{W}= Q_W^{\rm eff} - Q_W^{SM}$, 
\begin{eqnarray}
\label{deltaqsogeral}
\Delta Q_{W} & = & -\delta^2 Q_{W}^{\rm SM}-\delta^24 Z\sin\theta_W\cos\theta_W\dfrac{\epsilon}{\epsilon_{Z}} -\delta^2\dfrac{(q+u)(2Z+N)}{Q_{x1}\cos^2_\beta+Q_{x2}\sin^2_\beta}\\\nonumber
 & - & \delta^2\dfrac{(q+d)(Z+2N)}{Q_{x1}\cos^2_\beta+Q_{x2}\sin^2_\beta}{\left(1-\dfrac{l-e}{Q_{x1}\cos^2_\beta+Q_{x2}\sin^2_\beta}\right)}. 
\end{eqnarray}
Currently, the SM prediction for the weak nuclear charge in the Cesium case is \cite{Marciano:1982mm}
\begin{equation}
Q_{W}^{\rm SM}=-73.16(5), 
\end{equation}
so that the general expression Eq.\ \eqref{deltaqsogeral} becomes:  
\begin{eqnarray}
\label{deltaqsogeralcesium}
\Delta Q_{W} & = & 73.16 \delta^2-220\delta\left(\epsilon\frac{M_Z}{m_Z'}\right)\sin\theta_{W}\cos\theta_{W} -\delta^2\dfrac{188(q+u)}{Q_{x1}\cos^2_\beta+Q_{x2}\sin^2_\beta}\\\nonumber
 & - & \delta^2\dfrac{211(q+d)}{Q_{x1}\cos^2_\beta+Q_{x2}\sin^2_\beta}{\left(1-\dfrac{l-e}{Q_{x1}\cos^2_\beta+Q_{x2}\sin^2_\beta}\right)}.
\end{eqnarray}
On the other hand the experimental value for the weak nuclear charge in the Cesium case is  \cite{Porsev:2010de,Bennett:1999pd} 
\begin{equation}
Q_{W}^{\rm exp}=-73.16(35),
\end{equation}
and the $90\%$ CL bound on the difference is \cite{Davoudiasl:2012qa}
\begin{equation}
\label{noventapercent}
|\Delta Q_{W}(Cs)|=|Q_{W}^{\rm exp}-Q_{W}^{\rm SM}|<0.6,
\end{equation}
which yields the general APV expression for $U(1)_X$ models for the Cesium nucleus:
\begin{eqnarray}
\label{deltaqsogeral1}
 &  & \left| 73.16 \delta^2-220\delta\left(\epsilon\frac{M_Z}{m_Z'}\right)\sin\theta_{W}\cos\theta_{W} -\delta^2\dfrac{188(q+u)}{Q_{x1}\cos^2_\beta+Q_{x2}\sin^2_\beta} \right. \\ \nonumber 
 & - & \left.  \delta^2\dfrac{211(q+d)}{Q_{x1}\cos^2_\beta+Q_{x2}\sin^2_\beta}{\left(1-\dfrac{l-e}{Q_{x1}\cos^2_\beta+Q_{x2}\sin^2_\beta}\right)}\right|\times K(Cs)<0.6.
\end{eqnarray}
The correction factor $K(Cs)$ is introduced for low values of $M_{Z'}$ where the local limit approximation is not valid. Different values for this correction factor are listed in Table I of reference \cite{Bouchiat:2004sp}. At first order, one can drop the terms proportional to $\delta^2$ in Eq.\ \eqref{deltaqsogeral1} and then solve it for $\epsilon$ in terms of $\delta$, using $220\delta\left(\epsilon\frac{M_Z}{m_Z'}\right)\sin\theta_{W}\cos\theta_{W}=0.6$.\\

Doing so, we find the bound shown in the first line of Table \ref{tab:1}. 
The numerical upper limit on the kinetic mixing as a function of the $Z^\prime$ mass for different values of $\delta$ taking into account the energy dependence on $K(Cs)$ is displayed in Fig.\ \ref{Fig:APV}. \\

It is useful again to apply our procedure to a well known model in the literature such as the $B-L$ model. In this case $q=u=d=1/3$, $Q_{x2}=0$, $Q_{x1}=2$, $\ell = e$. With these values the expression \eqref{deltaqsogeralcesium} becomes 
\begin{equation}
\Delta Q _W = -59.84 \delta ^2 - 220\delta\left(\epsilon\frac{M_Z}{m_Z'}\right)\sin\theta_{W}\cos\theta_{W} - 133 \delta ^2 \tan ^2 \beta,
\label{deltaQtanbeta}
\end{equation}
which coincides with the expressions obtained in \cite{Davoudiasl:2012ag,Davoudiasl:2012qa}, except for the last term, that  arises due to the non-zero $U(1)_{B-L}$ charges of the fermions. Applying the $90\%$ CL bound in Eq.\ \eqref{deltaqsogeral1} we get 
\begin{equation}
\left|-59.84 \delta ^2 - 220\delta\left(\epsilon\frac{M_Z}{M_Z'}\right)\sin_{W}\cos_{W} - 133 \delta ^2 \tan ^2 \beta\right| \times K(Cs)<0.6.
\label{deltaQW}
\end{equation}
From Eq.\ \eqref{deltaQW} we can see the term proportional to $\delta^2$ can not always be dropped as we did before to obtain the  limit in the first line of Table \ref{tab:1}. For sufficiently large $\tan\beta$ the last term in Eq.\ \eqref{deltaQW} might become relevant yielding changes for the upper limits on the kinetic mixing. Since the importance of this last term is rather model dependent we will not devote time to discuss its impact here.\\
 
Regardless, the conclusion that Cesium nucleus provides an interesting and orthogonal test for new physics stands, and depending on the $U(1)_X$ model under study it gives rise to restrictive limits on the kinetic mixing parameter following Table \ref{tab:1}.\\

Another observable in APV experiments is given by the value of $\sin \theta_W$ that is measured at low energies. The shift in  $\sin^2\theta_W$ caused by the presence of a new vector boson that mixes with the $Z$ boson is found to be \cite{Davoudiasl:2012qa}
\begin{equation}
\Delta \sin^2\theta_W = -0.42 \epsilon \delta \frac{m_Z}{ m_{Z^{\prime}} } \frac{ m_{Z^\prime}^2}{m_{Z^\prime}^2 + Q^2},
\label{Eq:deltasin}
\end{equation}where $Q$ is the energy at which $\sin\theta_W$ is measured and $\Delta \sin^2\theta_W$ refers to the error on the measurement of $\sin^2\theta_W$ as shown in Table \ref{tab:1}. By plugging the experimental error bar as displayed in the third column of Table \ref{tab:1} into Eq.\ \eqref{Eq:deltasin} one can derive upper limits on $\epsilon$  as a function of $\delta$ as shown in the fourth column of Table \ref{tab:1}. The first two rows in Table \ref{tab:1} refer to past experiments, whereas the remaining rows represent projected experimental sensitivities. \\

Anyways, one can see that the Qweak experiment is not expected to be as sensitive to the kinetic mixing as the first measurements, but both Moller and MESA experiments should be able to surpass previous experiments yielding tight bounds on the kinetic mixing \cite{Benesch:2014bas,Berger:2015aaa,Bucoveanu:2016bgx}.\\

\subsection{Muon Anomalous Magnetic Moment}
\label{sec4}

Any charged particle has a magnetic dipole moment ($\vec{\mu}$) defined as 
\begin{equation}
\vec{\mu} = g \left( \frac{q}{2m}\right) \vec{s},
\end{equation}
where $s$ is the spin of the particle, $g$ is the gyromagnetic ratio, $q =\pm e$ is the electric charge of a given charged particle, and $m$ its mass (see \cite{Lindner:2016bgg} for a recent and extensive review). Loop corrections induce deviations from the tree-level value $g=2$, which are parametrized for the muon in terms of $a_{\mu}=(g_{\mu}-2)/2$, referred to as the anomalous magnetic moment.  An enormous effort  has been dedicated to precisely determine the SM contribution to $g-2$ \cite{Garwin:1960zz,Burnett:1967zfb,Kinoshita:1967txv}. Interestingly, the SM prediction does not agree with recent measurements  leading to~\cite{Blum:2013xva} 
\begin{equation}
\begin{aligned}
\Delta a_{\mu} &= a_{\mu}^{exp} -a_{\mu}^{SM} = (287 \pm 80 ) \times 10^{-11}, 
\label{Eq:gmuanomaly}
\end{aligned}
\end{equation}
which implies a $3.6\sigma$ evidence for new physics. Therefore, it is definitely worthwhile to explore new physics models capable of giving rise to a positive contribution to $g-2$. In the $U(1)_X$ models under investigation, a particle that fulfills this role is the massive $Z^\prime$ that yields 
\cite{Lindner:2016bgg,Leveille:1977rc,Jegerlehner:2009ry} 
\begin{subequations}
  \begin{equation}
    \Delta a_\mu\left(f, Z^\prime\right) = \frac{1}{8\pi^2} \frac{m_\mu^2}{m_{Z^\prime}^2} \int_0^1 \mathrm{d}x \sum_f \frac{\left|g_{v}^{f\mu}\right|^2 F^+(x) + \left|g_{a}^{f\mu}\right|^2 F^-(x)}{(1-x)\left(1-\lambda^2 x\right)+\epsilon_f^2\lambda^2 x},
    \label{Eq:mucomplete}
  \end{equation}
  with
  \begin{equation}
    F^\pm = 2x(1-x) (x-2\pm 2\epsilon_f) + \lambda^2x^2(1 \mp \epsilon_f)^2(1-x \pm \epsilon_f)\label{eq:P4def}
  \end{equation}
\end{subequations}
and $\epsilon_f \equiv \frac{m_{f}}{m_\mu}$, $\lambda \equiv \frac{m_\mu}{m_{Z^\prime}}$. Here $f$ are charged leptons. Since we are not dealing with flavor changing interactions in this work, $\epsilon_f \equiv 1$ and $m_f=m_\mu$. Moreover, in the limit of a $Z^{\prime}$ much heavier than the muon, the 
contribution simplifies to
\begin{equation} 
	\Delta a_\mu\left(Z^\prime\right) \simeq \frac{1}{12\pi^2}\frac{m_\mu^2}{m_{Z^\prime}^2} (g_v^2 - 5 g_a^2),
	\label{Eq:mu}
\end{equation}
where $g_v$ and $g_a$ are the vector and axial vector couplings of the $Z^\prime$ with the muon. Notice that only models where the vector coupling is more than five times larger than the axial vector couplings are capable of addressing the $g-2$ anomaly in agreement with \cite{Queiroz:2014zfa}. This condition is satisfied only in the $U(1)_D$ and $U(1)_F$ models.\\

That said, the region that explain the $g-2$ anomaly is easily obtained through the equality 
\begin{equation}
\frac{g_v^2}{(m_{Z^\prime}[{\rm GeV}]) ^2} \simeq 3.3 \times 10^{-5}.
\end{equation}
For instance, in the $U(1)_D$ model $g_v=-1.75 g_X$. Keeping $g_X=1$, we need $ m_{Z^{\prime}} \sim$ $540$~GeV to accommodate the $g-2$ anomaly, which is way beyond the region of interest in this work.  Anyways, such heavy gauge bosons are subject to stringent limits from dimuon searches as shown in \cite{Profumo:2013sca,Allanach:2015gkd,Alves:2015mua,Alves:2015pea,Patra:2015bga,Khachatryan:2016zqb,Lindner:2016lpp,Altmannshofer:2016jzy,ATLAS:2017wce,Arcadi:2017kky}, preventing such gauge bosons to be a solution to the $g-2$ anomaly. However, if we set $g_X=10^{-4}$, then  $ m_{Z^{\prime}} \sim$ $54$~MeV is required, being potentially able to explain the $g-2$ anomaly, as long as the kinetic and mass mixing parameters are kept sufficiently small. A more thorough discussion of the possibility of explaining $g-2$ in each of these models will be made elsewhere. It is interesting to see though, that one might be able to cure 2HDM from flavor changing interactions, generate neutrino masses, while solving a relevant and long standing anomaly in particle physics.
junior

\subsection{Neutrino-Electron Scattering}

Intensity frontier constitutes a promising endeavor in the quest for new physics, being able to explore models inaccessible at high-energy
frontiers. One canonical example are the precise measurements on neutrino-electron scattering using different targets, as measured by several experiments such as TEXONO, GEMMA, BOREXINO, LSND and CHARM. Since neutrino interactions are purely leptonic, they are subject to small uncertainties. Moreover, interesting models such as the dark photon  and light $Z^{\prime}$ models such as ours, predict different signals at these experiments. Therefore, the use of neutrino-electron scattering to explore hints of new physics is both theoretically and experimentally well motivated. \\

\begin{figure}
\centering
\includegraphics[scale=0.5]{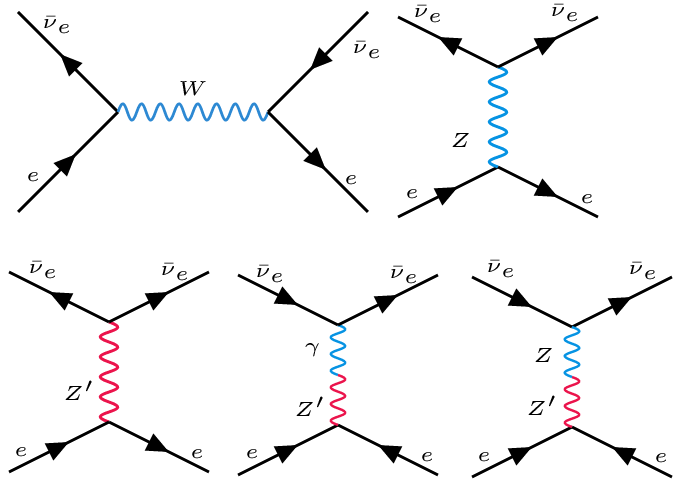}
\caption{Feynmann diagrams relevant for neutrino-electron scattering}
\label{Fig.nelectron}
\end{figure}

That said, several works have been done to place limits on new physics models based on neutrino-electron scattering data \cite{Miranda:1998ti,Ciechanowicz:2003ds,Kopp:2010qt,Harnik:2012ni,Liao:2013jwa,Rodejohann:2017vup}. Here we will briefly review the concept behind these and derive constraints the gauge couplings as a function of the $Z^{\prime}$ mass.\\

The physics behind these constraints lies on the computation of the neutrino-electron scattering due to new physics. In Fig.\ \ref{Fig.nelectron} we exhibit the SM diagram alongside the new physics ones. Following Ref.\ \cite{Bilmis:2015lja} the new physics neutrino-electron scattering cross section can be parametrized in terms of the $B-L$ model which is found to be \cite{Harnik:2012ni} 
\begin{equation}
\frac{d\sigma}{dE_R} = \frac{g_{B-L}^4 m_e}{4\pi E_\nu^2 (m_{Z^\prime}^2 + 2 m_e E_R)^2 } ( 2 E_\nu^2 + E_R^2 -e E_R E_\nu - m_e E_R)
\end{equation}where $E_R$ is the electron recoil energy, $E_\nu$ is the energy of the incoming neutrino, $m_e$ is the electron mass, and $G_F$ the Fermi constant. \\

\begin{figure}[t]
\centering
\includegraphics[scale=0.8]{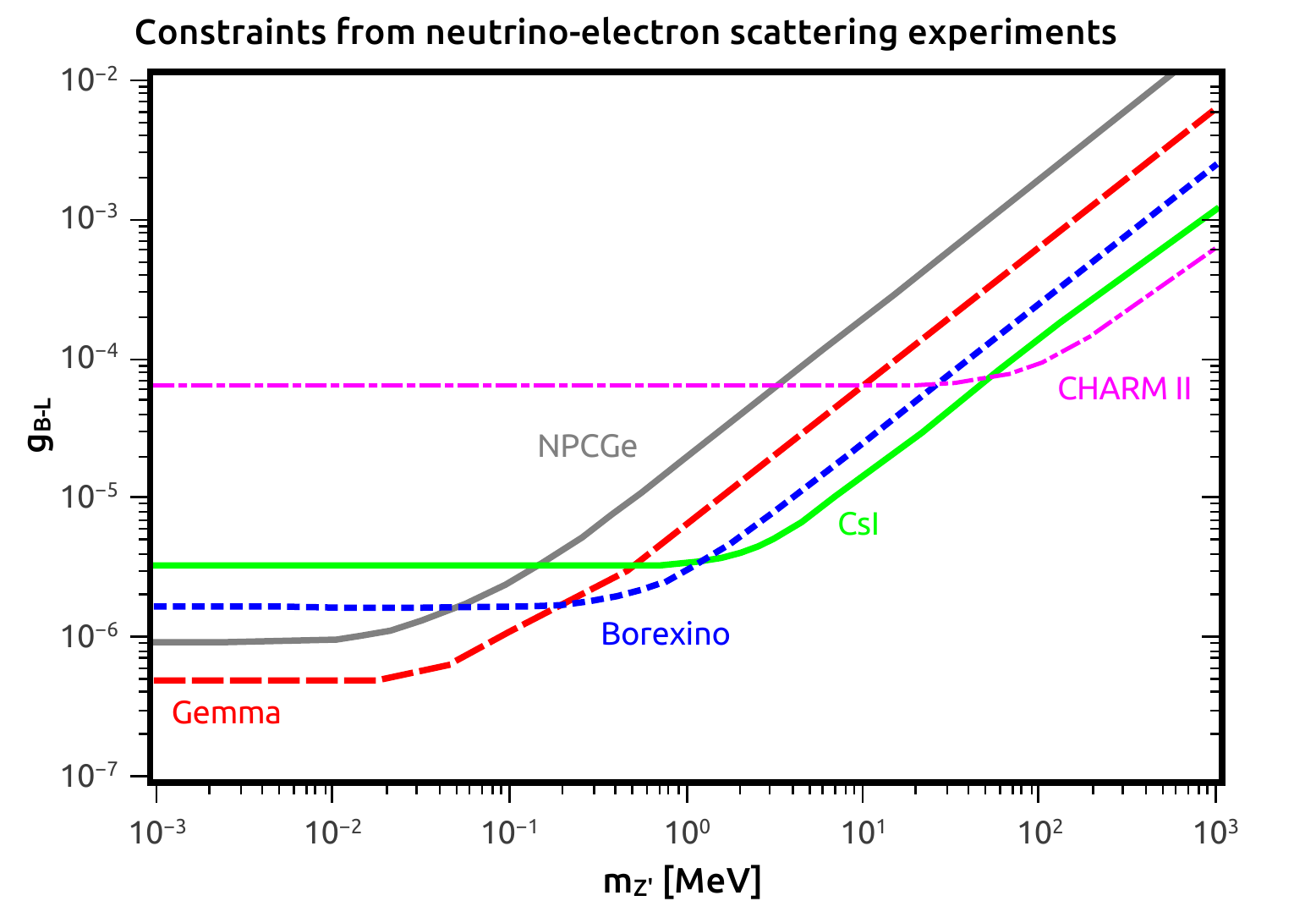}
\caption{Constraints on the $B-L$ model based on measurements of  neutrino-electron scattering.}
\label{fig:neutrinoelectronBLmodel}
\end{figure}

\begin{figure}[h]
\centering
\includegraphics[scale=0.8]{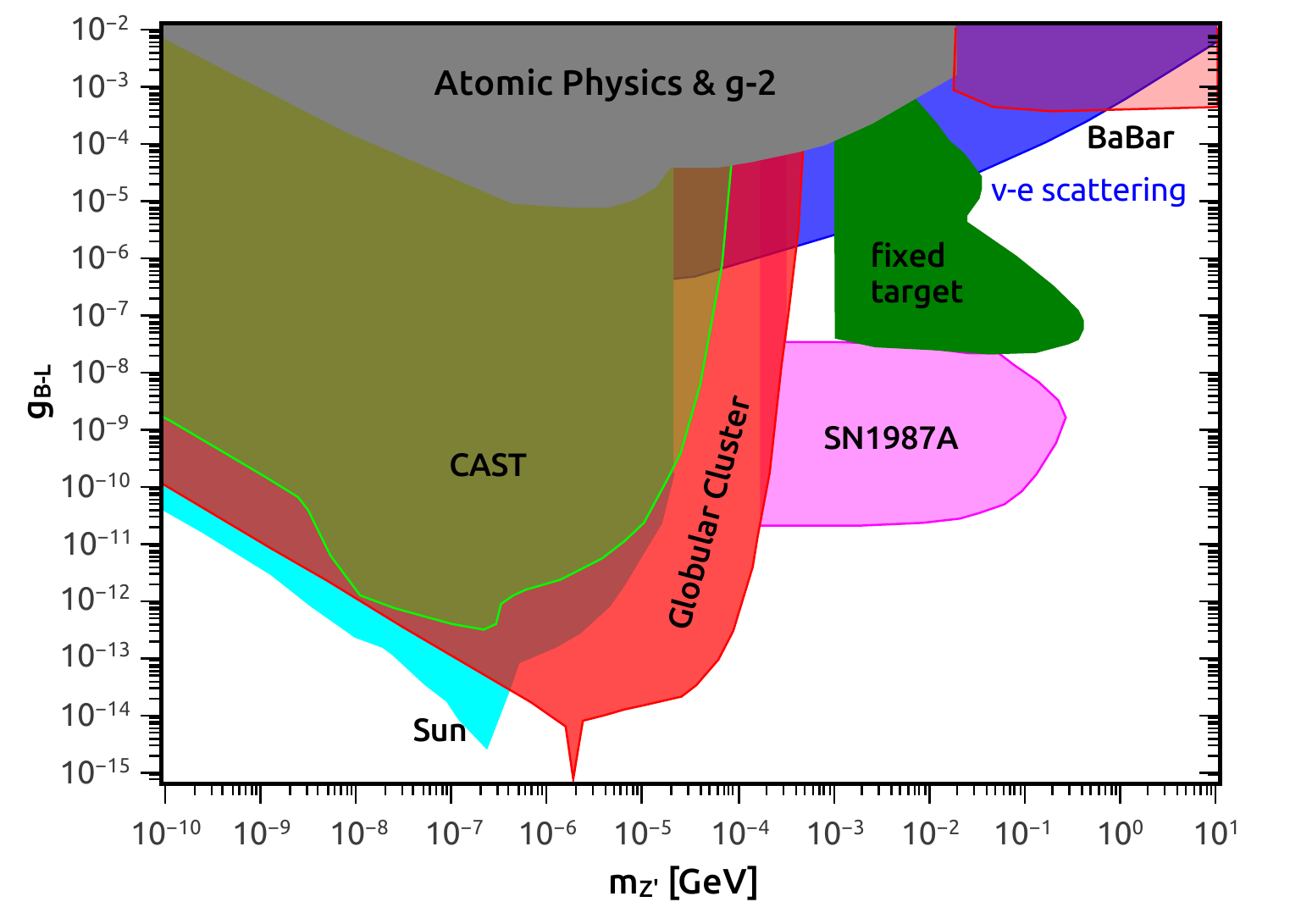}
\caption{Summary of constraints from neutrino-electron scattering on $U(1)_X$ models with very light $Z^{\prime}$ gauge bosons. These constraints have been interpreted from dark photon searches.}
\label{figneutrinoelectronglobal}
\end{figure}

The idea is to compute the expected neutrino-scattering rate from new physics, $(dR/dE_R)_{\rm NP}$, which is related to the neutrino-electron scattering through 
\begin{equation}
\left(\frac{dR}{dE_R}\right)_{NP} = t\, \rho_e \int_{E_\nu^{min}}^\infty \frac{d\Phi}{dE_\nu}\frac{d\sigma}{dE_R}dE_\nu , 
\end{equation}where $\Phi$ is the neutrino flux, $t$ is period of mock data taking, and $\rho_e$ is electron number density per kg of the target mass. Once that has been computed, one compares it with the measured rate and finds 90\% level limits applying a $\chi^2$ statistics as follows: 
\begin{equation}
\chi^2 = \sum_{i=1} \frac{\left(R_{\rm exp\,i} - (R_{\rm SM\, i}+ R_{\rm NP}) \right)^2}{\sigma_i}
\end{equation}where $R_{\rm exp}$, $R_{\rm SM}$ are the measured and SM predicted rates respectively, and $\sigma_i$ is the statistical error on the measurement of $R_{\rm exp}$. The index $i$ runs through energy bins.
Using data from several experiments subject to different energy threshold and type of incoming neutrino flavor as summarized in Table \ref{tab::experiment}, constraints on the new physics have been placed \cite{Bilmis:2015lja}. The limits were interpreted in terms of the $B-L$ model, as shown in Fig.\ \ref{fig:neutrinoelectronBLmodel}. These bounds are the most restrictive for $m_{Z^{\prime}} \sim 100\,{\rm MeV-1~GeV} $, as exhibited in Fig.\ \ref{figneutrinoelectronglobal} where all relevant constraints are put together. See \cite{Alexander:2016aln} for a recent review on neutrino-electron scattering experiments. \\

One needs to apply these constraints to the $U(1)_X$ models under study with care. Obviously, for the $B-L$ model in Table 2, the limits in Fig.\ \ref{figneutrinoelectronglobal} are directly applicable. For the remaining $U(1)_X$ models, one can estimate the limits through rescaling. Since the kinetic  and 
mass-mixing are constrained to be small, the leading diagram is the $t$-channel $Z^{\prime}$ exchange in Fig.\ \ref{Fig.nelectron}. Therefore, the scattering cross section scales with $g_{Z^{\prime}-\nu-\nu}^2 g_{Z^{\prime}-e-e}^2$, where $g_{Z^{\prime}-\nu-\nu}$, $g_{Z^{\prime}-e-e}$ are the $Z^\prime$ vectorial couplings with the neutrinos and electrons respectively. These are easily obtained knowing that the vector coupling with a given fermion field is $g_{fv} = g_X/2 (Q_{fL} +Q_{fR})$, where $Q_{fL}$ and $Q_{fR}$ are the charges of the left-handed and right-handed field components under $U(1)_X$ as displayed in Table \ref{cargas_u1_2hdm_tipoI}. In summary, there is a plot similar to Fig.\ \ref{fig:neutrinoelectronBLmodel} for each $U(1)_X$ model in this work. Clearly this exercise is outside the scope of this work. Anyways, it is clear that neutrino-electron scattering provides a competitive probe for new physics and is relevant for the $U(1)_X$ models under study. These bounds can be circumvented by tuning the kinetic mixing to sufficiently small values, similarly to the dark photon model.

\begin{table*} [t]
	\caption{Summary of experiments that constrained $\nu - e$ scattering.}
	\label{tab::experiment}
		\begin{tabular}{lcccc}
		 \hline 
			Experiment  &  
			Type of neutrino & $\left< E_{\nu} \right>$ & $T$ 
			\\ \hline 
			TEXONO-NPCGe~\cite{Chen:2014dsa} & $\nuebar$ &  1$-$2~MeV & 0.35$-$12 keV  
			\\
			TEXONO-HPGe~\cite{Li:2002pn,Wong:2006nx} & $\nuebar$ & 1$-$2~MeV & 12$-$60~keV 
			\\
			TEXONO-CsI(Tl)~\cite{Deniz:2009mu} & 
			$\nuebar$ & 1$-$2~MeV & 3$-$8~MeV
			\\
			LSND~\cite{Auerbach:2001wg} &
			$\nue$ & 36~MeV &18$-$50~MeV
			\\
			BOREXINO~\cite{Bellini:2011rx} &
			$\nue$ & 862 keV & 270$-$665~keV  
			\\
			GEMMA~\cite{Beda:2009kx} &
			$\nuebar$ & 1$-$2~MeV & 3$-$25~keV  
			\\
			CHARM II~\cite{Vilain:1993kd} &  
			$\nu_{\mu}$ &23.7~GeV &3-24~GeV
			\\
			CHARM II~\cite{Vilain:1993kd} & $\bar{\nu}_{\mu}$ &19.1~GeV &3-24~GeV  & 
			\\
			\hline
		\end{tabular}
\end{table*}

\subsection{Low Energy Accelerators}

\begin{figure}
\centering
\includegraphics[scale=0.7]{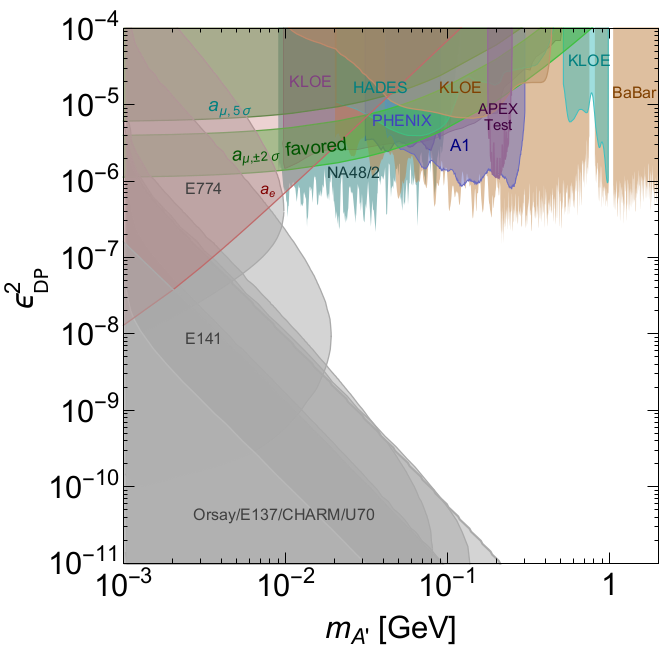}
\caption{Summary of bounds from low energy accelerators constraints on the dark photon model \cite{Alexander:2016aln}. After proper rescaling these constraints are also applicable to the $U(1)_X$ models in this work. In particular, the BaBar limits can be recast using the relation $\epsilon^2_{\rm DP} \rightarrow  (g_v^l)^2\, BR (Z^\prime \rightarrow l^+l^-)$. See text for details.}
\label{figdarkphoton}
\end{figure}

Low energy accelerators are capable of probing new physics models out of reach of high-energy colliders.  Models with light mediators, such as the dark photon model are considered a benchmark \cite{Pospelov:2007mp,Pospelov:2008zw}. The sensitivity of low energy accelerators is driven by high-intensity beams and/or high precision detectors. Such accelerators are usually divided into two classes: (i) collider; (ii) fixed-target experiments. In the former, high-intensity beams of $e^+e^-$ are capable of directly producing on-shell light mediators, whereas in the latter, light particles are produced as result of a decay chain created after the beam hits the target. In either case, the low-energy
accelerators are excellent laboratories to spot new physics effects. In Fig.\ \ref{figdarkphoton} we present a summary of current constraints on the dark photon model, with the dark photon, $A^\prime$, decaying into charged leptons. With care, the limits exhibited in Fig.\ \ref{figdarkphoton} can be applicable to the $U(1)_X$ are investigation.
For instance, the BaBar experiment searched for the $e^+e^- \rightarrow \gamma A^\prime$, with $A^\prime$ decaying into $l^+l^-$. The interaction of the dark photon with charged leptons reads $ \epsilon\, \bar{l} \gamma_\mu l\, A^{\prime \mu}$. Having in mind that the two important quantities are the production cross section and the branching ratio into electrons, one can recast the BaBar upper limits on the dark photon kinetic mixing ($\epsilon_{\rm DP}$) to other $U(1)_X$ models as follows
\begin{equation}
\epsilon^2_{\rm DP} \rightarrow  (g_v^l)^2\,  BR (Z^\prime \rightarrow l^+l^-),
\label{babareq}
\end{equation}where $g_v^l= g_X/2 (Q_L^l +Q_R^l)$ is the $Z^\prime$ vectorial coupling to charged leptons. Here $Q_L^l$ and $Q_R^l$ are the left-handed and right-handed charged lepton charges under $U(1)_X$.\\

In all $U(1)_X$ models that accommodate neutrino masses and are free from flavor changing interactions the $Z^\prime$ boson features a vectorial coupling with electrons. Since the SM fermions are charged under $U(1)_X$, in addition to the kinetic mixing term, a vectorial interaction proportional to $g_X$ also arises. Therefore, Eq.\ (\ref{babareq}) is valid when the term proportional to $g_X$ is dominant, otherwise, the bounds in Fig.\ \ref{figdarkphoton} are directly applicable. Hence, one can use Eq.\ \eqref{babareq} to obtain limits for each $U(1)_X$ model. A similar reasoning can be applied to other collider experiments. \\

As for fixed target experiments such as NA48/2, the rescaling is restricted to the branching ratio into charged leptons. Sometimes these experiments include both $e^+e^-$ and $\mu^+\mu^-$ decay modes in the analysis, while other times they consider only one of those. Our goal here is not to describe each one of these searches individually but rather present to the reader the existence of limits on the kinetic-mixing stemming from low energy accelerators. The precise bound on $\epsilon$ for each $U(1)_X$ model is not relevant for us, since they can all be evaded by simply tuning down the free kinetic mixing parameter.\\

Furthermore, it is worth pointing out that there is also a similar plot considering only invisible decays of the dark photon. However, as far as the $U(1)_X$ models go, the only possible invisible decay modes are the active neutrinos and right-handed neutrinos. Except in the case of the $U(1)_{B-L}$ model, this branching ratio is expected to be small, substantially weakening the limits on $\epsilon$. Thus the searches for visible decays are more constraining.\\

In summary, low energy accelerators yield very strong limits on the kinetic mixing parameter of the $U(1)_X$ models. 

\subsection{Discussion}

We have discussed a variety of limits on the parameters $\delta$ and the kinetic mixing $\epsilon$. They are model dependent. The bounds on $\delta$ were derived under the assumption that the fermions were uncharged under $U(1)_X$, where only the mass-mixing would dictate the $Z^\prime$ interactions with fermions. However, due to the presence of new interactions between the SM fermions and the $Z^{\prime}$ gauge boson these limits might be subject to changes by a factor of few depending on the value of $g_X$ and fermion charges under $U(1)_X$. As for the limits on the kinetic mixing, they were obtained assuming that the kinetic mixing alone dictates the observables since they were originally meant for the dark photon model. Since both $\delta$ and $\epsilon$, in principle, are arbitrarily small, the constraints presented in this work might be circumvented. A more detailed analysis incorporating precise bounds on the $U(1)_X$ models is left for the future. The goal of this work was to propose new 2HDM gauge models capable of accommodating neutrino masses and freeing the 2HDM from flavor changing interactions, as well as estimate what kind of phenomenology these models can generate. 

On a side note,  one more interesting avenue that is worth exploring is the near future is the possibility of accommodating a dark matter candidate. Naively, one can could have a Sub-GeV singlet fermion charged under $U(1)_X$ that interacts with SM fermions though the light $Z^{\prime}$ gauge boson, scenario which has been studied before in different contexts \cite{Arcadi:2013qia,Arcadi:2014lta,Arcadi:2015nea,Arcadi:2016qoz,Arcadi:2016kmk}.

\section{Conclusions}
\label{sec:concl}

Two Higgs Doublet Models are a natural extension of the Standard Model with interesting Higgs and collider phenomenology. These models are plagued with flavor changing interactions and for this reason a $Z_2$ symmetry has in the past been evoked to save  such flavor changing couplings from tight flavor constraints. In this work, we cure 2HDM model flavor changing interactions from gauge principles and additionally embed neutrino neutrino masses via the see-saw mechanism. In particular, we propose eight different models where neutrino masses and absence of flavor changing interactions are nicely explained. To do so, we gauge an Abelian gauge group that gives rise to a massive $Z^{\prime}$ via spontaneous symmetry breaking. We work in the light $Z^{\prime}$ regime, $m_{Z^\prime} \ll m_Z$, and investigate the associated phenomenology touching rare meson decays, Higgs physics, LEP precision data, neutrino-electron scattering, low energy accelerators and LHC probes.\\

In summary, we find that these models give rise to a rather rich phenomenology and can be made compatible with data while successfully generating neutrino masses and freeing 2HDM from flavor changing interactions.

\section*{Acknowledgments}

The authors are greatly thankful to Brian Batell, Rouven Essig and David Morrissey for correspondence. We are specially thankful to Rouven for proving the data files needed for Fig.10. FSQ is greatful to Carlos Pires, Paulo Rodrigues and Alvaro Ferraz for their hospitality at UFPB and UFRN where this project was partially conducted. FSQ thanks Fabio Iocco, Enrico Bertuzzo and Manuela Vecchi for the great Dark Matter Workshop in Sao Paulo where this project was finalized. D.C thanks Jo\~ao Silva and Jorge Rom\~ao from the CFTP Lisbon for useful discussions, time and motivation to develop this work. DC is partly supported by the Brazilian National Council for Scientific and Technological Development (CNPq), under grants 484157/2013-2 and 201066/2015-7. TM is supported by Coordena\c c\~ao de Aperfei\c c\~omento de Pessoal de N\'ivel Superior (Capes). WR is supported by the DFG with grant RO 2516/6-1 in the Heisenberg Programme. M.C was supported by the IMPRS-PTFS.

\appendix

\section{Conditions for Anomaly Freedom}
\label{sec:app1}

Generically we will call the $U(1)_{X}$ charges  $Y^{\prime}$, where $Y^{\prime} = l, q, e, u, d$. The anomaly free conditions can be read as:
\begin{description}

\item[\text{$ \left[ SU(3)_c \right] ^2 U(1)_X $}]: 
\begin{equation*}
\mathcal{A} = \text{Tr} \left[ \left\{ \frac{\lambda ^a}{2}, \frac{\lambda ^b}{2} \right\} Y ' _R \right] - \text{Tr} \left[ \left\{ \frac{\lambda ^a}{2}, \frac{\lambda ^b}{2} \right\} Y ' _L \right]
\end{equation*}
\begin{equation*}
\mathcal{A} \propto \sum _{\text{quarks}} Y ' _R - \sum _{\text{quarks}} Y ' _L = \left[ 3 u + 3 d \right] - \left[ 3 \cdot 2 q \right] = 0.
\end{equation*}
Therefore, 
\begin{equation}
u + d - 2 q = 0.
\label{anomalycond1}
\end{equation}

\item[\text{$ \left[ SU(2)_L \right] ^2 U(1)_X $}]: 
\begin{equation*}
\mathcal{A} = - \text{Tr} \left[ \left\{ \frac{\sigma ^a}{2}, \frac{\sigma ^b}{2} \right\} Y ' _L \right] \propto - \sum Y_L = - \left[ 2 l + 3 \cdot 2 q \right] = 0.
\end{equation*}
Therefore, 
\begin{equation}
l = - 3 q.
\label{anomalycond2}
\end{equation}

\item[\text{$ \left[ U(1)_Y \right] ^2 U(1)_X $}]: 
\begin{equation*}
\mathcal{A} = \text{Tr} \left[ \left\{ Y_R, Y_R \right\} Y ' _R \right] - \text{Tr} \left[ \left\{ Y_L, Y_L \right\} Y ' _L \right] \propto \sum  Y_R ^2 Y ' _R - \sum Y_L ^2 Y ' _L
\end{equation*}
\begin{equation*}
\mathcal{A} \propto \left[ \left( -2 \right) ^2 e + 3 \left( \frac{4}{3} \right) ^2 u + 3 \left( - \frac{2}{3} \right) ^2 d \right] - \left[ 2 \left( -1 \right) ^2 l + 3 \cdot 2 \left( \frac{1}{3} \right) ^2 q \right] = 0.
\end{equation*}
Therefore, 
\begin{equation}
6 e + 8 u + 2 d - 3 l - q = 0.
\label{anomalycond3}
\end{equation}

\item[\text{$ U(1)_Y \left[ U(1)_X \right] ^2 $}]: 
\begin{equation*}
\mathcal{A} = \text{Tr} \left[ \left\{ Y ' _R, Y ' _R \right\} Y _R \right] - \text{Tr} \left[ \left\{ Y ' _L, Y ' _L \right\} Y _L \right] \propto \sum  Y_R {Y ' _R} ^2 - \sum Y_L {Y ' _L} ^2
\end{equation*}
\begin{equation*}
\mathcal{A} \propto \left[ \left( -2 \right) e ^2 + 3 \left( \frac{4}{3} \right) u ^2 + 3 \left( - \frac{2}{3} \right) d ^2 \right] - \left[ 2 \left( -1 \right)  l ^2 + 3 \cdot 2 \left( \frac{1}{3} \right) q ^2 \right] = 0.
\end{equation*}
Therefore, 
\begin{equation}
- e ^2 + 2 u ^2 - d ^2 + l ^2 - q ^2 = 0.
\label{anomalycond4}
\end{equation}

\item[\text{$ \left[ U(1)_X \right] ^3 $}]: 
\begin{equation*}
\mathcal{A} = \text{Tr} \left[ \left\{ Y ' _R, Y ' _R \right\} Y ' _R \right] - \text{Tr} \left[ \left\{ Y ' _L, Y ' _L \right\} Y ' _L \right] \propto \sum {Y ' _R} ^3 - \sum {Y ' _L} ^3
\end{equation*}
\begin{equation*}
\mathcal{A} \propto \left[ e ^3 + 3 u ^3 + 3 d ^3 \right] - \left[ 2 l ^3 + 3 \cdot 2 q ^3 \right] = 0.
\end{equation*}
Therefore,
\begin{equation}
e ^3 + 3 u ^3 + 3 d ^3 - 2 l ^3 - 6 q ^3 = 0.
\label{anomalycond5}
\end{equation}

\end{description}

\section{Gauge bosons}
\label{sec:app2}
 
We will now derive the physical gauge boson spectrum, first of all let us write the covariant derivative Eq.\ \eqref{Dcovdiagonal} in terms of $\epsilon$ as 
\begin{equation}
\label{der_cov_u1_diag}
D_\mu = \partial _\mu + ig T^a W_\mu ^a + ig ' \frac{Q_Y}{2} B _{\mu} + \frac{i}{2} \left( g ' \frac{\epsilon Q_{Y}}{\cos \theta _W} + g_X Q_X \right) X_\mu ,
\end{equation}
or explicitly,
\begin{equation}
D_\mu = \partial _\mu + \frac{i}{2} \begin{pmatrix} g W_\mu ^3 + g ' Q_{Y} B_\mu + G_X X_\mu & g \sqrt{2} W_\mu ^+ \\ g \sqrt{2} W_\mu ^- & - g W_\mu ^3 + g ' Q_{Y} B_\mu + G_X X_\mu \end{pmatrix} ,
\end{equation}
where we defined for simplicity 
\begin{equation}
\label{GXeq}
G_{X_i} = \dfrac{g ' \epsilon Q_{Y_i}}{\cos \theta _W} + g_X Q_{X_i}  
\end{equation}with $Q_{Y_i}$ being the hypercharge of the scalar doublet under $SU(2)_L$, which in the 2HDM is taken to equal to $+1$ for both scalar doublets;  $Q_{X_i}$ is charge of the scalar doublet $i$ under $U(1)_X$.\\

We will use $D_\mu \Phi _i$ to refer to the action of the covariant derivative on the $i$ scalar doublet of $Y=1$ ($i=1,2$). Disregarding the term $\partial _{\mu}$ we have
\begin{equation}
D_\mu \Phi _i = \frac{i}{2 \sqrt{2}} \begin{pmatrix} g W_\mu ^3 + g ' B_\mu + G_{Xi} X_\mu & g \sqrt{2} W_\mu ^+ \\ g \sqrt{2} W_\mu ^- & - g W_\mu ^3 + g ' B_\mu + G_{Xi} X_\mu \end{pmatrix} \begin{pmatrix} 0 \\ v_i \end{pmatrix} ,
\end{equation}
\begin{equation}
D_\mu \Phi _i = \frac{i}{2 \sqrt{2}} v_i \begin{pmatrix} \sqrt{2} g W_\mu ^+ \\ - g W_\mu ^3 + g ' B_\mu + G_{Xi} X_\mu \end{pmatrix} . 
\end{equation}
Consequently, 
\begin{eqnarray}
\left( D_\mu \Phi _i \right) ^\dagger \left( D^\mu \Phi _i \right) &= \frac{1}{4} v_i ^2 g^2 W_\mu ^- W ^{+ \mu} + \frac{1}{8} v_i ^2 \left[ g^2 W_\mu^3 W^{3 \mu} + g^{'2} B_\mu B ^\mu + G_{Xi}^2 X_\mu X ^\mu \right]\nonumber \\
&+ \frac{1}{8} v_i ^2 \left[- 2 g g' W_\mu^3 B^\mu - 2 g G_{Xi} W_\mu^3 X^\mu + 2 g' G_{Xi}  B _\mu X^\mu \right] .
\end{eqnarray}
Carrying out the electroweak rotation as usual, 
\begin{eqnarray}
B_\mu & =  \cos \theta _W A_\mu - \sin \theta _W Z_\mu ^0  \nonumber\\
W_\mu ^3 & =  \sin \theta _W A_\mu + \cos \theta _W Z_\mu ^0 ,
\label{ewrotation}
\end{eqnarray}
we obtain 
\begin{equation}
\begin{split}
\left( D_\mu \Phi _i \right) ^\dagger \left( D^\mu \Phi _i \right) &= \frac{1}{4} v_i ^2 g^2 W_\mu ^- W ^{+ \mu} + \frac{1}{8} v_i ^2 \left[ g_Z ^2 Z_\mu ^0 Z^{0 \mu} + G_{Xi} ^2 X_\mu X ^\mu - 2 g_Z G_{Xi} Z_\mu ^0 X ^\mu \right] ,
\end{split}
\end{equation}
where $g_Z^2 = g^2 + g^{'2} = g^2 / \cos ^2 \theta _W$. As we can see, after the rotation Eq.\ \eqref{ewrotation} the field $A_\mu$ identified as the photon is massless, as it must be.

For the singlet $\Phi _S$ (with $ Q_{Y} = 0$ and $ T^a = 0 $ and disregarding the $\partial _\mu$ term) we obtain 
\begin{equation}
\begin{split}
D_\mu \Phi _S = \frac{i}{2 \sqrt{2}} v_s g_X q_X X_\mu ,
\end{split}
\end{equation}
so that
\begin{equation}
\left( D_\mu \Phi _S \right) ^\dagger \left( D^\mu \Phi _S \right) = \frac{1}{8} v_s ^2 g_X ^2 q_X ^2 X_\mu X^\mu .
\label{covdersingleto}
\end{equation}
Notice from Eq.\ \eqref{covdersingleto} that the singlet only contributes to the $U(1)_X$ gauge boson mass. Then: 
\begin{equation}
\begin{split}
\mathcal{L} _{\text{mass}} =& \left( D_\mu \Phi _1 \right) ^\dagger \left( D^\mu \Phi _1 \right) + \left( D_\mu \Phi _2 \right) ^\dagger \left( D^\mu \Phi _2 \right) + \left( D_\mu \Phi _S \right) ^\dagger \left( D^\mu \Phi _S \right)\\
 =& \frac{1}{4} g^2 v ^2 W_\mu ^- W ^{+ \mu} + \frac{1}{8} g_Z ^2 v ^2 Z_\mu ^0 Z^{0 \mu} - \frac{1}{4} g_Z \left( G_{X1} v_1 ^2 + G_{X2} v_2 ^2 \right) Z_\mu ^0 X ^\mu  \\
&+ \frac{1}{8} \left( v_1 ^2 G_{X1} ^2 + v_2 ^2 G_{X2} ^2 + v_s ^2 g_X ^2 q_X ^2 \right) X_\mu X ^\mu ,
\end{split}
\label{mixinggaugebosons}
\end{equation}
where $v ^2 = v_1 ^2 + v_2 ^2$. Finally Eq.\ \eqref{mixinggaugebosons} can be written as
\begin{equation}
\begin{split}
\mathcal{L} _{mass} &= m_W ^2 W_\mu ^- W ^{+ \mu} + \frac{1}{2} m_{Z^0} ^2 Z_\mu ^0 Z^{0 \mu} - \Delta ^2 Z_\mu ^0 X ^\mu + \frac{1}{2} m_X ^2 X_\mu X ^\mu,
\end{split}
\end{equation}
where
\begin{equation}
m_W ^2 = \frac{1}{4} g^2 v ^2,
\end{equation}
\begin{equation}
m_{Z^0} ^2 = \frac{1}{4} g_Z ^2 v ^2,
\label{Eq:MZ0}
\end{equation}
\begin{equation}
\Delta ^2 = \frac{1}{4} g_Z \left( G_{X1} v_1 ^2 + G_{X2} v_2 ^2 \right),
\label{Eq:Delta2}
\end{equation}

\begin{equation}
m_X ^2 = \frac{1}{4} \left( v_1 ^2 G_{X1} ^2 + v_2 ^2 G_{X2} ^2 + v_s ^2 g_X ^2 q_X ^2 \right) .
\label{Eq:MX}
\end{equation}

Summarizing, after the symmetry breaking pattern of this model we realize that there is a remaining mixing between $Z^0 _\mu$ and $X_\mu$, that may expressed through the matrix
\begin{equation}
m_{Z^0X} ^2 = \frac{1}{2} \begin{pmatrix} m_{Z^0} ^2 & - \Delta ^2 \\ - \Delta ^2 & m_X ^2 \end{pmatrix}, 
\end{equation}
or explicitly
\begin{equation}
m_{Z^0X} ^2 = \frac{1}{8} \begin{pmatrix} g_Z ^2 v ^2 & - g_Z \left( G_{X1} v_1 ^2 + G_{X2} v_2 ^2 \right) \\ - g_Z \left( G_{X1} v_1 ^2 + G_{X2} v_2 ^2 \right) & v_1 ^2 G_{X1} ^2 + v_2 ^2 G_{X2} ^2 + v_s ^2 g_X ^2 q_X ^2 \end{pmatrix}
\label{mixinzx}
\end{equation}
The above expression Eq.\ \eqref{mixinzx} for the mixing between the $Z^0 _\mu$ and $X_\mu$ bosons is given as function of arbitrary $U(1)_{X}$ charges of doublets/singlet scalars. It is important to note that when $Q_{X1}=Q_{X2}$, and there is not singlet contribution, the determinant of the matrix Eq.\eqref{mixinzx} is zero.

Eq.\ \eqref{mixinzx} is diagonalized through a rotation $O(\xi)$
\begin{equation}
\label{rotacao_zz_fisicos}
\begin{pmatrix} Z_\mu \\ Z ' _\mu \end{pmatrix} = \begin{pmatrix} \cos \xi & - \sin \xi \\ \sin \xi & \cos \xi \end{pmatrix} \begin{pmatrix} Z^0 _\mu \\ X_\mu \end{pmatrix},
\end{equation}
and its eigenvalues are:
\begin{equation}
\begin{split}
\label{autovalores_matriz_zz}
m_{Z} ^2 &= \frac{1}{2} \left[ m_{Z ^0} ^2 + m_X ^2 + \sqrt{ \left( m_{Z ^0} ^2 - m_X^2 \right) ^2 + 4 \left( \Delta ^2 \right) ^2} \right] \\
m_{Z '} ^2 &= \frac{1}{2} \left[ m_{Z ^0} ^2 + m_X ^2 - \sqrt{ \left( m_{Z ^0} ^2 - m_X^2 \right) ^2 + 4 \left( \Delta ^2 \right) ^2} \right] .
\end{split}
\end{equation}
The $\xi$ angle is given by
\begin{equation}
\tan 2\xi = \frac{2 \Delta ^2}{m^2 _{Z^0} - m^2 _{X}} .
\label{Eq:xi}
\end{equation}
The expressions for the gauge boson masses above are general but not very intuitive. We will simplify these equations by working in the limit in which the mass mixing is small and the $Z^{\prime}$ is much lighter than the $Z$ boson. That said, we can find a reduced formula for the masses as follows 
\begin{equation*}
\begin{split}
m_{Z} ^2 &\simeq  \frac{1}{2} \left[ m_{Z ^0} ^2 + \sqrt{ m_{Z ^0} ^4 + 4 \left( \Delta ^2 \right) ^2} \right] \simeq  \frac{1}{2} \left[ m_{Z ^0} ^2 + m_{Z ^0} ^2 \right]. 
\end{split}
\end{equation*}
In this case: 
\begin{equation}
m_{Z} ^2 \simeq  m_{Z ^0} ^2 = \frac{1}{4} g_Z ^2 v ^2,
\label{Eq:MZ}
\end{equation} {\bf being $g_Z=\frac{g}{\cos \theta_W}$.}
Similarly for the $Z '$ one finds 
\begin{equation}
\begin{split}
m_{Z '} ^2 &= \frac{1}{2} \left[ m_{Z ^0} ^2 + m_X ^2 - \sqrt{ \left( m_{Z ^0} ^2 - m_X^2 \right) ^2 + 4 \left( \Delta ^2 \right) ^2} \right] \\
&= \frac{1}{2} \left\{ m_{Z ^0} ^2 + m_X ^2 - \left( m_{Z ^0} ^2 - m_X^2 \right) \left[ 1 + \frac{4 \left( \Delta ^2 \right) ^2}{\left( m_{Z ^0} ^2 - m_X^2 \right) ^2} \right] ^{\frac{1}{2}} \right\} \\
&\simeq \frac{1}{2} \left\{ m_{Z ^0} ^2 + m_X ^2 - \left( m_{Z ^0} ^2 - m_X^2 \right) \left[ 1 + \frac{2 \left( \Delta ^2 \right) ^2}{\left( m_{Z ^0} ^2 - m_X^2 \right) ^2} \right] \right\} \\
&\simeq \frac{1}{2} \left[ m_{Z ^0} ^2 + m_X ^2 - m_{Z ^0} ^2 + m_X^2 - \frac{2 \left( \Delta ^2 \right) ^2}{m_{Z ^0} ^2} \right] \\
&\simeq m_X ^2 - \frac{\left( \Delta ^2 \right) ^2}{m_{Z ^0} ^2} ,
\end{split}
\label{uuu}
\end{equation}
We may also further simplify Eq.\ \eqref{uuu} by working out explicitly $\Delta$ in the small-mixing regime of interest. The mixing angle must satisfy $\xi \ll 1$ by the measurements of LEP experiment, i.e. 
\begin{equation}
\tan 2\xi \simeq  \sin 2\xi \simeq  2\xi 
\end{equation}
with which one gets 
\begin{equation}
\xi \simeq \frac{\Delta ^2}{m^2 _{Z^0} - m^2 _{X}} .
\end{equation}
For the case $m^2 _{Z^0} \gg  m^2 _{X}$ we find
\begin{equation}
\label{xi_delta}
\xi \simeq \frac{\Delta ^2}{m^2 _{Z^0}}=\dfrac{1}{g_{z}}(G_{X1}\cos^{2}\beta+G_{X2}\sin^{2}\beta).
\end{equation}
Substituting the Eq.\ \eqref{GXeq} into Eq.\ \eqref{xi_delta} we obtain
\begin{equation}
\xi \simeq \frac{1}{g_Z} \left[ \left(\frac{g^\prime \epsilon Q_{Y_1}}{\cos\theta_W} + g_X Q_{X1} \right) \cos^2\beta  + \left(\frac{g^\prime \epsilon Q_{Y_2}}{\cos\theta_W} + g_X Q_{X2} \right) \sin^2\beta \right].
\end{equation}
which simplifies to
\begin{equation}
\xi \simeq \frac{1}{g_Z} \left[  \left( g_X Q_{X1}\cos^2\beta + g_X Q_{X2}\sin^2\beta \right) + \left(\frac{g^\prime \epsilon Q_{Y_1}}{\cos\theta_W}  \cos^2\beta + \frac{g^\prime \epsilon Q_{Y_2}}{\cos\theta_W} \sin^2\beta \right)\right].
\label{xiEq2}
\end{equation}
Since both Higgs doublets have the same hypercharge equal to $+1$, $g^\prime=e/\sin\theta_W$ and $g=e/\cos\theta_W$, we further reduce Eq.\ \eqref{xiEq2} to
\begin{equation}
\xi \simeq \frac{\Delta ^2}{m^2 _{Z^0}}=\dfrac{g_{X}}{g_{Z}}(Q_{X1}\cos^{2}\beta+Q_{X2}\sin^{2}\beta) + \epsilon \tan \theta _W ,
\label{xisimplifiedEq}
\end{equation} 
which can also be written as
\begin{equation}
\xi = \epsilon_Z + \epsilon \tan\theta_W
\end{equation}
where
\begin{equation}
\epsilon_Z \equiv \dfrac{g_{X}}{g_{Z}}(Q_{X1}\cos^{2}\beta+Q_{X2}\sin^{2}\beta).
\label{Eq:epsilonZ}
\end{equation}
Eq.\ \eqref{xisimplifiedEq} is the general expression for the mass-mixing between the $Z$ boson and the $Z^{\prime}$ stemming from an arbritarry $U(1)_X$ symmetry in the limit $m_{Z^{\prime}} \ll m_Z$. 

In particular, for the $B-L$ case it is straightforward to prove that Eq.\ \eqref{xisimplifiedEq} becomes
\begin{equation}
\label{angulo_xi1}
\xi \simeq \dfrac{\Delta^2}{m^2 _{Z^0}} \simeq 2 \frac{g_X}{g_Z} \cos ^2 \beta + \epsilon \tan \theta _W =\epsilon _Z+ \epsilon \tan \theta _W,
\end{equation}
where 
\begin{equation}
\label{epsilon_z}
\epsilon _Z = 2 \frac{g_X}{g_Z} \cos ^2 \beta,
\end{equation}
in agreement with \cite{Lee:2013fda}. 
The parameter $\epsilon_Z$ appears often throughout the manuscript via its connection to the $\xi$ in Eq.\ \eqref{xisimplifiedEq}. 

Anyways, with Eq.\ \eqref{xisimplifiedEq} we can obtain the general expression for the $Z^{\prime}$ mass. To do so, we need a few ingredients. Firstly, notice that
\begin{eqnarray}
\frac{\Delta ^4}{m^2 _{Z}}  = && \frac{g_X^2 v^2}{4} Q_{X1} \cos^2\beta (1-\sin^2\beta)+ \frac{g_X^2 v^2}{2} Q_{X1} Q_{X2} \cos^2\beta\sin^2\beta \nonumber\\
&& + \frac{g_X^2 v^2}{4} Q_{X2}^2 \sin^2\beta (1-\cos^2\beta) + \frac{ g_Z^2 v^2 \epsilon^2}{4}\tan^2\theta_W \nonumber\\
&& \frac{g_X g_Z v^2}{2} ( Q_{X1} \cos^2\beta + Q_{X2} \sin^2\beta) \epsilon \tan\theta_W,
\label{Eq:Deltaexpanded}
\end{eqnarray}
with $m_Z^2$ defined in Eq.\ \eqref{Eq:MZ0}. 
Secondly, expanding Eq.\ \eqref{Eq:MX} we get
\begin{equation}
m_X^2=\frac{1}{4}\left[ v_1^2 \left(g_X Q_{X1}+ g_Z \epsilon \tan\theta_W Q_{Y1} \right)^2 + v_2^2 \left(g_X Q_{X2} +g_Z \epsilon \tan\theta_W Q_{Y2} \right)^2 + v_s^2 g_X^2 q_X^2 \right]
\end{equation}
which simplifies to
\begin{equation}
\begin{split}
m_X^2= & \frac{g_Z^2 \epsilon^2 \tan^2\theta_W v^2}{4} +\frac{g_X^2}{4}( Q_{X1}^2 v_1^2 + Q_{X2}^2 v_2^2)\\ 
& + \frac{g_X g_Z \epsilon \tan\theta_W}{2} (Q_{X1} v_1^2 + Q_{X2} v_2^2 )+ \frac{v_S^2 g_X^2 q_X^2}{4}.
\label{Eq:MX2simplified}
\end{split}
\end{equation}
Now substituting  Eq.\ (\ref{Eq:Deltaexpanded}) and Eq.\ \eqref{Eq:MX2simplified} into Eq.\ \eqref{uuu} we find
\begin{equation}
\begin{split}
m_{Z^\prime}^2 = & \frac{v_s^2}{4} g_X^2 q_X^2 + \frac{g_X^2 v^2}{4} Q_{X1}^2 \sin^2\beta \cos^2\beta + \frac{g_X^2 v^2}{4} Q_{X2}^2\cos^2\beta\sin^2\beta \\ 
& -\frac{g_X^2 v^2}{2} Q_{X1} Q_{X2} \cos^2\beta \sin^2\beta
\end{split}
\end{equation}
which reduces to
\begin{eqnarray}
m_{Z^\prime}^2=  \frac{v_s^2}{4} g_X^2 q_X^2 + \frac{g_X^2 v^2 \cos^2\beta \sin^2\beta}{4}(Q_{X1} - Q_{X2})^2.
\label{Eq:MZprimegeneral}
\end{eqnarray}
We emphasize that $q_X$, $Q_{X1}$, $Q_{X2}$ are the charges under $U(1)_X$ of the singlet scalar, Higgs doublets $\Phi_1$ and $\Phi_2$ respectively, $\tan\beta=v_2/v_1$, $v=246$~GeV, $v_s$ sets the $U(1)_X$ scale of spontaneous symmetry breaking, and $g_X$ is the coupling constant of the $U(1)_X$ symmetry. Eq.\ \eqref{Eq:MZprimegeneral} accounts for the $Z^{\prime}$ mass for every single $U(1)_X$ models studied in this work. 

A few remarks are in order:
\begin{itemize}
\item[(i)] The $Z^{\prime}$ mass is controlled by $g_X$. Thus in order to achieve $m_{Z^{\prime}} \ll m_Z$ one needs to sufficiently suppress this coupling. 

\item[(ii)] The $Z^{\prime}$ mass is generated via spontaneous symmetry breaking and for this reason it depends on the $v_s$  which sets the $U(1)_X$ breaking and $v$ due to the $Z-Z^{\prime}$ mass mixing.

\item[(iii)] The $Z^{\prime}$ mass as expected depends on the $U(1)_X$ charges of the scalar doublets and the singlet scalar since they all enter into the covariant derivative of the respective scalar field from which the $Z$ and $Z^{\prime}$ are obtained. 

\item[(iv)]  If $(Q_{X1}- Q_{X2})^2$ is not much larger than four as occurs for many $U(1)_X$ models in Table \ref{cargas_u1_2hdm_tipoI}, then $m_{Z^{\prime}}$ is approximately
\begin{equation}
m_{Z^{\prime}}^2 = \frac{v_{s}^2}{4}\, g_{X}^2 q_X^2.
\label{mZprimeapproximate}
\end{equation}
For instance, in the $B-L$ model, $Q_{X1}=2,q_X=2, Q_{X2}=0$, implying that 
\begin{equation}
 B-L: m^{2}_{Z^{\prime}} = v^{2}_{s}g_{X}^{2} + g_{X}^{2}v^{2}\cos^{2} \beta\sin^{2} \beta.
\label{mZprimeBL}
\end{equation}
\end{itemize}
Setting $v_s= 1$~TeV, we need $g_X=10^{-3} - 10^{-6}$ to achieve $m_{Z^{\prime}} = 1\, {\rm MeV}-1\, {\rm GeV}$.  Notice that this small coupling constant is a feature common to all dark photon-like models such as ours.

\section{ $\delta$ Parameter}
\label{sec:app3}
Defining $\tan \beta_{d} = \dfrac{v_s}{v_1}$, we can write $m_{Z '}$ from \eqref{Eq:MZprimegeneral} as:
\begin{equation}
\begin{aligned}
m_{Z^\prime}^2
&=  \frac{g_X^2 v^2 \cos^2\beta\left[\sin^2\beta(Q_{X1} - Q_{X2})^2+\tan^2\beta_d  q_X^2\right]}{4},\\
&=  \frac{g_X^2 v^2 \cos^2\beta\left[q_X^2 +\cos^2\beta_d\left(\sin^2\beta(Q_{X1} - Q_{X2})^2-q_X^2\right)\right]}{4\cos^2\beta_d},
\end{aligned}
\end{equation}

\begin{equation}
\begin{aligned}
\Rightarrow m_{Z^\prime}
&=  g_X v \cos^2\beta\dfrac{\sqrt{\left[q_X^2 +\cos^2\beta_d\left(\sin^2\beta(Q_{X1} - Q_{X2})^2-q_X^2\right)\right]}}{2\cos\beta\cos\beta_d},
\\
&=  \dfrac{g_X v \cos^2\beta}{\delta},
\end{aligned}
\end{equation}
with 
\begin{equation}
\delta = \dfrac{2\cos \beta \cos \beta_d}{\sqrt{\left[q_X^2 +\cos^2\beta_d\left(\sin^2\beta(Q_{X1} - Q_{X2})^2-q_X^2\right)\right]}}.
\label{deltinhageral}
\end{equation}
Even in this general scenario we realize that there is a relation among the masses of the neutral gauge bosons and $\epsilon_{Z}$ from Eq.\ (\ref{Eq:epsilonZ}): 
\begin{equation}
\delta=\dfrac{m_Z}{m_{Z^{\prime}}}\epsilon_{Z}.
\label{deltaepsilonz}
\end{equation}
In the $B-L$ model where $Q_{X1}=2,q_X=2, Q_{X2}=0$, $\delta$ from Eq.\ \eqref{deltinhageral} is reduced to
\begin{equation}
\delta = \dfrac{\cos \beta \cos \beta_d}{\sqrt{1-\cos^2 \beta \cos^2 \beta_d}},
\label{deltinha}
\end{equation}
and Eq.\ \eqref{deltaepsilonz} is reproduced for the $m_{Z'}$ and $\delta$ values given in equations Eq.\ \eqref{mZprimeBL} and Eq.\ \eqref{deltinha}, respectively. In the doublets only case, $v_{s}=0$, $\cos \beta_{d}=1$ and consequently \eqref{deltinha} becomes 
\begin{equation}
\delta \tan \beta =1.
\end{equation}
On the other hand, in the limit $v_{2} \gg v_{1}$ and $v_{s} \gg v_{1}$: 
\begin{eqnarray}
\delta \simeq & \cos \beta \cos \beta_{d} 
\simeq  \dfrac{1}{\tan \beta \tan \beta_{d}}.
\end{eqnarray}

\section{Currents for $Z$ and $Z^{\prime}$}
\label{sec:app4}

In this section we will derive the generalized interactions among fermions and gauge bosons from the following Lagrangian:
\begin{equation}
\mathcal{L} _{\text{fermion}} = \sum _{\text{fermions}} \bar{\Psi} ^L i \gamma ^\mu D_\mu \Psi ^L + \bar{\Psi} ^R i \gamma ^\mu D_\mu \Psi ^R .
\label{ldiracu1}
\end{equation}
After the electroweak rotation \eqref{ewrotation} the covariant derivative Eq.\  \eqref{Dcovdiagonal} for neutral gauge bosons becomes (the charged interactions are the same as those of the SM):
\begin{equation}
\begin{split}
D_\mu ^L = & ig T^3 \left( \sin \theta _W A_\mu + \cos \theta _W Z_\mu ^0 \right) + ig ' \frac{Q_Y}{2} \left( \cos \theta _W A_\mu - \sin \theta _W Z_\mu ^0 \right) \\
& + \frac{i}{2} \left( g ' Q_Y \frac{\epsilon}{\cos \theta _W} + g_X Q_X \right)X_{\mu}. 
\end{split}
\end{equation}
In Appendix \ref{sec:app2} we have demonstrated that after the final SSB process a mixing between $Z_\mu ^{0}$ and $X_{\mu}$ remains, and this is the origin of $\delta$. Replacing $Z^{0}_{\mu}$ and $X_{\mu}$ as function of the physical bosons $Z_{\mu}$ and $Z^{\prime}_{\mu}$, Eq.\ \eqref{rotacao_zz_fisicos}, we obtain
\begin{equation}
\begin{split}
\bar{\Psi} ^L i \gamma ^\mu D_\mu ^L \Psi ^L = & - e Q_f \bar{\psi} _f ^L \gamma ^\mu \psi _f ^L A_\mu \\
&- \left[ g_Z \left( T^L _{3f} - Q_f \sin ^2 \theta _W \right) \cos \xi - \frac{1}{2} \left( \epsilon g _Z Q_{Yf} ^L \tan \theta _W + g_X Q_{Xf} ^L \right) \sin \xi \right] \bar{\psi} _f ^L \gamma ^\mu \psi _f ^L Z_\mu \\
&- \left[ g_Z \left( T^L _{3f} - Q_f \sin ^2 \theta _W \right) \sin \xi + \frac{1}{2} \left( \epsilon g _Z Q_{Yf} ^L \tan \theta _W + g_X Q_{Xf} ^L \right)\cos \xi \right] \bar{\psi} _f ^L \gamma ^\mu \psi _f ^L Z' _\mu,
\end{split}
\label{diracLu1}
\end{equation}
where the relations $g \sin \theta _W = g' \cos \theta _W = e$, $g_Z = g / \cos \theta _W$, $g ' = g_Z \sin \theta _W$ and $T^3 + Q_Y / 2 = Q_f$ have been used. For the right-handed fields it suffices to replace $T^L _{3f}$ for $T^R _{3f}=0$, in which case:
\begin{equation}
\begin{split}
\bar{\Psi} ^R i \gamma ^\mu D_\mu ^R \Psi ^R = & - e Q_f \bar{\psi} _f ^R \gamma ^\mu \psi _f ^R A_\mu\\
&- \left[ - g_Z Q_f \sin ^2 \theta _W \cos \xi - \frac{1}{2} \left( \epsilon g _Z Q_{Yf} ^R \tan \theta _W + g_X Q_{Xf} ^R \right) \sin \xi \right] \bar{\psi} _f ^R \gamma ^\mu \psi _f ^R Z_\mu \\
&- \left[ - g_Z Q_f \sin ^2 \theta _W \sin \xi + \frac{1}{2} \left( \epsilon g _Z Q_{Yf} ^R \tan \theta _W + g_X Q_{Xf} ^R \right)\cos \xi \right] \bar{\psi} _f ^R \gamma ^\mu \psi _f ^R Z' _\mu.
\end{split}
\label{diracRu1}
\end{equation}
The generalized interactions among fermions and gauge bosons, Eq.\ \eqref{ldiracu1}, is the sum of the contributions \eqref{diracLu1} and \eqref{diracRu1}, and can be written as follows:
\begin{equation}
\begin{split}
\mathcal{L} _{\text{fermion}} = &- e Q_f \bar{\psi} _f \gamma ^\mu \psi _f A_\mu \\
&- \left[ g_Z \left( T_{3f} - Q_f \sin ^2 \theta _W \right) \cos \xi - \frac{1}{2} \epsilon g _Z Q_{Yf} ^L \tan \theta _W \sin \xi \right] \bar{\psi} _f ^L \gamma ^\mu \psi _f ^L Z_\mu \\
&- \left[ - g_Z Q_f \sin ^2 \theta _W \cos \xi - \frac{1}{2} \epsilon g _Z Q_{Yf} ^R \tan \theta _W \sin \xi \right] \bar{\psi} _f ^R \gamma ^\mu \psi _f ^R Z_\mu \\
&- \left[ g_Z \left( T_{3f} - Q_f \sin ^2 \theta _W \right) \sin \xi + \frac{1}{2} \epsilon g _Z Q_{Yf} ^L \tan \theta _W \cos \xi \right] \bar{\psi} _f ^L \gamma ^\mu \psi _f ^L Z' _\mu \\
&- \left[ - g_Z Q_f \sin ^2 \theta _W \sin \xi + \frac{1}{2} \epsilon g _Z Q_{Yf} ^R \tan \theta _W \cos \xi \right] \bar{\psi} _f ^R \gamma ^\mu \psi _f ^R Z' _\mu \\
&+ \frac{1}{2} g_X Q_{Xf} ^L \sin \xi \bar{\psi} _f ^L \gamma ^\mu \psi _f ^L Z_\mu + \frac{1}{2} g_X Q_{Xf} ^R \sin \xi \bar{\psi} _f ^R \gamma ^\mu \psi _f ^R Z_\mu - \frac{1}{2} g_X Q_{Xf} ^L \cos \xi \bar{\psi} _f ^L \gamma ^\mu \psi _f ^L Z' _\mu \\
&- \frac{1}{2} g_X Q_{Xf} ^R \cos \xi \bar{\psi} _f ^R \gamma ^\mu \psi _f ^R Z' _\mu. 
\end{split}
\label{generaldiracu1}
\end{equation}
The last two lines of \eqref{generaldiracu1} are the contributions introduced when the charges of the fermions under $U(1)_{X}$ are non-zero. In Appendix \ref{sec:app5} we  derive explicitly the neutral currents of both $Z$ and Dark $Z$ bosons of reference \cite{Lee:2013fda}  ($Q^{L,R}_{X}=0$ case). After that Eq.\ \eqref{generaldiracu1} can be written as
\begin{equation}
\begin{split}
\mathcal{L} = &- e J ^\mu _{em} A_\mu - \dfrac{g_Z}{2} J ^\mu _{NC} Z_\mu - \left( \epsilon e J^\mu _{em} + \dfrac{\epsilon _Z g_Z}{2} J^\mu _{NC} \right) Z' _\mu \\
&+ \frac{1}{4} g_X \sin \xi \left[ \left( Q_{Xf} ^R + Q_{Xf} ^L \right) \bar{\psi} _f \gamma ^\mu \psi _f + \left( Q_{Xf} ^R - Q_{Xf} ^L \right) \bar{\psi} _f \gamma ^\mu \gamma _5 \psi _f \right] Z_\mu \\
&- \frac{1}{4} g_X \cos \xi \left[ \left( Q_{Xf} ^R + Q_{Xf} ^L \right) \bar{\psi} _f \gamma ^\mu \psi _f - \left( Q_{Xf} ^L - Q_{Xf} ^R \right) \bar{\psi} _f \gamma ^\mu \gamma _5 \psi _f \right] Z' _\mu.
\end{split}
\label{www}
\end{equation}
Eq.\ \eqref{www} is the general neutral current for all $U(1)_X$ models studied in this work.  Since we are interested in the regime in which the mixing angle is much smaller than one, $\xi \ll 1$,  and $g_X \ll 1$, then $Z$ properties will be kept unmodified.  

For concreteness, we shall obtain again the neutral current for a well-known model, such as the $U(1)_{B-L}$ model. In this case, we find 
\begin{equation}
\begin{split}
\mathcal{L} = &- e J ^\mu _{em} A_\mu - \dfrac{g_Z}{2} J ^\mu _{NC} Z_\mu - \left( \epsilon e J^\mu _{em} + \dfrac{\epsilon _Z g_Z}{2} J^\mu _{NC} \right) Z' _\mu \\
&- \dfrac{\epsilon _{Z} g_{Z}}{2} \left[ \dfrac{a}{4 \cos^{2} \beta} \bar{\psi} _f \gamma ^\mu \psi _f \right] Z' _\mu,
\end{split}
\label{zzgeralcoma}
\end{equation}
Here $a=-2$ for charged leptons and $a=2/3$ for quarks. Notice that in our case we have  a new vector coupling for $Z^{\prime}$ when compared to the $Z^{\prime}$ of the Dark 2HDM \cite{Lee:2013fda}. 

\section{Comparison with the 2HDM with Gauged $U(1)_N$}
\label{sec:app5}

It is important to cross-check our findings with the existing literature. In \cite{Lee:2013fda} a 2HDM similar to the $U(1)_N$ model in Table \ref{cargas_u1_2hdm_tipoI} was studied. Therefore, in this setup all fermions are uncharged under the $U(1)_X$ symmetry, i.e.\ $Q^{L,R}_{X}=0$. Using Eq.\ \eqref{generaldiracu1} the neutral current involving the $Z$ boson reads
\begin{equation}
\begin{split}
\mathcal{L} _Z &= - \dfrac{g_Z}{2} \cos \xi J ^\mu _{NC} Z_\mu - \epsilon g _Z \tan \theta _W \sin \xi \left[ \left( \frac{T_{3f}}{2} - Q_f \right) \bar{\psi} _f \gamma ^\mu \psi _f - \frac{T_{3f}}{2} \bar{\psi} _f \gamma ^\mu \gamma _5 \psi _f \right] Z_\mu.
\end{split}
\label{generalzu1}
\end{equation}
Since the mixing angle ($\xi$) and the kinetic mixing $(\epsilon)$ are much smaller than one, only the SM neutral current, the first term of Eq.\ \eqref{generalzu1} is left. In other words, the $Z$ properties are kept identical to the SM.

As for the neutral current of the $Z^\prime$ boson,  we get from Eq.\ \eqref{generaldiracu1} that 
\begin{equation}
\begin{split}
\mathcal{L} _{Z'} &= - g_Z \sin \xi \left[ \left( \frac{T_{3f}}{2} - Q_f \sin ^2 \theta _W \right) \bar{\psi} _f \gamma ^\mu \psi _f - \frac{T_{3f}}{2} \bar{\psi} _f \gamma ^\mu \gamma _5 \psi _f \right] Z' _\mu\\
&+ \epsilon g _Z \tan \theta _W \cos \xi \left[ \left( \frac{T_{3f}}{2} - Q_f \right) \bar{\psi} _f \gamma ^\mu \psi _f - \frac{1}{2} T_{3f} \bar{\psi} _f \gamma ^\mu \gamma _5 \psi _f \right] Z' _\mu.
\end{split}
\end{equation}
Using Eq.\ \eqref{angulo_xi1} and taking $\xi\ll 1$, we find

\begin{equation}
\mathcal{L} _{Z'} = - \epsilon e Q_f \bar{\psi} _f \gamma ^\mu \psi _f Z' _\mu - \dfrac{\epsilon _Z g_Z}{2} \left[ \left( T_{3f} - 2Q_f \sin ^2 \theta _W \right) \bar{\psi} _f \gamma ^\mu \psi _f - T_{3f} \bar{\psi} _f \gamma ^\mu \gamma _5 \psi _f \right] Z' _\mu 
\end{equation}which simplifies to 
\begin{equation}
\mathcal{L} _{Z'} = - \left( \epsilon e J^\mu _{em} + \dfrac{\epsilon _Z g_Z}{2} J^\mu _{NC} \right) Z' _\mu.
\end{equation}

Our limiting case of the $U(1)_N$ model matches the result of 
\cite{Lee:2013fda}, once again validating our findings.

\section{Higgs Interactions to Vector Bosons}
\label{sec:app6}
In this section we summarize the Higgs-gauge boson vertices under the assumption that the mixing between the Higgs doublets and the singlet scalar is suppressed. We find that 
\begin{equation}
{\cal C}_{H-Z-Z}= \frac{g_Z^2 v}{2}\cos(\beta-\alpha),
\end{equation}
\begin{equation}
{\cal C}_{H-Z-Z^\prime}= - g_Z g_X v \cos\beta \sin\beta \sin(\beta-\alpha),
\end{equation}
\begin{equation}
{\cal C}_{H-Z^\prime-Z^\prime}= 2 g_X^{2} v \cos\beta \sin\beta ( \cos^3\beta \sin\alpha +\sin^3\beta \cos\alpha),
\end{equation}
\begin{equation}
{\cal C}_{h-Z-Z}= \frac{g_Z^2 v}{2}\sin(\beta-\alpha),
\end{equation}
\begin{equation}
{\cal C}_{h-Z-Z^\prime}= - g_Z g_X v \cos\beta \sin\beta \cos(\beta-\alpha),
\end{equation}
\begin{equation}
{\cal C}_{h-Z^\prime-Z^\prime}= 2 g_X^{2} v \cos\beta \sin\beta ( \cos^3\beta \sin\alpha - \sin^3\beta \cos\alpha).
\end{equation}

\bibliographystyle{JHEPfixed}
\bibliography{combined2}

\end{document}